# SILICON IN THE QUANTUM LIMIT:
# QUANTUM COMPUTING AND DECOHERENCE
# IN SILICON ARCHITECTURES

by

Charles George Tahan

A dissertation submitted in partial fulfillment of

the requirements for the degree of

Doctor of Philosophy

(Physics)

at the

UNIVERSITY OF WISCONSIN - MADISON

2005

# Acknowledgments

There are many who deserve the credit for where I am today. First and foremost are my parents, who supported and helped me with all my dreams. My thesis advisor, Robert Joynt, and the University of Wisconsin-Madison Physics Department deserve great credit for giving me a chance to prove myself as a scientist. Bob has been a great advisor. Mark Friesen, who I've worked with and bothered extensively over the last four years has been a life saver. Mark Eriksson has been a second advisor to me. The many people I've worked with and have had fun doing so deserve attention as well: Keith Slinker, Srijit Goswami, Don Savage, Max Lagally, Ryan Toonen, Hua Qin, Robert Blick, Jim Truitt, Dan van der Weide, Sue Coppersmith, Alex Rimberg, Gerhard Klimeck, and many others, including all my previous mentors and friends.

Thank you all very much.

# Abstract

The pursuit of spin and quantum entanglement-based devices in solid-state systems has become a global endeavor. The approach of the quantum size limit in computer electronics, the many recent advances in nanofabrication, and the rediscovery that information is physical (and thus based on quantum physics) have started a worldwide race to understand and control quantum systems in a coherent and useful way. Semiconductor architectures hold promise for quantum information processing (QIP) applications due to their large industrial base and perceived scalability potential. Electron spins in silicon in particular may be an excellent architecture for QIP and also for spin electronics (spintronics) applications. While the charge of an electron is easily manipulated by charged gates, the spin degree of freedom is well isolated from charge fluctuations. This leads to very good spin quantum bit (qubit) stability or quantum coherence properties. Inherently small spin-orbit coupling and the existence of a spin-zero Si isotope also facilitate long single spin coherence times. Here we consider the relaxation properties of localized electronic states in silicon due to donors, quantum wells, and quantum dots. Our analysis is impeded by the complicated, many-valley band structure of silicon and previously unaddressed physics in silicon quantum wells. We find that electron spins in silicon and especially strained silicon have excellent decoherence properties. Where possible we compare with experiment to test our theories. We go beyond issues of coherence in a quantum computer to problems of control and measurement. Precisely what makes spin relaxation so long in semiconductor architectures makes spin measurement so difficult. To address this, we propose a new scheme for spin readout which has the added benefit of automatic spin initialization, a vital component of quantum computing and quantum error correction. Our results represent important practical milestones on the way to the design and construction of a silicon-based quantum computer.

# Contents

# List of Figures

# List of Tables

# Chapter 1

# Introduction

Silicon may be the best understood material on earth. The wild success of the complementary metal oxide semiconductor (CMOS) processes have produced a manufacturing and technical infrastructure that measures as a significant percentage of the global economy. Solid state physicists, materials scientists, crystallagraphers, and electrical engineers have devoted over fifty years to growing, understanding, and extending silicon-based electronic devices. But as this thesis will show and in part address, our current prowess in silicon and related technologies comes up short when we contemplate building even a few-qubit quantum computer. This level of knowledge and control of electronic states in silicon at the kind of sensitivity needed for quantum information processing is what we refer to in the title of this work as silicon in the quantum limit.

Quantum computing began with the hopeful idea of simulating one quantum system with another, man-made quantum system. Even a system of a few hundred entangled qubits (quantum two-level systems) is thought to be impossible to simulate exactly on a classical computer. Simulation would involve dealing with exponentially large matrices, determined by the Hilbert space of the quantum system. A quantum computer, Feynman noted, could use this quantum parallelism to simulate other, more interesting quantum systems or perhaps solve difficult computational problems.[1] Only in the last decade or so has technology begun to catch up

---

[1]Most people attribute the classic talk that Richard Feynman gave on December 29th, 1959 at the annual meeting of the American Physical Society at the California Institute of Technology (Caltech), *There's plenty of room at the bottom*, with the introduction of both nanotechnology (unnamed at the time) and quantum

to Feynman's musings. Advances in nanomanufacturing and quantum optics have given us control over single, fundamental excitations of nature. From a physicist's perspective, this is very exciting. Controlling a single electron, a single photon (or phonon or plasmon for that matter) allows new avenues to test the foundations of quantum theory and new potentials for amazing devices. The field of quantum information began with just such a consideration. In the 1980s David Deutsch, Charles Bennett, Feynman, and others started thinking about how an operation or program run with quantum components might be more powerful than its classical counterpart. The discovery of a quantum algorithm that beats its classical equivalent [18] and of the quantum teleportation procedure [10] lead to the realization that quantum physics was a more natural—and more powerful—foundation for computation and communication. See Figure 1.1 for a few examples of quantum circuits. Zeros and ones were out, quantum bits—able to exist as a linear superposition of states—were in. But two specific milestones, as it turns out both discovered by the same man, made tackling the immense challenge of building a quantum computer fundable and (perhaps) worth devoting a career to. In 1991 Peter Shor developed a quantum algorithm for prime factorization [75] instantly making RSA public key encryption (which depends on the hardness of the prime factorization problem for security) vulnerable. He also answered the principle worry of quantum computer nay-sayers, error correction for delicate and unclonable [20, 91] quantum states is possible [76].

This thesis focuses on some specific technical details pertaining to one possible implementation scheme for a quantum information processor. Progress and discoveries in quantum information theory (QIT), however, give us direction in the design process. Figure 1.1 summarizes an amended list of criteria a quantum computer architecture should meet, though this list is still evolving. We will use QIT as a resource in making design decisions and pull information as we need it. The field has grown vast in the last fifteen or so years and many excellent introductory texts have been written where more detailed information can be found [60].

Many architectures have been proposed for QIP. Early success was achieved in NMR systems [13]. A 7-qubit equivalent QC was constructed and Shor's algorithm performed to factor the number 15 [85]. The qubits were encoded on nuclear spins of a designer molecule. The system proved unscalable. Atomic [14] and superconducting systems [58], as examples, also hold computing. In fact, Feynman introduced his quantum computer much later, in the 80s.

promise, where the quantum information is stored in specific and somewhat arbitrary quantum levels.

Spins in semiconductors have some advantages over other schemes for quantum information processing. Spin-1/2 particles are natural 2 level quantum systems. The spin degree of freedom is often well isolated from other degrees of freedom, leading to good quantum coherence properties. A large toolbox of magnetic resonance techniques for electrons (ESR) and nuclei (NMR) have been developed over the last century for qubit control, measurement, and interaction. These and other considerations lead Loss and DiVincenzo to propose an electron spin-based quantum dot quantum computer.[55] Quantum dots have a long history and recent demonstration of single electron control [26, 25, 43] and single-spin readout [44] in GaAs quantum dots gives promise to the use of these systems as qubit processors. Quantum dot architectures benefit from a very fast two-qubit operation, via the Heisenberg spin exchange, which can be controlled by lithographically defined top-gates (see Figure 1.2). Modern advances in nano-lithography and semiconductor processing in both silicon and GaAs may allow for quantum dot (qubit) scalability into the millions.

Silicon, especially, has attractive properties for large scale quantum dot quantum computers. Silicon has better spin coherence properties than GaAs. Silicon benefits from (1) a low bulk spin-orbit coupling and (2) the availability of a spin-0 silicon isotope (allowing for isotopic purification) which eliminates the dominant decoherence mechanism in these architectures (nuclear spectral diffusion) [17]. In addition, silicon electronics has a proven record of fast operation and scalable integration as yet unmatched in any other system. These conditions have resulted in several proposals for spin-based QC in silicon [48, 87, 32]. Ours, quantum well quantum dots in a SiGe-Si-SiGe heterostructure with lithographically defined top-gates, is outlined in Ref. [32] and Figure 1.4.

Many challenges exist for QIP in silicon quantum well quantum dots. Silicon quantum wells remain immature relative to GaAs quantum well technology. Since silicon and germanium have different lattice constants, lattice dislocations, threads, and roughness are the norm in present day devices. This not only limits mobility but has implications for exchanged-based quantum computing. Silicon has other complexities. Unlike GaAs, its principle competitor for QDQC, the conduction band of silicon does not have its minimum at the $\Gamma$ point. Instead, it has

1. A scalable physical system with well-characterized qubits.
2. The ability to initialize the state of the qubits to a simple fudicial state.
3. Long relevant decoherence times, much longer than the gate operation time.
4. A "universal" set of gates.
5. A qubit specific measurement capability.
6. *The ability to interconvert stationary and flying qubits.*
7. *The ability to faithfully transmit flying qubits between specified locations.*
8. *A constant supply of initialized qubits for error correction.*

Table 1.1: The so-called DiVincenzo criteria for an "ordinary" quantum computer. Items in italics are updates from the original list.

minima 0.85 of the way to the brillouin zone edge along the principle cubic axes. Therefore, the electron exists as a linear combination of these six minima, or valleys, in bulk silicon. This has implications from decoherence to exchange, virtually all aspects of spin-based quantum dot quantum computing in silicon QC. Figure 1.3 introduces some of silicon's complexities.

We tackle a small fraction of the challenges that are faced in building a silicon QC. We primarily devote our attention to the electronic spin and orbital relaxation of trapped electrons in silicon devices. Understanding these time scales and how they follow principle environmental parameters is vital to designing, constructing, and measuring electrons in silicon nanodevices. Despite a half decade of investigation into the behavior of electrons in silicon, the physics of spins has been remarkably neglected and now must be revisited.

Chapter 2 introduces the basics of spin decoherence and builds off historical results to calculate the spin-flip times of donors and quantum dots in strained silicon devices. In Chapter 3, we characterize the physical system of our SiGe quantum well quantum dot and estimate the spin-orbit coupling that emerges in these types of asymmetric devices, which will be vital to calculating spin relaxation in a silicon quantum dot. By calculating the spin relaxation behavior of electrons in present-day silicon 2DEGs and comparing to experiment, we can extract the needed value and compare it to our estimate. Chapter 4 directly addresses the relaxation of electronic states in a silicon quantum dot. We calculate the orbital and spin relaxation of proposed qubit devices. This is the most important chapter for future experiments in silicon quantum dot quantum computing. Finally, in Chapter 5, we propose a new readout and initialization scheme for electrons in semiconductor dots. We use the results of the previous chapters to analyze its feasibility. We end with conclusions and an outlook for future work.

## (a) Quantum Parallelism

f(x) is a binary function: $f(\{0,1\}) = \{0,1\}$

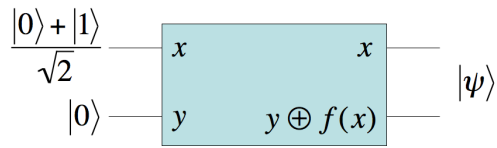

$$|\psi\rangle = \frac{|0, f(0)\rangle + |1, f(1)\rangle}{\sqrt{2}}$$

Measurement will only choose one.

## (b) Deutsch Algorithm

Ask a global question: Is the function f(x) constant or not?

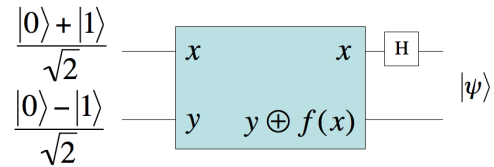

$$|\psi\rangle = \pm \left| f(0) \oplus f(1) \right\rangle \left[ \frac{|0\rangle - |1\rangle}{\sqrt{2}} \right]$$

Qubit 1 encodes the answer to the global question.

## (c) Quantum Teleportation

Want to send the state psi from alice to bob.

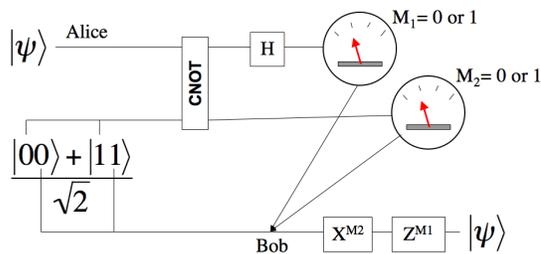

The qubit state is transferred from Alice to Bob utilizing the entanglement of the Bell state as a resource. It is not copied.

## (d) Copying

Classical copying circuit:

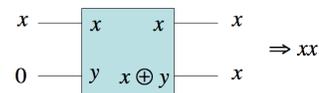

Quantum version

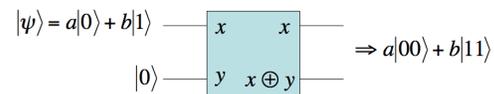

But, $|\psi\rangle|\psi\rangle = a^2|00\rangle + ab|01\rangle + ab|10\rangle + b^2|11\rangle$

Figure 1.1: Examples of basic quantum computation circuits. (a) Quantum parallelism allows a quantum computer to calculate multiple inputs to a function in one shot. (b) The Deutsch algorithm takes only one run to answer the question of whether the function is constant or not, whereas a classical computer would need two. (c) Quantum teleportation uses a maximally entangled EPR pair to send an arbitrary quantum state between two parties. (d) Quantum states cannot be copied.

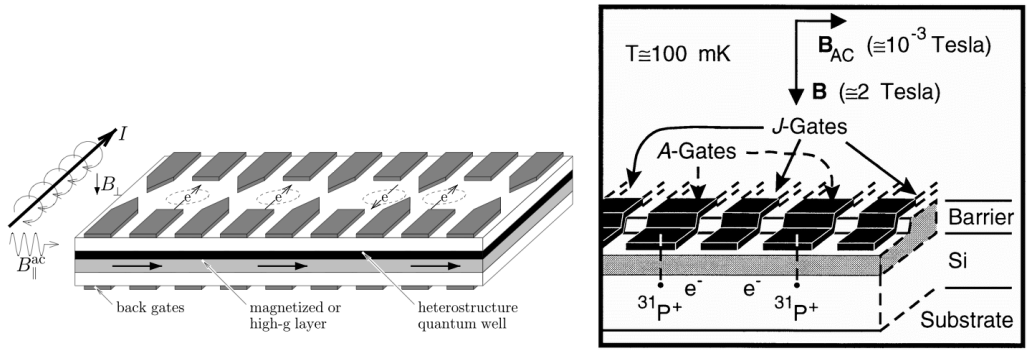

Figure 1.2: Architectures for exchange-based quantum computing with electrons. (Left) The Loss DiVincenzo quantum dot quantum computer proposal [55]. Single electrons are trapped vertically in a semiconductor quantum well heterostructure and horizontally by charge top gates. Manipulating the potential on the plunger gates in between electrons can allow for their wave functions to overlap. Because of the Pauli exclusion principle, the spins of the electron pair undergo spin exchange, $H = J\mathbf{S}_1 \cdot \mathbf{S}_2$. This two-qubit interaction together with single spin rotation has been shown to be a universal set of operations for any quantum computation. (Right) The Kane proposal for exchange-based quantum computation [48]. Electrons are trapped on P donors in a silicon matrix and are allowed to exchange via manipulation by top gates.

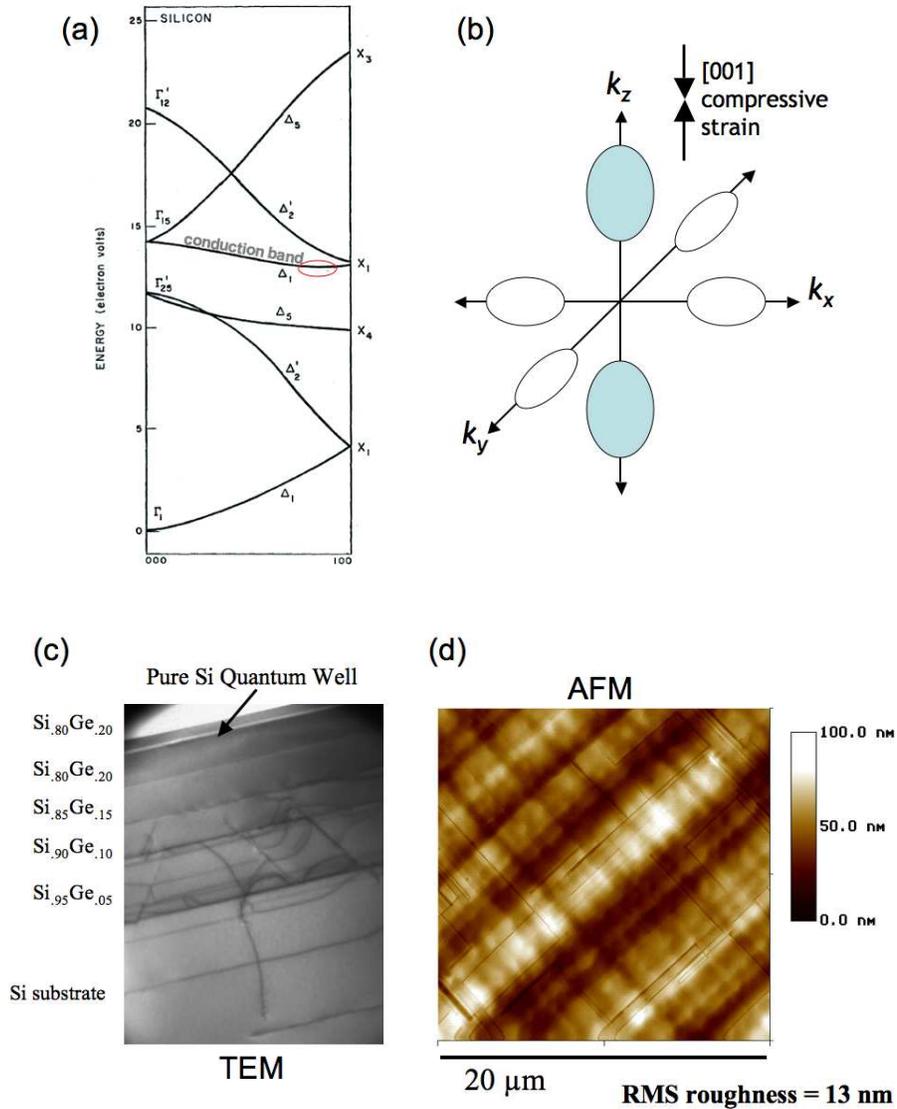

Figure 1.3: Working with silicon heterostructures. (a) Electron band diagram of bulk silicon. Along a principle axis, the conduction band has a minimum slightly before the Brillouin zone boundary. (b) The six valleys of silicon portrayed iconically. With compressive strain in the [001] direction—as in a quantum well—the electron exists increasingly in the $\pm z$ valleys, which are lowered in energy relative to the other four. (c) A modern step-graded SiGe heterostructure with a pure strained silicon quantum well. Note the threading dislocations. (d) The so-called crosshatch on the surface of SiGe-Si-SiGe quantum well heterostructure. This is fundamentally due to the lattice mismatch between Si and Ge.

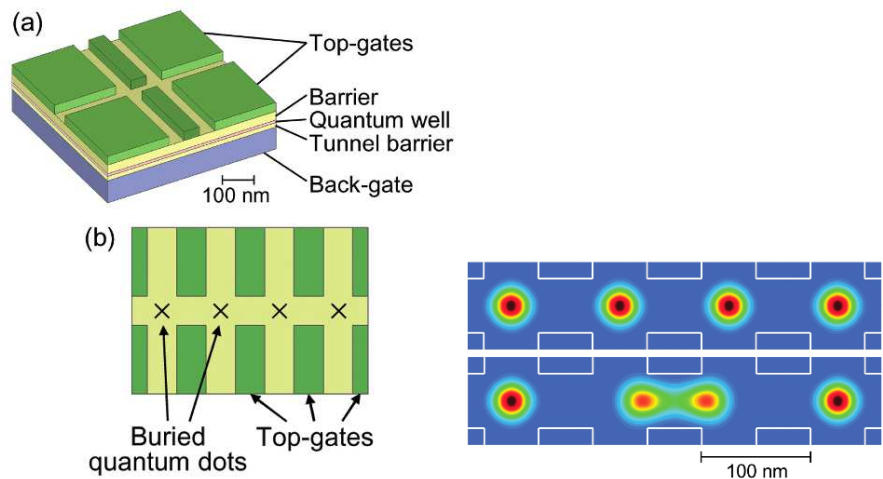

1. Single electron isolation via top gates and Si quantum well. The qubit is a single spin.

2. Two-qubit entanglement operations via Heisenberg spin exchange.

3. One-qubit operations via electron spin resonance (ac magnetic microwave pulses) or via encoded qubits and exchange.

4. Readout via spin-charge transduction and rf-SET charge detection.

5. Initialization by thermalization or optical pumping.

Figure 1.4: (Left) Silicon-Germanium architectures for quantum dot quantum computing [32]. (Right) Quantum dot electron qubits isolated (top) and undergoing an exchange operation (bottom).

# Chapter 2

# Donor spin relaxation in strained Si

Donors in semiconductors can in many ways be thought of as natural quantum dots. In silicon, decoherence mechanisms of donor spins have a long history of investigation going back some sixty years. Building on this prior theory and understanding is a good first step in our considerations of spin qubit decoherence in SiGe heterostructure quantum dots. Consider phosphorous (P) donors in silicon. With its five valence electrons, substituting a P atom in a silicon matrix leaves one electron free which, attracted to the positive nucleus of the atom, forms a single electron localized state. The physics that governs this system is very similar to that of a single electron quantum dot, our principal system of interest. Below we show how the dominant spin relaxation mechanism goes away with increasing [001] strain for both donors in silicon and for lateral silicon quantum dots.

In this chapter direct phonon spin-lattice relaxation of an electron qubit bound by a donor impurity or quantum dot heterostructure is investigated. The aim is to evaluate the importance of decoherence from this mechanism in several important solid-state quantum computer designs operating at low temperatures. We calculate the relaxation rate $1/T_1$ as a function of [100] uniaxial strain, temperature, magnetic field, and silicon/germanium content for Si:P bound electrons and quantum dots. The quantum dot potential is smoother, leading to smaller splittings of the valley degeneracies. We have estimated these splittings in order to obtain upper bounds for the relaxation rate. In general, we find that the relaxation rate is strongly decreased by uniaxial compressive strain in a SiGe-Si-SiGe quantum well, making this strain an important positive design feature. Ge concentrations (particularly over 85%) increases the rate, making Si-rich materials preferable. We conclude that SiGe bound electron qubits must meet certain conditions to minimize decoherence but that spin-relaxation does not rule out the solid-state implementation of error-tolerant quantum computing.

## 2.1 Introduction

The prospect of quantum computing (QC) has caused great excitement in condensed-matter physics. If a set of qubits can be maintained in a coherent, controllable many-body state, certain very difficult computational problems become tractable. In particular, successful QC would mean a revolution in the areas of cryptography [75] and data-base searching [38]. In addition, it would mean a great advance in general technical capabilities, since the control of individual quantum systems and their interactions would represent a new era in nanotechnology.

However, from a practical point of view, a dilemma presents itself immediately. On the one hand, one wishes to control quantum degrees of freedom using external influences, since that is how a quantum algorithm is implemented, and to measure them, since that is the output step. On the other hand, the system must be isolated from the environment, since random perturbations will destroy the quantum coherence that is the whole advantage of QC. This is the isolation-control dilemma, and it leads to a very rough figure of merit $F$ for any quantum computer. If we define the decoherence time $\tau$ as the time it takes to lose quantum coherence, and the clock speed $s$ (roughly the inverse of the time to run a logic gate), then the figure of merit is $F = s\tau$ and a practical machine should satisfy $F > 10^4 - 10^5$ at least. If the clock speed is limited only by standard electronics, then we may be able to achieve $s \approx 10^9$ Hz. This would imply that $\tau = 1$ ms is a lower limit for the decoherence time.

The dilemma has not yet been solved, though a number of solutions have been proposed. A particularly attractive solution is to use spin degrees of freedom as qubits. Nuclear spins interact relatively weakly with their environment because the coupling, proportional to the magnetic moment, is small. Yet there has grown up a sophisticated technology for the manipulation of nuclear spins, and some rudimentary computations have been performed [85]. Readout is the main difficulty with this approach, since the field created by a single moment is tiny, and pure states cannot be achieved in the macroscopic samples used.

Electron spins also interact weakly with the environment in some circumstances: relaxation times in excess of $10^3$ s have been measured for donor bound states of phosphorus-doped silicon (Si:P) [28, 29, 89]. The corresponding manipulation technology (ESR) has also reached a high level of sophistication, but the magnetic moment of the electron exceeds that of the nucleus by three orders of magnitude, which presents problems of isolation. Readout should be easier than for nuclei, since detection of single electron charges is certainly possible. Electron spin detection has recently been achieved using differing techniques [44, 67].

Solid-state implementations of QC are particularly attractive because of the possibility of using existing computer technology to scale small numbers of qubits up to the $10^5$ or so that would be needed for nontrivial computations. The first paper to propose using the electron spin in a quantum dot subjected to a strong dc magnetic field was that of Loss and DiVincenzo [55]. Kane [48] proposed employing the nuclear spin in the Si:P system as the qubit. A specific structure consisting of silicon-germanium (SiGe) layers was proposed by Vrijen *et al.* [87]. This structure incorporates the idea that the g-factor of an electron can be changed by moving it in a Ge concentration gradient, allowing individual electron to be addressed by the external ac field. A different SiGe structure has been proposed by Friesen *et al.* [32]. This structure is designed so that the electron number on the dots, and the coupling between the dots, can be carefully controlled.

Solid-state implementations must also face the isolation control dilemma. Decoherence times must exceed the 1 ms number in the actual physical structures that are needed for the operation of quantum algorithms. In this thesis, we examine whether this can be the case for some of the existing proposals based on electron-spin qubits. In the process, we hope to learn something about modifications to these structures that can increase $\tau$. We shall focus on low-temperature operation, since, as we shall see, this will probably be necessary in order to obtain sufficiently large $\tau$.

We can build on a large body of work, both theoretical and experimental, from the 1950s and 1960s on ESR in doped semiconductors. In a series of papers, Feher and co-workers [28, 29, 89] investigated the relaxation time for the spin of electrons bound on donor sites in lightly doped Si. At sufficiently low temperatures, the relaxation time $T_1$ is dominated by single-phonon emission and absorption. In the presence of spin-orbit coupling, this can relax the spin, causing decoherence. The theory was worked out by Hasegawa [39] and Roth [66, 65].

It must of course be recognized that this spin-lattice relaxation time is not necessarily to be identified with the decoherence time. The decoherence time is the shortest time for any process to permanently erase the phase information in the wave function. This may mean the phase for a single spin, but it also means that the relative phases of the wave functions of different spins must also be preserved, so that processes that cause mutual decoherence must also be taken into account. The actual decoherence time is the minimum of all of these times. A spin relaxation time in excess of 1 ms is a necessary, not a sufficient, condition for the viability of a solid-state electron-spin QC proposal.

A QC must have precise input as well as an accurate algorithm. Preparation of the spin state is often proposed to be done by a thermalization of the spin system at a low temperature. The time to do this actually sets an upper limit on the relaxation time of whatever processes thermalize the spins to the lattice. A limit of perhaps 1-10 s is a reasonable requirement.

This paper focuses on $T_1$, the time for relaxation of the longitudinal component of the magnetization, by spin-phonon interactions. These processes cause real spin-flip transitions. They occur at random times and thus indubitably cause decoherence. In addition to these processes characterized fully by $T_1$, there are processes which introduce random phase changes in the spin wavefunctions. To characterize all such processes by a single "dephasing time" $T_2$ will usually not be sufficient for understanding the operation of a multi-qubit system.[12] Difficulties of definition arise, and care must be taken to specify which phase is involved and to what extent it is randomized. For a single spin system there is no ambiguity. The $2 \times 2$ density matrix $\rho_{ij}$ for the qubit with cylindrical symmetry involves only two independent parameters $\rho_{11} - \rho_{22}$, and $\rho_{12}$. The time dependence of $\rho_{11} - \rho_{22}$ after a system preparation is exponential with a decay constant $T_1$, and represents the return of the longitudinal component of the magnetization to its equilibrium value. The decay is due to inelastic transitions of the type calculated in this paper. $\rho_{12}$, on the other hand, is nonzero only if the preparation of the spin state has a transverse component: $S_x(t = 0) \neq 0$. The decay of this quantity represents the irreversible conversion of this state to an incoherent mixture of "up" and "down" states. Again, this is genuine decoherence of the spin state, since the phase information cannot be recovered.

The time dependence of $\rho_{12}$ when the spin is in a strong field was calculated by Mozyrsky *et al.* [59] using a Markovian approximate master equation. They found that the time scale of the decay due to spin-phonon coupling is very short, of the order of the time for a phonon to cross the electron's wave function, which is about $10^{-10}$ s. But the decay is incomplete, with $\rho_{12}$ retaining all but $10^{-8}$ of its original value. Their calculation was for Si:P, but a very similar result should hold for the dot case. Decoherence at this level is certainly acceptable for quantum computing. These authors also pointed out that the decay of the remainder of $\rho_{12}$ is due to the spin-flip processes computed in this paper. If $T_2$ is defined as the dominant decay time of the off-diagonal density matrix element, then $T_1 = T_2$ for spin-phonon processes.

A quite different source of decoherence is the hyperfine coupling to nuclear spins. The nuclear spins produce an effective random magnetic field on the electrons. Recent calculations using semi-classical averaging techniques [41] obtained a very short relaxation time $T_\Delta \approx 1$ ns for GaAs based dot systems in a strong field. This represents the decay of the transverse magnetization of an ensemble of dots. This is a dephasing time, but not a decoherence time. The electron spins precess in what is effectively the frozen field of the nuclei. This field is spatially random, and the differential precession of the electron spins leads to the magnetization decay. However, this is not an irreversible loss of the phase information of the collective wave function. Spin echo experiments are very beautiful demonstrations of precisely this point. This "inhomogeneous broadening" presents challenges for the calibration and operation of quantum computers, but does not destroy coherence.

Finally, in any implementation based on electron-spin qubits, there will certainly exist small interactions between the spins themselves. The dipole-dipole interaction, for one, cannot be avoided, and there may be indirect spin-spin interactions mediated by the gates. A recent paper suggests that these interactions set the fundamental time scale $T_M$ for Si quantum dot implementations of QC [17]. These interactions do produce experimental broadening of ESR lines in experiments on bulk systems, and this might be taken as decoherence. In our view, however, these interactions do not destroy the coherence of a state. The system is a set of all the qubits. During the course of a quantum algorithm they are collectively in a pure state (in principle). Any decoherence that destroys the purity of the state comes from averaging over the unknown states of the environment. The broadening that comes from dipole-dipole interactions comes, in NMR and ESR calculations, from averaging over the states of the system itself, which is not an appropriate method for calculating decoherence. The effect of qubit-qubit interactions that cannot be turned off is to complicate the quantum algorithm. A quantum algorithm is a unitary transformation that must always include the effect of the system Hamiltonian (including dipole-dipole interactions) in addition to external operations. In every case except for very simple ones, this algorithm must first be computed, presumably with the help of a classical computer. This step in QC may be termed "quantum compilation." The issue that qubit-qubit interactions raise is not one of decoherence, but rather whether the determination of the algorithm, the compilation step, becomes prohibitively difficult. This could happen for two reasons. One is that the interactions are so poorly known that they cannot be corrected for. It seems likely that quantum error correction can resolve this difficulty. A second and more interesting possibility is that the interactions convert the computation of the algorithm itself into a problem that grows exponentially with the size of the system. We regard this as an open question and a deep one, that combines many-body theory with algorithm design and error correction. We note that in NMR implementations the interaction between the qubits also cannot be turned off, but it can be canceled by refocusing [60].

In this chapter, our aim is to evaluate the importance of spin-phonon coupling as a source of decoherence in quantum dot qubits. Fundamentally, the issue is whether the long relaxation times $T_1$ observed at low temperatures in bulk Si:P carry over to SiGe dots proposed for QC.

## 2.2   Strained Si Quantum Well

Si-Ge heterostructures are utilized widely in the digital electronics industry, and presently have the shortest switching times of any device. One reason for their success lies in the ability to engineer structures of near perfect purity, with control over thicknesses and interfaces that approaches atomic precision—a technological tour de force. An equally key achievement has been the harnessing of strain as a tool to control band offsets in heterostructure devices. We present calculations of spin relaxation for real SiGe structures such as those proposed by Vrijen *et al.* [8] and Friesen *et al.* [32] Accordingly, we have calculated the electron wavefunctions in quantum wells, which is needed as input for these calculations. Details of these calculations were presented in Ref. [32], and will not be repeated here. In this section we only describe those aspects of the calculations that are germane to spin relaxation.

Quantum wells are constructed by sandwiching a very thin layer of one material between two others. Electrons can be confined in the quantum well layer when the conduction band offsets produce a potential well. The key to this technology is therefore to understand the band structure of the various layers. In this section we will consider a particular class of wells formed of pure Si, sandwiched between barrier layers of SiGe. We will find that this is optimal from the standpoint of spin coherence. Metallic gates or impurities create zero-dimensional bound states that define a quantum dot. We first review briefly effective mass theory for dots in pure, unstrained Si, then unstrained SiGe, and finally strained Si.

In pure Si, the $\Delta$ conduction band minima occur near the symmetry points X, in the directions {001}. In a perfect Si crystal these minima are six-fold degenerate, the valleys being equivalent. In the dot the electron feels a potential $V_g(\vec{r})$ in addition to the atomic potential, which lifts the degeneracy, though the splittings are not large. The spatial variation of $V_g(\vec{r})$ is on length scales generally much longer than the lattice spacing. For the moment, we shall assume that the electron is in the ground state of $V_g(\vec{r})$ and ignore mixing with any excited states. To the extent that the scale of variation of $V_g(\vec{r})$ is much longer than the atomic spacing there are six nearly degenerate ground states. This is referred to as the "valley degeneracy." The wavefunctions can be written as [51]

$$\Phi_n(\vec{r}) = \sum_{j=1}^{6} \alpha_n^{(j)} F_j(\vec{r}) \phi_j(\vec{r}). \qquad (2.1)$$

Here $\phi_j$ is a Bloch function of the form

$$\phi_j(\vec{r}) = u_j(\vec{r}) e^{i\vec{k}\cdot\vec{r}}, \qquad (2.2)$$

where $\vec{k}_j$ are the six $\Delta$ minima $\{+k_0\hat{x}, -k_0\hat{x}, +k_0\hat{y}, -k_0\hat{y}, +k_0\hat{z}, -k_0\hat{z}\}$ (we shall always use this ordering), and $u_j(\vec{r})$ are periodic functions with the same periodicity as the crystal potential

$V_p(\vec{r})$. The $F_{\pm z}$ are envelope functions that satisfy the Schroedinger-like equation

$$\left[-\frac{\hbar^2}{2m_l}\frac{\partial^2}{\partial z^2} - \frac{\hbar^2}{2m_t}\left(\frac{\partial^2}{\partial x^2} + \frac{\partial^2}{\partial y^2}\right) + V_g(\vec{r})\right] F_{\pm z}(\vec{r}) = (E - E_{k_z}^{(\Delta)}) F_{\pm z}(\vec{r}), \qquad (2.3)$$

and are independently normalized to unity, similar to wavefunctions. Analogous equations can be given for the $\pm\hat{\mathbf{x}}$ and $\pm\hat{\mathbf{y}}$ minima. We see that $F_x = F_{-x}$, $F_y = F_{-y}$, and $F_z = F_{-z}$, so only three independent envelope functions must be computed. $m_l$ and $m_t$ are the longitudinal and transverse effective masses associated with the anisotropic conduction band valleys. $E_{k_z}^{(\Delta)}$ is the $\Delta$ conduction-band edge at $\mathbf{k}_z$. The splitting of the degeneracy comes from corrections to this envelope-function approximation. Different choices of the constants $\alpha_n^{(j)}$ determine the six states. Their values will be discussed in Sec. III. This formalism is a good approximation for both dot and impurity bound states, as the valley splittings are much smaller than the energy scales in Eq. 2.3.

Germanium is completely miscible in Si, forming a random alloy. For a variable Ge content $x$, $\mathrm{Si}_{1-x}\mathrm{Ge}_x$ exhibits materials properties that vary gradually over the composition range. The alloy lattice constant $a_0(x)$ follows a linear interpolation between pure Si and Ge, known as Vegard's law, quite accurately for all $x$: $a_0(x) = (1-x)a_{Si} + xa_{Ge}$ [90]. Electronic properties show an abrupt change in behavior near $x = 0.85$, where the Si-like $\Delta$ minima cross over to fourfold-degenerate, Ge-like, L minima. In this work we focus on the range $x \lesssim 0.5$, which is strictly Si-like, though we will have some remarks below on Ge-rich structures. Throughout this range, properties such as effective mass and the dielectric constant vary only slightly from pure Si values. For our calculations, the most important parameter is the conduction band edge, $E^{(\Delta)}(x)$, which remains sixfold degenerate in the range $x \lesssim 0.5$. The theory of the variation of $E^{(\Delta)}$ with $x$ is not germane to the present work, and we simply quote the empirical result, linear in $x$, which is consistent with Ref. [64]:

$$\Delta E^{(\Delta)}(x) = E^{(\Delta)}(x) - E^{(\Delta)}(0) = 0.23x \text{ (eV)}. \qquad (2.4)$$

[We note, however, that Ref. [68] suggested a slope for $E^{(\Delta)}(x)$ of opposite sign.] The relatively weak variations of the effective mass and the dielectric constant will be ignored here.

We consider thin Si wells in which the Si layer grows pseudomorphically. The in-plane lattice constant $a_\parallel$ must be the same for all layers, causing a tetragonal distortion in the strained layer(s). Here we consider the case of strained Si grown on the (001) surface of relaxed $Si_{1-x}Ge_x$. The in-plane Si lattice constant depends on $x$ as

$$a_\parallel(x) = (1-x)a_{Si} + xa_{Ge}. \tag{2.5}$$

Since $a_{Ge} > a_{Si}$, the Si is under tensile strain in the plane. Hence the out-of-plane Si lattice constant $a_\perp$ is reduced according to continuum elastic theory,

$$a_\perp(x) = a_{Si}\left[1 - 2\frac{c_{12}}{c_{11}}\frac{a_\parallel(x) - a_{Si}}{a_{Si}}\right], \tag{2.6}$$

where $c_{11}$ and $c_{12}$ are elastic constants for pure Si.

Strain produces shifts of the $\Delta$ band proportional to the strain variables

$$\varepsilon_\parallel(x) = \frac{a_\parallel(x) - a_{Si}}{a_{Si}} \text{ and } \varepsilon_\perp(x) = \frac{a_\perp(x) - a_{Si}}{a_{Si}}, \tag{2.7}$$

with proportionality constants called the dilational and uniaxial deformation potentials, $\Xi_d^{(\Delta)}$ and $\Xi_u^{(\Delta)}$, respectively. Because of the anisotropic nature of the strain, the two $\hat{\mathbf{z}}$ minima are shifted down relative to the $\hat{\mathbf{x}}$ and $\hat{\mathbf{y}}$ minima, resulting in a splitting of the $\Delta$ conduction band. The net shifts with respect to the unstrained Si $\Delta$ band are given by [64]

$$\Delta E^{(\Delta_\perp)}(x) = \left(\Xi_d^{(\Delta)} + \frac{1}{3}\Xi_u^{(\Delta)}\right)[2\varepsilon_\parallel(x) + \varepsilon_\perp(x)] + \frac{2}{3}\Xi_u^{(\Delta)}[\varepsilon_\perp(x) - \varepsilon_\parallel(x)], \tag{2.8}$$

$$\Delta E^{(\Delta_\parallel)}(x) = \left(\Xi_d^{(\Delta)} + \frac{1}{3}\Xi_u^{(\Delta)}\right)[2\varepsilon_\parallel(x) + \varepsilon_\perp(x)] - \frac{2}{3}\Xi_u^{(\Delta)}[\varepsilon_\perp(x) - \varepsilon_\parallel(x)]. \tag{2.9}$$

The first terms in Eqs. (8) and (9) are hydrostatic strain terms, which lower the conduction edge compared to unstrained Si. The second terms in Eqs. (8) and (9) produce the splitting, associated with uniaxial strain. To perform our calculations, we use the materials parameters given in Table I. However we note that the deformation potentials, particularly $\Xi_d^{(\Delta)}$, are very difficult to measure experimentally. Considerable disagreement exists in the literature as to the value and even the sign of $\Xi_d^{(\Delta)}$ [30]. The value given in Table I was reported (but not endorsed) in Ref. [30], and provides energy-band variations in general agreement with Refs. 18 and 19. We arrive at the following strain-induced shifts of the conduction band edge for pure Si:

$$\Delta E^{(\Delta_\perp)} = -0.86x \text{ and } \Delta E^{(\Delta_\parallel)} = -0.16x. \tag{2.10}$$

The corresponding shift in the relaxed barrier layers, due to the presence of Ge, was given in Eq. (4). Together, these results describe the conduction band offsets for the quantum well that are used in our simulations.

We now apply our results to two specific quantum well designs of interest for quantum computing. Design 1, shown in the inset of Fig. 1, is a version of that proposed by Vrijen *et al.* [87], in which electrons are trapped on donor ions (usually P), implanted in a semiconductor matrix. In that work, the quantum well is split into Ge- and Si-rich regions to facilitate single qubit operations. For simplicity, we consider here a uniform quantum well, formed of pure Si, with a single dopant ion located at the center of the well. In such a device, single qubit operations can be accomplished using a coded qubit scheme [21]. Design 2, proposed by Friesen *et al.* [32], is shown in Fig. 2. The confinement potential for the electrons is much softer than in design 1. Electrons are trapped vertically by the quantum well, and laterally by the electrostatic potential arising from lithographically patterned, metallic top-gates. Additionally, the quantum dot is tunnel-coupled to a degenerate doped back gate. The dimensions for both designs are given in the figures.

The wave function of the bound electron is computed in the envelope-function formalism (Eq. 2.3). Couplings between the different valleys are introduced through the perturbation theory described in Sec. IV. This procedure provides the specific values of $\alpha_g^{(i)}$ for the ground state, which we use in our calculations.

The abrupt conduction-band offsets are handled by matching the ground state wave function $\Phi_g(\vec{r})$ and $\partial_z \Phi_g(\vec{r})$ at the interfaces. (Remember that we have equated effective masses on both sides of the interfaces.) Due to the linear independence of the Bloch functions, the boundary conditions do not cause a mixing of the envelope functions. Solutions of Eq. 2.3 and the analogous $F_x$ equation are obtained, using commercial three-dimensional finite-element

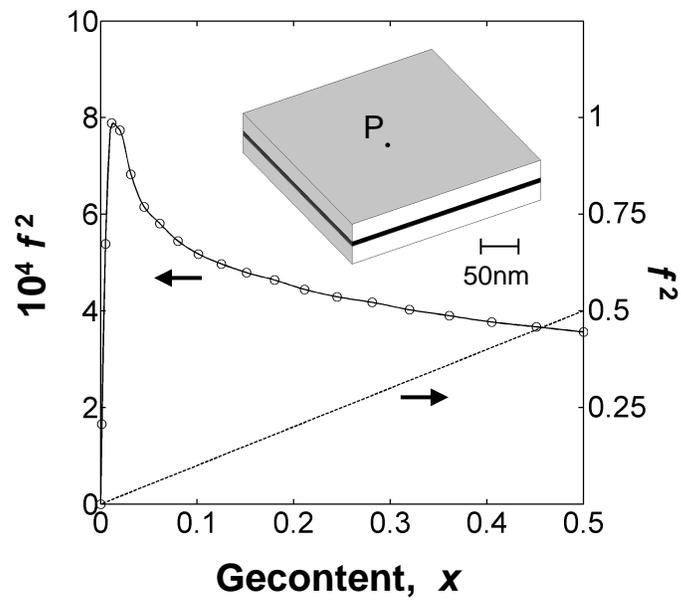

Figure 2.1: Probability $f^2$ for finding a donor-bound electron at a Ge site, as a function of Ge content $x$. The simulated structure, design 1, is shown in the inset. A strained Si quantum well of thickness 6 nm is sandwiched between relaxed $Si_{1-x}Ge_x$ barrier regions of thickness 20 nm. The electron is bound to a $P^{1+}$ ion embedded in the center of the quantum well.

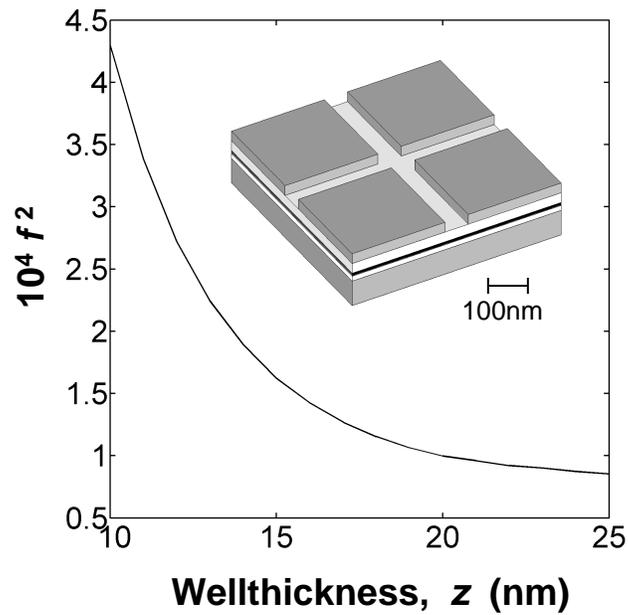

Figure 2.2: Probability $f^2$ for finding an electro-statically bound electron at a Ge site, as a function of quantum-well thickness $z$. The inset shows the heterostructure layers for design 2, beginning at bottom: a thick, doped semiconductor back gate, a relaxed $Si_{1-x}Ge_x$ barrier layer, a strained Si quantum well, a thick, relaxed $Si_{1-x}Ge_x$ barrier layer, and lithographically patterned metallic top gates. The distance between back and top gates is held fixed at 40 nm, while the quantum well, of variable thickness $z$ is centered 15 nm above the back gate.

software.

As will be seen in Sec. V, the key quantity for the computation of $T_1$ is $f^2$, which describes the probability for the bound electron to be on a Ge atom. Ge is associated with reduced coherence times, by virtue of its large spin-orbit coupling. Referring to Eq. (1) one deduces that this probability may be expressed as

$$f^2 = x \left[ 4(\alpha_g^{(x)})^2 \int_{\Omega_b} d^3r F_x^2(\mathbf{r}) + 2(\alpha_g^{(z)})^2 \int_{\Omega_b} d^3r F_z^2(\mathbf{r}) \right], \qquad (2.11)$$

where $\Omega_b$ is the volume outside the quantum well, if the well is pure Si. The subscript $g$ refers to the ground state. The term in the square brackets reflects the probability of finding the bound electron in a barrier region, while $x$ gives the probability that the electron is on a Ge site.

Figure 1 shows the results of our calculations for $f^2$ in design 1, as a function of the Ge concentration $x$ in the $Si_{1-x}Ge_x$ barriers. For $x > 0.02$, $f^2$ decreases with $x$ for two reasons. First, as $x$ increases, the conduction-band offset at the quantum well also increases, allowing less of the wave function to penetrate the barrier. Second, the spatial extent the electron in the $\hat{\mathbf{z}}$ direction is greater for $F_x$ than $F_z$, because of the anisotropic effective mass.

However, less of $F_x$ is mixed into the wave function for large $x$, since $\alpha_x$ becomes very small. For $x \leq 0.02$, $f^2$ drops quickly to zero, due to the absence of Ge in the barriers. In the actual design of Vrijen *et al.* [87], there is Ge in the active layer. To give an idea of the effect of this, we include an equivalent value of $f^2$ for such a structure.

Figure 2 shows results of the $f^2$ calculation for design 2, as a function of the quantum well thickness, $z$. To perform the calculations, we have considered a fixed Ge concentration, $x = 0.05$, and taken the limit of large strain, so that $\alpha_g^{(\pm x)} \simeq \alpha_g^{(\pm y)} \simeq 0$ and $\alpha_g^{(\pm z)} \simeq 1/\sqrt{2}$ for the ground state. As $z$ increases, less of the wave function penetrates the barrier regions, causing $f^2$ to decrease.

## 2.3 Spin relaxation due to coupling to phonons

In this section we give the method for calculating $T_1$, the spin-flip time of a spin qubit in the ground orbital state due to emission or absorption of a phonon, following the logic used by Hasegawa [39] and Roth [66] for bulk Si. Consider a single impurity with a unit positive charge, such as a phosphorus atom, at the origin. In the absence of central cell corrections, there is a 12-fold-degenerate ground state, including spin. This valley degeneracy of the ground state is reduced to 2 by these corrections, and the splitting between the two fold spin-degenerate ground state and the higher states is of order $\Delta E \sim 10$ meV. We shall discuss the detailed linear combinations ($\alpha$ values) of the states below, as the coefficients giving the various valley amplitudes play an important role in the calculation of matrix elements. These 12 states may all be thought of as hydrogenic 1s states. The splitting of 1s and 2s is about 30 meV, larger than the 10 meV valley splittings. Let us now split the twofold degenerate ground state by applying a dc magnetic field in the $z$-direction. The transition rates between these states are denoted by $W_{\uparrow\downarrow}$ and $W_{\downarrow\uparrow}$. The relaxation time $T_1$ is defined by $1/T_1 = W_{\uparrow\downarrow} + W_{\downarrow\uparrow}$.

The transitions are caused by phonons, but there are important approximate symmetries that suppress these transitions. These are the following (1) Spin rotation symmetry, meaning that the electron spin cannot be flipped by a phonon; this symmetry is broken by spin-orbit coupling (SOC) (2) Time-reversal symmetry, meaning that one state cannot be changed into its time-reversed partner by emission or absorption of a phonon; this symmetry is broken by the external magnetic field (3) Point-group symmetries; these are partially broken when strain is applied.

The spin rotation symmetry would rule out phonon mediated transitions between the two states entirely if there were no SOC. This means that the effects of SOC on the wave functions, even though these effects are small in relatively low-Z Si, must be taken into account. When we refer to a state as $\uparrow$ or $\downarrow$, these symbols must be taken to refer to the majority-spin content of the state, not to a pure spin state. Transition rates are roughly proportional to $(g-2)^2$ [more precisely $(g_l - g_t)^2$, where $g_t(g_l)$ is the transverse (longitudinal) $g$ factor, see below for definitions].

The time-reversal symmetry implies that transitions cannot take place directly between Kramers-degenerate states even in the presence of spin-orbit coupling. The direct phonon-mediated transitions between the two states of interest to us are strongly suppressed by this approximate symmetry. It is broken only by the external field $H$. The fastest processes then involve a virtual excitation to higher-energy states that are mixed into the ground state by $H$. Hence $1/T_1$ involves a factor $(\mu_B H/\Delta E)^2$. There is an additional factor of $H^2$ from the phonon density of states, giving an overall rate $1/T_1 \sim H^4$ in the limit of small $H$.

The point group symmetry is reduced from cubic to tetragonal under strain. This has complex effects that we will explain below.

Before giving actual calculations, we summarize those differences between the electrons in donor impurity states and in an artificial dot that affect $T_1$. The most obvious is the single-particle potential that binds the electron. The gate potential is much smoother than the hydrogenic potential of the impurity. This implies that the corrections to the effective-mass approximation are much weaker, and $\Delta E$ will be much reduced. It is difficult to compute the energy splittings precisely, but considerations based on the method of Sham and Nakayama [71] give splittings in the range $0.5 - 0.1$ meV in the structure of Friesen $et\ al.$ [32]. This increases the relaxation rate. (In fact, a naive estimate of the enhancement is a factor of 400.) (More recent calculations put the valley splitting in the range 0.1-1 meV, still ten or more times below that of the donor case [6].) On the other hand, the structures we consider have strong lattice strain. This partly lifts the valley degeneracy and also reduces the matrix elements, which decreases the rate. Another aspect of some of the proposed designs is the presence of Ge with its much stronger SOC. This will act to decrease the spin relaxation time.

## 2.4   Pure Si Quantum Dots

We first consider the case of pure Si under uniaxial strain. The ingredients of the calculation are as follows.

From Sec. II we have the solutions to the Schroedinger equation $[\mathcal{H}_0 + V_g(\vec{r})]\Phi_{n0}(\vec{r}) = E_n \Phi_{n0}(\vec{r})$. $\mathcal{H}_0$ is the unperturbed crystal Hamiltonian without SOC and it has a full space group symmetry. $V_g(\vec{r})$ is the gate and/or impurity potential.

To calculate $T_1$, we must also include SOC, which we treat as a perturbation: $H_{SOC} =$

$\lambda_{Si} \sum_{\vec{R}} \vec{L}_{\vec{R}} \cdot \vec{S}_{\vec{R}}$. The resulting states $\Phi_n(\vec{r})$ are twofold degenerate because of time-reversal symmetry. Let us denote these states as $\Phi_{n\uparrow}(\vec{r})$ and $\Phi_{n\downarrow}(\vec{r})$. They are not eigenstates of spin, so the arrows denote a pseudo-spin. We may define the pseudo-up state as that which evolves from the spin-up state as spin-orbit coupling is turned on adiabatically, and similarly for the pseudo-down state. Because of valley and pseudo-spin degeneracy, there are two ground states $\Phi_{gs}(\vec{r})$ and ten excited states $\Phi_{rs}(\vec{r})$. The 12-fold degeneracy in the effective-mass approximation is broken by central-cell corrections in the impurity case and smaller corrections in the quantum dot.

There is also the Zeeman Hamiltonian of the external field $\mathcal{H}_Z = \mu_B \vec{B} \cdot (\vec{L} + 2\vec{S})$. In the field $\Phi_{n\uparrow}(\vec{r})$ and $\Phi_{n\downarrow}(\vec{r})$ are no longer degenerate. Note that the energy splitting may depend on the direction of $\vec{B}$.

Finally, we have the electron-phonon coupling Hamiltonian $\mathcal{H}_{ep}$. A phonon represents a time-dependent perturbation. This will create transitions whose rate is given by the Fermi golden rule. We are interested in the transitions between $\Phi_{g\uparrow}(\vec{r})$ and $\Phi_{g\downarrow}(\vec{r})$. However, $\langle \Phi_{g\uparrow}(\vec{r}) | \mathcal{H}_{ep} | \Phi_{g\downarrow}(\vec{r}) \rangle = 0$ in the absence of the external field. Thus we need to calculate in next order in perturbation theory using an effective Hamiltonian

$$\mathcal{H}' = \sum_{rs} \frac{1}{E_g - E_r} \left\{ [\mathcal{H}_Z | \Phi_{rs}(\vec{r}) \rangle \langle \Phi_{rs}(\vec{r}) | \mathcal{H}_{ep}] + [\mathcal{H}_{ep} | \Phi_{rs}(\vec{r}) \rangle \langle \Phi_{rs}(\vec{r}) | \mathcal{H}_Z] \right\}. \qquad (2.12)$$

Here $r$ runs over the excited states, $r = 2...6$, $s = \uparrow, \downarrow$.

The relaxation time is given by

$$1/T_1 = W_{\uparrow\downarrow} + W_{\downarrow\uparrow},$$

where

$$W_{\downarrow\uparrow} = (2\pi/\hbar) \times \Sigma_{\vec{q}\lambda} |\langle \Phi_{g\uparrow}(\vec{r}) | \mathcal{H}' | \Phi_{g\downarrow}(\vec{r}) \rangle|^2 \delta(E_{g\downarrow} - E_{g\uparrow} - \hbar\omega_{\vec{q}\lambda})[1 + n(\omega_{\vec{q}\lambda})]$$

is the rate for transitions from the higher-energy (pseudo-spin down) state to the lower energy

(pseudo-spin-up) state and

$$W_{\uparrow\downarrow} = (2\pi/\hbar) \times \Sigma_{\vec{q}\lambda} |\langle \Phi_{g\downarrow}(\vec{r})|\mathcal{H}'|\Phi_{g\uparrow}(\vec{r})\rangle|^2 \delta(E_{g\downarrow} - E_{g\uparrow} - \hbar\omega_{\vec{q}\lambda}) n(\omega_{\vec{q}\lambda})$$

is the rate for transitions from the lower-energy (pseudo-spin-up) state to the higher-energy (pseudo-spin-down) state where the sum is over phonon modes $\vec{q}\lambda$ with energies $\omega_{\vec{q}\lambda}$. A thermodynamic average over the lattice states has been taken. It yields the Bose occupation factors $n(\omega_{\vec{q}\lambda})$ for the phonons.

The matrix elements of $\mathcal{H}'$ are computed as follows. The expectation value of the external field $H_Z$ can be written as

$$
\begin{aligned}
\langle \Phi_{ns}(\vec{r})|H_Z|\Phi_{n's'}(\vec{r})\rangle &= \sum_{i=1}^{6} \alpha_n^{(i)} \sum_{j=1}^{6} \alpha_{n'}^{(j)} \mu_B \vec{B} \cdot g^{(i)} \cdot \vec{\sigma}_{s,s'} \delta_{ij} \\
&= \mu_B \vec{B} \cdot \left[ \sum_{i=1}^{6} \alpha_n^{(i)} \alpha_{n'}^{(i)} g^{(i)} \right] \cdot \vec{\sigma}_{s,s'} \\
&= \mu_B \vec{B} \cdot \mathbf{D}_{nn'} \cdot \vec{\sigma}_{s,s'} \qquad (2.13)
\end{aligned}
$$

and the tensor $g^{(i)}$ is the effective $g$ factor at the $i$th valley. This equation defines the tensor $\mathbf{D}_{nn'}$ that characterizes the coupling of the various states by the external field. The principal axes of the $g$ tensor are the same as that of the effective mass tensor at the $i$th valley. It has the form

$$
g^{(\pm x)} = \begin{pmatrix} g_l & 0 & 0 \\ 0 & g_t & 0 \\ 0 & 0 & g_t \end{pmatrix}, \quad g^{(\pm y)} = \begin{pmatrix} g_t & 0 & 0 \\ 0 & g_l & 0 \\ 0 & 0 & g_t \end{pmatrix}, \quad g^{(\pm z)} = \begin{pmatrix} g_t & 0 & 0 \\ 0 & g_t & 0 \\ 0 & 0 & g_l \end{pmatrix}, \quad (2.14)
$$

There are only two independent constants.

A simple example of the diagonal part of the $\mathbf{D}$ tensor is that for the ground state of an impurity in the unstrained lattice when the central cell corrections are included. Then we have $\alpha_g^{(i)} = 1/\sqrt{6}$, and $\mathbf{D}_{gg}$ is proportional to the unit matrix,

$$\langle \Phi_{gs}(\vec{r})|H_Z|\Phi_{gs'}(\vec{r})\rangle = g_g \mu_B \vec{B} \cdot \vec{\sigma}_{ss'}, \qquad (2.15)$$

with

$$g_g = \frac{2}{3} g_t + \frac{1}{3} g_l. \tag{2.16}$$

The matrix elements between ground and excited states have the form

$$\langle \Phi_{gs}(\vec{r}) | H_Z | \Phi_{rs'}(\vec{r}) \rangle = g' \mu_B \vec{B} \cdot \mathbf{D}_{gr} \cdot \vec{\sigma}_{ss'}, \tag{2.17}$$

with

$$g' = \frac{1}{3} (g_l - g_t), \tag{2.18}$$

and the tensor $\mathbf{D}_r$ is defined by

$$\mathbf{D}_{gr} = 3 \sum_{i=1}^{6} \alpha_g^{(i)} \alpha_r^{(i)} \hat{k}^{(i)} \hat{k}^{(i)} \tag{2.19}$$

where $\hat{k}^{(i)}$ is the "local" anisotropy axis. If the original $g$ were isotropic, then $g' = 0$ and there would be no coupling between different states and no spin relaxation.

If the lattice is strained, then the $\alpha$ coefficients become strain-dependent and the general expression for $\mathbf{D}$ from Eq. (13) must be used. Uniaxial strain lifts the degeneracy of the valleys. We include this effect in the Hamiltonian and it determines the proper combinations of the $\alpha_n^{(i)}$ defined in Eq. (1). These then feed into $\mathbf{D}_{gr}$. As a function of strain $\alpha_n^{(g)}$ for the ground state cross over from the completely symmetric combination $\alpha_g^{(i)} = 1/\sqrt{6}$ to the combination $\alpha_g^{(\pm x)} = \alpha_g^{(\pm y)} = 0$, $\alpha_g^{(\pm z)} = 1/\sqrt{2}$ in the limit of large strain [28, 29, 89].

The phonons involved are just the acoustic ones, one longitudinal and two transverse—these are the only ones with low enough energy to play a role in relaxing the spins. The matrix elements of the electron-phonon interaction are only nonzero within one valley and for a single phonon mode they are conventionally parameterized as

$$\langle \Psi_{c\vec{k}s}^{(i)}(\vec{r}) | H_{ep}^{(\vec{q}\lambda)} | \Psi_{c\vec{k}'s}^{(i)}(\vec{r}) \rangle = i b_{\vec{q}\lambda} \hat{e}_\lambda(\vec{q}) \cdot (\Xi_d 1 + \Xi_u \hat{k}^{(i)} \hat{k}^{(i)}) \cdot \vec{q} + H.c. \tag{2.20}$$

near the $i$th valley, where $\vec{q} = \vec{k} - \vec{k}'$ and $\hat{e}_\lambda$ is the polarization vector. $\vec{b}_{\vec{q}\lambda}$ destroys a phonon with wave vector $\vec{q}$ and polarization $\lambda$. Once again, we see that the interaction can be characterized by just two parameters, in this case $\Xi_d$ and $\Xi_u$, as already defined in Eq. (7). Performing the integration over the envelope function at wave vector $\vec{q}$ now gives

$$\langle\Phi_{gs}(\vec{r})|H_{ep}^{(\vec{q}\lambda)}|\Phi_{gs'}(\vec{r})\rangle_{\vec{q}} = \left(\Xi_d + \frac{1}{3}\Xi_u\right)A(\vec{q})(b_{\vec{q}\lambda} + b_{\vec{q}\lambda})\delta_{s,s'}, \tag{2.21}$$

$$\langle\Phi_{gs}(\vec{r})|H_{ep}^{(\vec{q}\lambda)}|\Phi_{rs'}(\vec{r})\rangle_{\vec{q}} = \frac{1}{3}\Xi_u A(\vec{q})[i\hat{e}_\lambda(\vec{q})\cdot\mathbf{D}_{gr}\cdot\vec{q}b_{\vec{q}\lambda} - i\hat{e}_\lambda^*(\vec{q})\cdot\mathbf{D}_{gr}\cdot\vec{q}b_{\vec{q}\lambda}^*]\delta_{s,s'}, \tag{2.22}$$

where

$$A^{(i)}(\vec{q}) = \sum_{\vec{k}} F^{(i)*}(\vec{k}+\vec{q})F^{(i)}(\vec{k}) = \int d^3r F^2(r)e^{i\vec{q}\cdot\vec{r}}. \tag{2.23}$$

Thus the electron-phonon interaction involves a form factor for the bound states. Since $F$ is normalized, we have $A^{(i)}(\vec{q}) \approx 1$ when the wavelength of the phonon is much longer than the spatial extent $a^*$ of the bound state: $qa^* \ll 1$. The calculations of Sec. II indicate that this is the case. $A^{(i)}(\vec{q})$ is also independent of $(i)$.

In the golden rule calculation, the energy denominator $(E_g - E_r)^{-2}$ will suppress contributions from the excited states of $V_g$. Thus we will keep only states that are split off from the ground state by corrections to the effective mass approximation. This approach works very well in Si:P and should be even better for the quantum dot.

This produces the golden-rule transition rate

$$W_{\uparrow\downarrow} = \frac{2\pi}{\hbar}\left[\frac{1}{3}\Xi_u g'\mu_B\right]^2\sum_{\vec{q}\lambda}A^2(\vec{q})\delta(E_{g\uparrow}-E_{g\downarrow}-\hbar\omega_{\vec{q}\lambda})\langle a_{\vec{q}\lambda}a_{\vec{q}\lambda}^*\rangle\left|\sum_{i=2}^{6}\frac{\vec{B}\cdot\mathbf{D}_{gr}\cdot\vec{\sigma}_{\uparrow\downarrow}\hat{e}_\lambda(\vec{q})\cdot\mathbf{D}_{gr}\cdot\vec{q}\delta_{ss'}}{E_g-E_r}\right|. \tag{2.24}$$

We approximate the phonon dispersion as $\omega_{\vec{q}\lambda} = v_\lambda q$. Setting $A = 1$, performing the integral over the magnitude of $\vec{q}$, and repeating the calculation for $W_{\downarrow\uparrow}$, we obtain a total spin relaxation time

$$\begin{aligned}
\frac{1}{T_s} &= W_{\uparrow\downarrow} + W_{\downarrow\uparrow} \\
&= \frac{1}{8\pi^2\rho\hbar^4}\left(\frac{g'\mu_B B\Xi_u}{3}\right)^2[2n(g\mu_B B)+1]g_g^3\mu_B^3 B^3\sum_{\lambda=1}^{3}\frac{1}{v_\lambda^5}\int_0^{2\pi}d\phi'\int_0^{\pi}\sin\theta'd\theta'
\end{aligned}$$

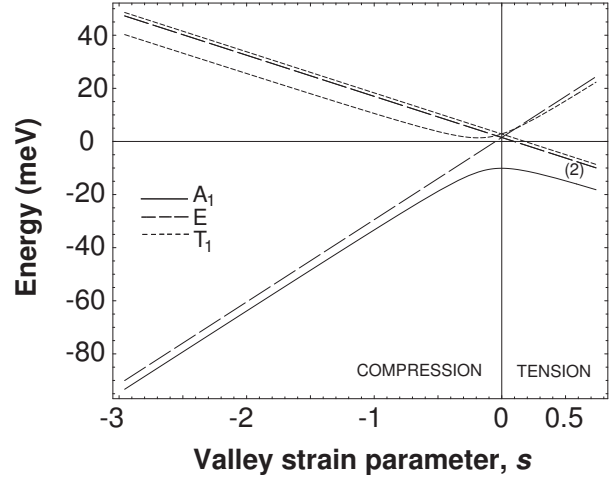

Figure 2.3: Energy of the six 1s-like donor levels of an electron bound by a P impurity in silicon (with respect to the energy center of gravity) vs the strain parameter $s$ with an uniaxial stress applied in the [100] direction. (It is important to note for clarity that what we call compression relative to the [100] or growth direction is equivalent to tension in the plane of the quantum well. This label convention differs across fields.) For reference, $s = -3$ corresponds to the compressive ($s$ negative) strain caused in a pure silicon layer by a $Si_{0.8}Ge_{0.2}$ sublayer. The energies are expressed in eV and the numbers in parenthesis indicate the degeneracy of the level. The $A_1$ (ground state, solid line) and an E level (which both go up with strain) are mixed by spin-orbit coupling allowing spin relaxation proportional to $1/\Delta E$. Thus, with all else equal, an increase in strain causes the relaxation rate to decrease. For a detailed analysis see Wilson and Feher (Ref. 4), where $s$ here is their $s$ times 100, or Koiller *et al.* (Ref. [52]) whose notation was $s = 100(6\chi\Delta_c/\Xi_c)$.

$$\times \left| \sum_{r=2}^{6} [\vec{B}(\theta, \phi) \cdot \mathbf{D}_{gr} \cdot \vec{\sigma}_{\uparrow\downarrow} \left[ \frac{\hat{e}_\lambda(\vec{q}) \cdot \mathbf{D}_{gr} \cdot \vec{q}}{E_g - E_r}(\theta', \phi') \right] \right|^2. \tag{2.25}$$

Here $(\theta, \phi)$ are the polar axes of the direction of $\vec{B}$ measured from the [100] direction of the crystal. $\rho$ is the mass density. This is our basic result, in a form very similar to that given by Hasegawa [39].

In Fig. 3 the effect of strain on the lowest energy levels is shown. Here the zero-strain splittings are those of a P impurity. A similar plot was given in Ref. [28, 29, 89], but at that time the correct zero-strain splittings were not known. In the dot, the energy splittings for the unstrained case would be one to three orders of magnitude smaller (see below for a discussion).

In Fig. [28, 29, 89], we show the effect of strain on the relaxation time of an electron in a P impurity potential. There are two effects: the overall increase of the energy denominators and the change of the ground state to a less symmetric valley weighting. This leads to a nonmonotonic dependence of $T_1$ on strain, but in the region of interest, the effect of strain is to greatly increase $T_1$, since most proposed structures have $s < -1$. At large strain, only one energy denominator remains small, that between the ground state, symmetric in the $\pm z$ valleys, and the first excited state, antisymmetric in the $\pm z$ valleys. The overlap matrix $\mathbf{D}$ is very small between these two states. This reduction of the matrix element is the dominant effect.

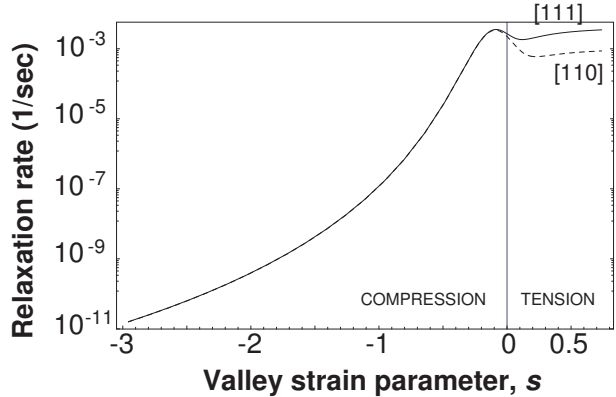

Figure 2.4: Relaxation rates of a P impurity bound electron in [100] uniaxially strained silicon vs strain parameter $s$ for a temperature of 3 K and a magnetic field $\vec{H} = H(\theta, \phi)$ of 1 T in the [111] ($\theta = \phi = \pi/4$) and [110] ($\theta = \pi/4$, $\phi = 0$) directions respectively. $s = -3$ corresponds to the strain caused in the pure silicon layer by a $Si_{0.8}Ge_{0.2}$ sublayer.

Also of interest is the dependence of $T_1$ on the angle of the external field, since this may serve as a diagnostic tool in experiments to verify that the relaxation process is really due to spin-phonon coupling. As a function of strain, this dependence becomes highly anisotropic, as seen in Fig. 5. This is due to the elimination of all but the $\pm z$ valleys from the problem at high strain. We note that the limiting value $T_1 \to \infty$ when the external field is along a crystal axis is cut off by intervalley scattering effects not included in the present calculation [28, 29, 89].

The change in the confining potential reduces the corrections to the effective-mass approximation, as discussed in Sec. III which in turn reduces the energy denominators, leading to a decrease in $T_1$. In Friesen *et al.*'s structure, we have estimated the splittings $E_r - E_g$ using the method of Sham and Nakayama [70], and they range from 0.05 to 0.1 meV, depending on the gate voltages that produce the potential. This may be incorporated into the calculation of the $\alpha_n^{(i)}$ by introducing a variable coupling $\Delta_c$ that mixes the $\pm x$ and $\pm y$ valley wave functions

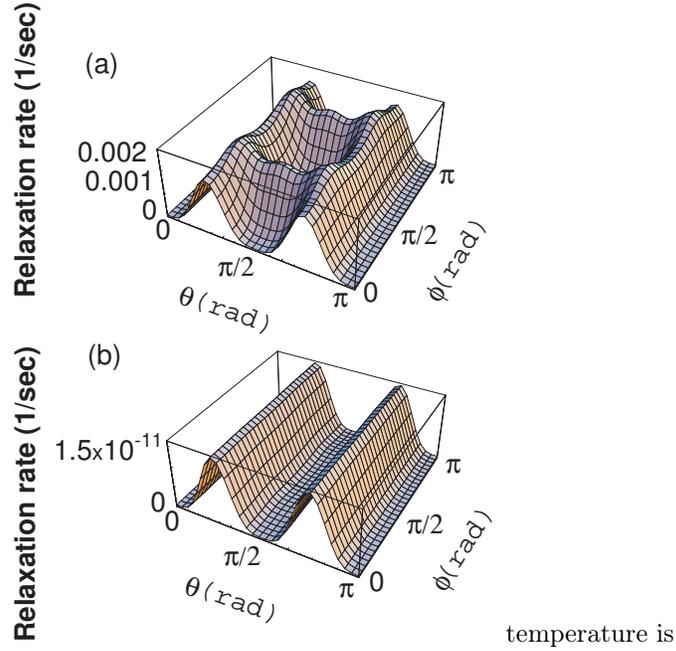

Relaxation rate (1/sec)

(a)

0.002
0.001
0

$\phi(rad)$

$\pi$

$\pi/2$

0

$\theta(rad)$

0

$\pi/2$

$\pi$

Relaxation rate (1/sec)

(b)

$1.5\times10^{-11}$

0

$\phi(rad)$

$\pi$

$\pi/2$

0

$\theta(rad)$

0

$\pi/2$

$\pi$

temperature is

Figure 2.5: Dependence of the relaxation rate on the magnetic-field direction, with $\vec{H} = H(\theta, \phi)$, for (a) unstrained Si and (b) compressively strained Si (corresponding to a $Si_{0.8}Ge_{0.2}$ sublayer). The magnetic field is set at 1 T, and the temperature is 3 K.

with the $\pm z$ valley wave functions. For a precise definition of $\Delta_c$, see Ref. [28, 29, 89]. Approximately, however, $\Delta_c \approx 0.2(E_r - E_g) < 10^{-24}$ eV. The results of the calculations are shown in Fig. 6. The graphs show the absolutely crucial role that strain plays in the determination of $T_1$. Because of the small energy denominators in the dot, the rate is extremely fast at small $\Delta_c$ for the unstrained case: $1/T_1 \approx \Delta_c^{-2}$.

We also point out that in structures where $T_1$ appears to be too long for efficient preparation of the spins (as discussed in Sec. I) the deficiency can be made up by increasing the temperature $T$. As seen in Fig. 7, $T_1$ decreases very rapidly as $T$ is increased. Actually, this calculation even underestimates the decrease, since multi-phonon processes begin to contribute at about 3 K. However, it must be borne in mind that the temperature must be small enough that the initial state is essentially pure (all spins up, for example). A more practical method of decreasing $T_1$ would be to increase the magnetic field.

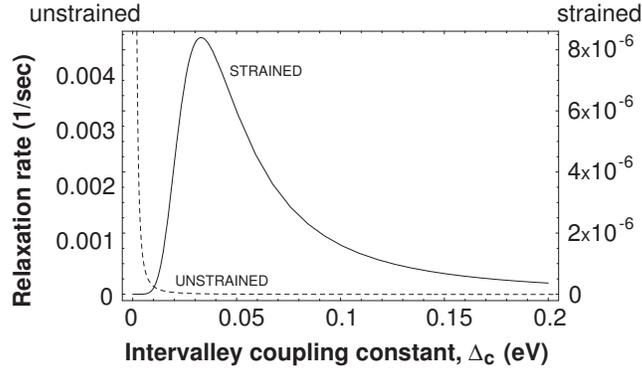

Figure 2.6: Relaxation rate of a bound electron in silicon. The solid line is for a fixed [100] uniaxial strain $s = -3$, corresponding to the strain caused in the pure silicon layer by a $Si_{0.8}Ge_{0.2}$ sublayer. The dashed line is the unstrained case. The rate is plotted for varying intervalley coupling constant $\Delta_c$. $\Delta_c$ corresponds to 1/6th of the energy splitting between the singlet (ground state) and triplet (next excited state) valley energies *in unstrained Si*. As a result, it controls the mixing between the $\pm x$ and $\pm y$ valley wave functions with the $\pm z$ valley wave functions. The magnetic field is set at 1 T along the [111] direction, and the temperature is 3 K.

## 2.5   Structures containing Ge

The effect of alloying with Ge is to increase the SOC and hence to increase $|g - 2|$. First-principles calculations of Ge impurities in Si have shown that there is little effect on the states near the bottom of the conduction band, though this is not necessarily the case for higher energy states in the band [72]. This is in accord with the isoelectronic character of the atoms in the alloy. It is then reasonable to employ the virtual-crystal approximation (VCA). The approximation should be quantitatively accurate for small, (say <10%) concentrations of Ge, but may be taken as a good qualitative guide also to higher concentrations. The Bloch functions in the absence of SOC satisfy

$$H_0 \Psi_{n\vec{k}s} = E_{n\vec{k}} \Psi_{n\vec{k}s}, \tag{2.26}$$

where

$$\Psi_{n\vec{k}s} = \frac{1}{\sqrt{N_c}} \sum_{j=1}^{N_c} \sum_{l=1}^{2} \exp(i\vec{k} \cdot r_{jl}) \sum_{m=0}^{3} a_{lm}(n\vec{k}) \phi_{ms}(\vec{r} - r_{jl}), \tag{2.27}$$

where $N_c$ is the number of unit cells, $j$ labels the unit cells, $l$ labels the two positions in the unit cell, $m$ labels the four atomic states, $\phi_{ms}$ are the atomic orbitals, and the coefficients $a_{lm}(n\vec{k})$

—

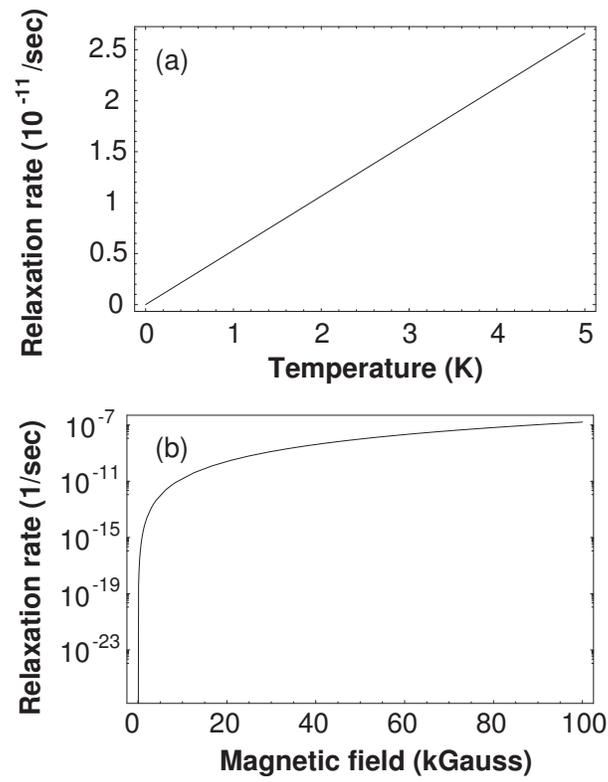

Figure 2.7: Relaxation rates of a P impurity bound electron in [100] uniaxially strained silicon ($s = -3$, corresponding to the strain caused in the pure silicon layer by a $Si_{0.8}Ge_{0.2}$ sublayer) vs (a) a temperature with the magnetic field set at 1 T and (b) a magnetic-field strength with the temperature set at 3 K.

give the proper linear combination of atomic orbitals for the state at momentum $\vec{k}$ and band $n$. The normalization condition is

$$\sum_{lm} |a_{lm}(n\vec{k})|^2 = 1. \tag{2.28}$$

In the VCA, the $a_{lm}(n\vec{k})$ are the same for the Si and Ge sites. The SOC Hamiltonian for the alloy is written as

$$H_{SOC} = \sum_{\vec{R}} \lambda_{\vec{R}} \vec{L}_{\vec{R}} \cdot \vec{S}_{\vec{R}}, \tag{2.29}$$

where $\lambda_{\vec{R}}$ is the SOC strength for Si (Ge) when $\vec{R}$ is a Si (Ge) site. The energy shift of an electron in the external field $\vec{B} = (B_x, B_y, B_z)$ can be written as

$$\Delta E_{n\vec{k}s}(\vec{B}) = -\mu_B \sum_{i=x,y,z} g_i(n\vec{k}) B_i s, \tag{2.30}$$

where

$$g_i(n\vec{k}) = 2 - 2\hbar^2 \left( \sum_{\vec{R}} \lambda_{\vec{R}}/N_c \right) \sum_{n' \neq n, l, m, m'} |a_{lm}(n\vec{k})|^2 |a_{lm'}(n'\vec{k})|^2 (E_{n\vec{k}} - E_{n'\vec{k}})^{-1} (\varepsilon_{imm'})^2 \tag{2.31}$$

and $\varepsilon_{imm'}$ is the completely antisymmetric symbol with $\varepsilon_{123} = 1$.

Our approximation now consists in regarding the dependence on the Ge concentration as coming only in the term $(\Sigma_{\vec{R}} \lambda_{\vec{R}}/N_c)$. Hence any component of the $g$ tensor is proportional to a weighted average of $\lambda_{Si}$ and $\lambda_{Ge}$. In particular, consider the conduction band minimum at $\vec{k} = (0, 0, k_0)$ in the compound $Si_{1-x}Ge_x$. Then we have $g_z(0) - 2 = [g_z(0) - 2][(1 - x) + x(\lambda_{Ge}/\lambda_{Si})]$, where $g_z(0)$ is the value for pure silicon. The numbers, known from atomic physics, are $\lambda_{Ge}/\lambda_{Si} = 0.295/0.044 = 6.71$, so $g_z(x) - 2 = [g_z(0) - 2)][(1-x) + 6.71x]$. Similarly, we have $g_y(0) - 2 = [g_y(0) - 2][(1 - x) + 6.71x]$, etc. Only the overall scale of $g$ changes, not the anisotropy in the VCA.

The presence of Ge in the lattice breaks the translation symmetry, an effect that is neglected in the VCA. Taking this into account would lead to a $g$ factor that is averaged over momentum space rather than one that is evaluated only at the Si conduction-band minima. If we take the

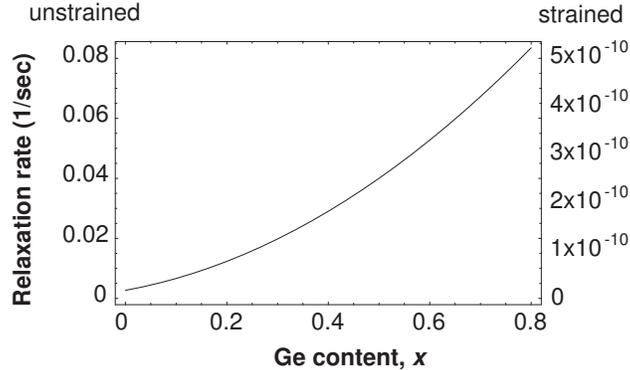

unstrained                                                    strained

Figure 2.8: Relaxation rate of a P impurity bound electron in [100] uniaxially strained silicon vs the concentration of germanium within unstrained (left vertical axis) and compressively strained (right vertical axis, corresponding to a $Si_{0.8}Ge_{0.2}$ sublayer) pure silicon. The magnetic field is set at 1 T along the [111] direction, and the temperature is 3 K. Here we assume that the addition of germanium does not affect the strain within the bulk and only acts to increase the g factor and spin-orbit coupling. We expect this approximation to hold for small concentrations of germanium and to break down with increasing concentration as the conduction-band valley minima switch from X-type silicon to L-type germanium.

L minimum of pure Ge as an example, we would expect that the corrections to $|g - 2|$ would be greater if there is some averaging over momentum space. Thus a calculation going beyond the VCA would give corrections that would reduce $T_1$.

We show the results for differing Ge concentrations in the active region of the well in Fig. 8. In the Ge-poor regime $x < 0.85$, the relaxation rate $1/T_1$ decreases fairly slowly. Thus designs with a Si-rich well will not suffer from very fast spin relaxation on the minority Ge sites.

By contrast, if we cross the critical point around $x = 0.85$ where the conduction band minima switch positions, then there is a very rapid decrease of $T_1$. This is due to the much greater g-factor anisotropy of the [111] minima: $g' \approx -0.4$, (as compared to $g' < 0.001$ in Si). The results for the Ge-rich region are much simpler than in the Si-rich region. Uniaxial strain does not affect the relative valley energies or the matrix elements parameterized by **D**. The only effect of going to the dot from the impurity is to reduce corrections to the effective mass approximation, which will strongly decrease the relaxation time. Indeed, if we take the impurity result for Ge:P from Ref. [39], which is $T_1 = 2.3 \times 10^{-3}$ s, and estimate the decrease in $\Delta E$ as about a factor of 100, then we obtain $T_1 \simeq 10^{-6}$ to $10^{-7}$ s, which is vastly shorter than that of Si-rich structures. The physics is the same as that which governs the divergence of the rate in Fig. 6 for the unstrained case, namely the much smaller energy denominators.

To make contact with Sec. II note that if there is Ge in the barrier regions, as is the case in most SiGe designs, then we will have an envelope function that weights Si and Ge differently, then $x$ in the VCA formulas must be replaced by $f^2 = \Sigma_{\vec{R}} |f_{\vec{R}}|^2$, where $|f_{\vec{R}}|^2$ is the amplitude that the electron is on a Ge site and the sum runs only over Ge sites, as described in Sec. II. This is the product of the Ge concentration in the layer, times the probability that the electron is on the layer, summed over layers. Referring to Figs. 1 and 2, we see that $f^2$ is always less than $10^{-3}$ for structures that have pure Si active layers and Ge in the barrier layers. The effect on $T_1$ is small, so these structures are "phonon safe."

## 2.6   Discussion and Implications

Decoherence due to spin relaxation by emission and absorption of single phonons does not pose a major obstacle to quantum computation in quantum dot SiGe heterostructures using spin qubits, if these qubits are properly designed. This spin relaxation mechanism is the dominant one in the case of electrons in donor bound states, such as in Si:P.

In order that this source of decoherence be kept under control, however, certain conditions must be satisfied in the design of the structure. Unstrained SiGe alloys generally have relatively short spin relaxation times due to valley degeneracy. This degeneracy produces a strong spin mixing in the Kramers-degenerate ground states of the quantum dot once the field is turned on. This in turn increases the electron-phonon matrix element that connects the two states and decreases the relaxation time. This effect, fundamentally due to spin-orbit coupling, is stronger in the quantum dot than in the impurity state. The corrections to effective-mass theory, stronger for the impurity potential, tend to lift the degeneracy. The crucial role that strain plays, in Si-rich structures, is to lift the valley degeneracy. In this respect compressive uniaxial strain is the best, since the residual degeneracy is reduced to two valleys. The wave function for the ground state is then symmetric in the valley index, and the first excited state, the only one with a small energy denominator, is antisymmetric in this index. This fact strongly suppresses the spin mixing effect of the external field. Thus a workable design should include uniaxial strain that exceeds a critical value. This value is small enough that there will be sufficient strain in most practical structures in which the active layer is Si rich and sandwiched between relatively Ge-rich layers. The active layer must be thin enough to avoid dislocation-mediated relaxation.

Spin-orbit coupling is stronger in Ge than in Si due to the higher atomic number. This suggests that Si-rich structures are to be preferred, and the calculations bear this out. The relaxation time decreases with increasing $x$ in a $Si_{1-x}Ge_x$ alloy. In fact, if $x$ increases beyond $x = 0.85$, decoherence becomes very strong, and quantum dots in these Ge-rich structures, such as those proposed by Vrijen *et al.* [87], may well run into difficulties for this reason. A key point for these structures is that uniaxial stress does not lift the valley degeneracy, since the valleys have moved off the crystal axes. A strain-induced suppression of the spin relaxation cannot occur.

Beyond design issues, fabrication quality is also important. Even a small concentration of magnetic impurities will negatively impact the spin relaxation. In modern semiconductor technology, however, concentrations of magnetic impurities much less than 0.1 ppm are routine. The impact of lattice imperfections is less clear. These will generally act to lower the symmetry of the system, and to lessen the accuracy of the effective-mass approximation. As we have seen in the context of the impurity calculations, these effects generally increase the spin relaxation time because of reduced degeneracy. On the other hand, localized bound states can form at such imperfections. If this results in free-electron spins, it could have seriously negative effects on the relaxation time. Similar effects would result from lattice defects that produce localized phonon modes. Finally, it appears that quantum dot designs will need rather low temperatures in order to operate. The impurity experiments clearly indicate that multi-phonon relaxation increases rapidly with temperature [11, 12], with a crossover to this regime at about 3 K. We speculate that this crossover temperature is pushed up when the effects of strain are included, but we have not yet done the requisite calculations.

# Chapter 3

# Spin-orbit coupling and relaxation in Si 2DEGs

The Rashba spin-orbit coupling coefficient is a key parameter not only for two-dimensional electron gas (2DEG) spin relaxation calculations but also for calculations of decoherence in lateral silicon quantum dots. This chapter paves the way for the next in that it derives and verifies the dominant spin-orbit coupling term relevant to spin relaxation in quantum wells and dots. This term is due to structural inversion asymmetry and was not pertinent to calculations in the previous chapter. Understanding spin relaxation in silicon 2DEGs is also key because it is possible to measure currently via generic bulk electron spin resonance techniques, a natural checkpoint on the road to single spin relaxation measurements in a quantum dot.

Silicon is a leading candidate material for spin-based devices, and 2DEGs formed in silicon heterostructures have been proposed for both spin transport and quantum dot quantum computing applications. The key parameter for these applications is the spin relaxation time. Here we apply the theory of D'yakonov and Perel' (DP) to calculate the electron spin resonance linewidth of a silicon 2DEG due to structural inversion asymmetry for arbitrary static magnetic field direction at low temperatures. We estimate the Rashba spin-orbit coupling coefficient in silicon quantum wells and find the $T_1$ and $T_2$ times of the spins from this mechanism as a function of momentum scattering time, magnetic field, and device-specific parameters. We obtain agreement with some existing samples (though not all) for the angular dependence of the relaxation times and show that the magnitudes are consistent with the DP mechanism in general. We suggest how to increase the relaxation times by appropriate device design.

## 3.1 Introduction

Electron spins in silicon have been proposed as an attractive architecture for spintronics and quantum information devices. The inherently low and tunable spin-orbit coupling (SOC) in silicon heterostructures and the possibility of eliminating hyperfine couplings by isotopic purification bodes well for quantum coherent spin-based qubits and spin transport. Early experiments together with theory have shown that coherence times can be upwards of three orders of magnitude longer than in GaAs.[45, 84, 83, 17]

Energy relaxation of localized spin states has attracted theoretical attention [66, 39, 50, 61] and experimental effort [89, 83] for decades, and this activity has recently revived in the context of quantum computation. The idea is to store quantum information in the spin of a single electron confined in a semiconductor structure, either attached to a donor atom or confined electrostatically in a quantum dot. Spin transport, also of great interest, encodes information in the spin states of an ensemble of electrons. In both cases, electron spin resonance (ESR) measurements of spin relaxation provide a key and available measure of spin coherence properties of electrons in silicon quantum wells, though not a one-to-one correspondence. Our aim in this paper is to explain some existing ESR results for silicon 2DEGs at low temperatures and to make predictions for future experiments.

The structures that concern us here are layered semiconductor devices of Si and SiGe. The active layer is the quantum well (QW) that confines the electrons in the growth direction. This layer will be assumed to be composed of pure, [001] strained silicon. We shall also neglect any roughness or miscut at the the Si/SiGe interfaces. Devices made in this way are commonly referred to in the semiconductor industry as Modulation-Doped Field Effect Transistors (MODFETs) and are designed to maximize mobility. Figure 3.1 introduces two example structures.

Extensive theoretical work has been done on spin relaxation in GaAs and other III-V materials. The developments that began with the theory of D'yakonov and Perel' [23, 22, 93] are most relevant for our purposes. These authors found that fluctuating effective magnetic fields due to momentum scattering in the presence of SOC are the dominant spin relaxation mechanism in semiconductor 2DEGs for low temperatures. Here, we start with this assumption and use a general spin-density matrix approach to calculate the relaxation times of a 2DEG in the presence of a static magnetic field, including explicitly the angular dependence. Understanding the angular dependence of the linewidth is important for comparison with ESR experiments and the extraction of physically relevant parameters such as the momentum relaxation time. We also calculate separately in a $k \cdot p$ formalism the Rashba spin-orbit coupling parameter in silicon quantum wells. This is a key parameter in our calculation as well as for other spin control considerations, both negative and positive.

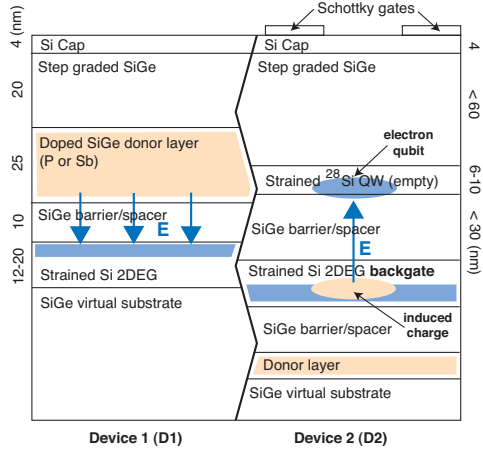

Figure 3.1: Strong internal electric fields in the growth direction ($z$) are common in silicon quantum well devices. *D1:* A typical, high-mobility SiGe heterostructure uses a donor layer to populate a high-density 2DEG. The charge separation results in an $E_z \sim 10^6$ V/m. *D2:* A proposed quantum dot quantum computer [32] which utilizes a tunnel-coupled backgate to populate the quantum well without the need for a nearby donor layer. Here, $E_z > 10^5$ V/m due to the image potential formed on the backgate.

In the next section we discuss the origin and magnitude of the SOC in realistic heterostructures. The section following that presents our calculation. Lastly, we compare with experiment and discuss the implications for device design.

## 3.2    Spin-orbit Hamiltonians

The SO Hamiltonian comes naturally out of the Dirac equation and to first order in momentum is given in an arbitrary potential $V$ by

$$H_{SO} = \frac{-\hbar \vec{\sigma} \cdot \mathbf{p} \times \nabla \mathbf{V}}{4m^2c^2} = \frac{\hbar \nabla \mathbf{V} \cdot (\vec{\sigma} \times \mathbf{p})}{4m^2c^2}$$

where $\vec{\sigma}$ are the Pauli matrices. In reduced dimensional systems such as semiconductor heterostructures, spin-orbit coupling is thought to be dominated by inversion asymmetry within the device. Basically this refers to internal electric fields within the semiconductor which are either due to the asymmetry within the crystal unit cell itself (called bulk inversion asymmetry (BIA) or Dresselhaus) common to non-centrosymmetric crystals such as GaAs or to interface electric field/band offset effects at a heterojunction or induced externally through an electric field (called structural inversion asymmetry or Rashba). The electron moving in these internal electric fields feels an effective magnetic field and its spin rotates about it. The bulk or Dresselhaus term can be derived for GaAs in the Kane model and is given by [93]

$$H_D = \frac{2\Delta}{3\sqrt{2mE_g m_{cv} E_g}} \vec{\sigma} \cdot \vec{\kappa}$$

where $\Delta$ is the SO splitting of the conduction band in the direct band gap material, $E_g$ is the band gap, $m_{cv}$ is the parameter of the Kane model, and $\kappa_x = p_x(p_y^2 - p_z^2)$ (and cyclic permutations thereof). In 2D, this reduces to the more familiar approximation

$$H_D^{2D} = \beta(p_y \sigma_y - p_x \sigma_x)$$

where

$$\beta = \frac{2}{3} \langle p_z^2 \rangle \frac{\Delta}{(2mE_g)^{1/2} m_{cv} E_g}.$$

The Rashba term with corresponding coefficient $\alpha$ is more straightforward. It comes almost explicitly from the usual SO Hamiltonian. Notice that

$$H_R = \frac{\hbar}{4m^2c^2} \nabla \mathbf{V} \cdot (\vec{\sigma} \times \mathbf{p}) = \frac{\hbar e}{4m^2c^2} \mathbf{E}(\mathbf{r}) \cdot (\vec{\sigma} \times \mathbf{p}) = \alpha(\mathbf{r}) \cdot (\vec{\sigma} \times \mathbf{p}).$$

The prefactor $\hbar/4m^2c^2$ is only for free electrons, the actual constant in a semiconductor is enhanced by band effects. To get the familiar Rashba expression, simply assume that confinement in one direction is much stronger than the other two. In a 2DEG, this would be the $z$-direction and thus the $z$-component of the above dot-product is selected:

$$H_R^{2D} = \alpha(p_x\sigma_y - p_y\sigma_x) \propto eE_z(\vec{\sigma} \times \mathbf{p})_z.$$

In reality, calculating the Rashba coefficient $\alpha$ is more complicated than simply knowing the electric field at the interface. Band effects also play a role.

## 3.3   Spin-orbit coupling

The strong macroscopic electric fields inside heterostructure QWs are important for understanding SOC, especially in silicon. These fields are also device-specific, so we carry out our calculations on the two representative structures in Figure 3.1. Both devices have square QWs, with equal barriers on the top and bottom interface. The first is typical of MODFETs and employs a donor layer above the QW in order to populate it. This charge separation produces an electric field between the two layers (across the barrier or spacer layer) which can be approximated by

$$E_z \approx \frac{en_s}{\epsilon_0\epsilon_{Si}} = -6 \times 10^6 \text{ V/m}, \tag{3.1}$$

where $n_s = 4 \times 10^{15}$ m$^{-2}$ is the density of electrons in the 2DEG for Device 1, $e$ is the charge of an electron, and $\epsilon_i$ are the dielectric constants. We assume that *the QW is populated only by donor-layer electrons*, leaving an equal amount of positive charge behind. The second structure is one that has been proposed for use in a quantum computer device.[32] It utilizes a near-lying, tunnel-coupled backgate 2DEG ($< 30$ nm away) together with Schottky top-gates to populate the QW selectively. This situation also results in a strong electric field due to the image potential on the back gate. For one qubit, this can be estimated as

$$E_z \approx \frac{e}{4\pi\epsilon_0\epsilon_{Si}d^2} = 3 \times 10^5 \text{ V/m}, \tag{3.2}$$

where $d = 20$ nm is the distance from the QW to the back gate for Device 2. Schottky top-gates and other device parameters can augment or reduce this growth-direction electric field nominally up to the breakdown field of silicon, $3 \times 10^7$ V/m, or the ionization energy of the electron.[47] Indeed, this field can actually be smaller than that due to the top-gates in certain dot configurations.

The shift of the electron g-factor from its free-electron value $g_0 = 2.00232$ is one measure of SOC in a system. It is quite small in bulk silicon and depends on the magnetic field direction in the (elliptical) conduction band minima ($\Delta g_{\parallel} \approx -0.003$, $\Delta g_{\perp} \approx -0.004$).[89] However, it is difficult to reliably extract the SOC strength in a 2DEG from $\Delta g$. Many parameters (e.g., strain, barrier penetration, Ge content ($g_{Ge} = 1.4$), non-parabolicity of the band minima) influence the magnitude and sign of $\Delta g$ and it may show considerable sample dependence. The non-parabolicity effects are especially sensitive to the electron density within the QW and can hide the magnitude of SOC within a system.[46]

In these silicon heterostructures, SOC is dominated by inversion asymmetry within the device. The spin-orbit (SO) Hamiltonian to first order in momentum is given in an arbitrary electrostatic field $E_z$ by

$$H_{SO} = \frac{\hbar}{4m^2c^2} E_z \vec{\sigma} \cdot (\hat{z} \times \mathbf{p}),$$

where $\sigma_i$ are the Pauli matrices. Note that the effective magnetic field that acts on the spin is in the plane of the layer. In Si heterostructures, the macroscopic fields, which do not average out, are more important than the atomic electric fields. In the non-centrosymmetric III-V materials such as GaAs, this is not necessarily the case and the resulting Dresselhaus or *bulk* inversion asymmetry fields are usually dominant. The asymmetry considered here, due either to an interface, charge distribution, or external potential, is usually called Rashba or *structural* inversion asymmetry.

The Rashba term comes directly from the SO Hamiltonian if we assume one, dominant symmetry-breaking electric field in the structure and average over a momentum state. In a QW, as we have pointed out above, the electric field is in the growth ($z$) direction and thus the

$z$-component of the above dot-product is selected and we obtain

$$H_R^{2D} = \alpha(p_x\sigma_y - p_y\sigma_x) \propto E_z(\vec{\sigma} \times \mathbf{p})_z, \tag{3.3}$$

which is then the Rashba-Bychkov Hamiltonian.[9]

Strictly speaking, as de Andrade e Silva *et. al.* point out [16], the conduction-band-edge profile, $E_c$, and the space charge separation (or applied electrostatic field), $E_z$, contribute separately and sometimes dissimilarly to the SOC. For example, the wave function discontinuity (band offset) across a material-interface can cause Rashba spin-splitting itself. However, in devices of the type considered here, the macroscopic field should be the main contribution. These same authors have derived an expression for $\alpha$ in the Kane model for GaAs. We have adapted their work for Si, using a 5-parameter 8-band Kane model. This is 8 bands including spin, which means just the lowest conduction band and the three highest valence bands. By calculating the breaking of the degeneracy between the spin-up and spin-down states of the lowest conduction band, we find

$$\alpha = \frac{2PP_z\Delta_d}{\sqrt{2}\hbar E_{v1}E_{v2}} \left(\frac{1}{E_{v1}} + \frac{1}{E_{v2}}\right) e\left\langle E_z \right\rangle, \tag{3.4}$$

where we have taken the average of the electric field in the $z$-direction. Here $P = \hbar\langle X|p_x|S\rangle/im$, $P_z = \hbar\langle Z|p_z|S\rangle/im$, $\Delta_d = 0.044$ eV is the spin-orbit splitting of the two highest conduction bands, $E_{v1} = 3.1$ eV is the direct gap of the strained sample, and $E_{v2} = 7$ eV is the gap between the conduction band minimum and the lowest of the three valence bands. (These are the 5 parameters mentioned above.) $m$ is the bare electron mass. The matrix elements that define $P$ and $P_z$ are to be taken between the cell-periodic functions of the indicated symmetry at the position of the conduction-band minimum. Unfortunately, these are not well known in Si, since other bands contribute. We may note that $P$ and $P_z$ are examples of momentum matrix elements that don't vary too much in III-V materials and Ge,[54] and are given approximately by $2mP_{(z)}^2/\hbar^2 \approx 22$ eV. With these values we find that for devices of type 1,

$$\alpha \approx 1.66 \times 10^{-6} \left\langle E_z \right\rangle \text{ m/s}. \tag{3.5}$$

Previous Kane models for GaAs involve matrix elements at $k = 0$. Our theory is new since it takes into account the proper symmetry of silicon with its minima well away from the zone center.

Wilamowski *et. al.* [88], using conduction electron spin resonance (CESR), have measured $\alpha \approx 5.94$ m/s ($\alpha e/\hbar = 0.55/\sqrt{2} \times 10^{-12}$ eV·cm in their units) [1] in a $Si_{0.75}Ge_{0.25}/Si/Si_{0.75}Ge_{0.25}$ QW where the strained-silicon layer was roughly $12 - 20$ nm. Carrier concentrations were $n_s \approx 4 \times 10^{15} \, m^{-2}$. These numbers correspond to our Device 1 parameters. Our equations then give, using Eq. 3.5,

$$\alpha^{D1} \approx 5.1 \text{ m/s for } \langle E_z \rangle = 3 \times 10^6 \text{ V/m.} \tag{3.6}$$

Theory compares in order of magnitude and we believe that our estimation has some utility as a guide for device design.

For Device 2, $\alpha^{D2} \approx 0.25$ m/s for $\langle E_z \rangle = 1.5 \times 10^5$ V/m. This device remains to be built. For Device 1, we can also predict the zero magnetic field spin-splitting in a silicon 2DEG using $|\epsilon_+ - \epsilon_-| \leq 2\alpha p_F$,

$$2\alpha p_F = 2\alpha\hbar\sqrt{\frac{4\pi n_s}{4}} \approx 0.75 \, \mu\text{eV,}$$

where 4 is the degeneracy factor in silicon (spin+valley). This has not yet been directly measured to our knowledge. Taking a Zeeman splitting of $g\mu_B B = 0.75\mu$eV with a g-factor of 2, this implies an internal, in-plane, effective magnetic field—the so-called *Rashba field*—of roughly 62 Gauss, which is the direct result of SOC in the silicon 2DEG of Device 1.

Let's consider the relevance of our silicon SOC results to QC and spintronics. We note that the magnitude of the Rashba coefficient is much smaller than for GaAs. For the same electric field and 2DEG density as Device 1, a similar $\vec{k} \cdot \vec{p}$ theory for GaAs arrives at $\alpha_{GaAs} \approx 230$ m/s.[16] But GaAs itself is not a high Rashba III-V semiconductor and is thought to be be dominated by Dresselhaus SOC ($\beta_{GaAs} \approx 1000$ m/s).[49] InAs-based heterojuctions, for example, may have orders of magnitude higher Rashba values.[15] This means that SOC effects in silicon devices will be much smaller, including decoherence and gating errors that are SOC-based.

---

[1]Note that [88] uses the wrong expression for the Fermi wave vector, failing to include the valley degeneracy, so their data analysis is off by a factor of sqrt(2).

## 3.4 Spin relaxation

We wish to consider the combined effects of the SO Hamiltonian

$$H_R^{2D} = \alpha(p_x\sigma_y - p_y\sigma_x)$$

and the scattering Hamiltonian. The scattering may be from phonons or from static disorder. We take the semiclassical approach, in which the effect of scattering is to cause transitions at random intervals from one wavepacket centered at $\vec{p}$ with $\varepsilon_{\vec{p}} = \varepsilon_F$ to another centered at $\vec{p'}$ with $\varepsilon_{\vec{p'}} = \varepsilon_F$, where $\varepsilon_F$ is the Fermi energy. This corresponds to a random switching in the direction of the effective magnetic field that acts on the spin degree of freedom. This is the D'yakonov-Perel' mechanism of spin relaxation.[23, 22] The measured quantity in the continuous-wave experiments carried out on 2DEGs is $T_2$, the transverse relaxation time, while pulsed experiments can also measure $T_1$, the longitudinal relaxation time. For our purposes, a density matrix approach is the natural one, since we will eventually want to perform an ensemble average over all possible scattering sequences. Since the physical model of spins in a random time-dependent magnetic field is the same as that for relaxation of nuclear spins in liquids, the Redfield technique may be used.

We outline the calculation only briefly, since the details are parallel to the discussion in standard texts.[78] The $2 \times 2$ density matrix $\rho$ allows us to compute the expectation values of the spin by $\langle \sigma_i \rangle = Tr \ \{\sigma_i\rho\}$. For a single system described by the Hamiltonian $H$, we have the equation of motion $d\rho/dt = \frac{i}{\hbar} [\rho, H]$. In the case of

$$H = H_0 + H_1(t), \tag{3.7}$$

where $H_1$ is small, it is convenient to go to the interaction representation

$$\rho^{int} = \exp(-iH_0t/\hbar)\rho \exp(iH_0t/\hbar), \tag{3.8}$$

and then we get

$$\frac{d\rho^{int}}{dt} = \frac{i}{\hbar} \left[ \rho^{int}, H_1^{int}(t) \right], \tag{3.9}$$

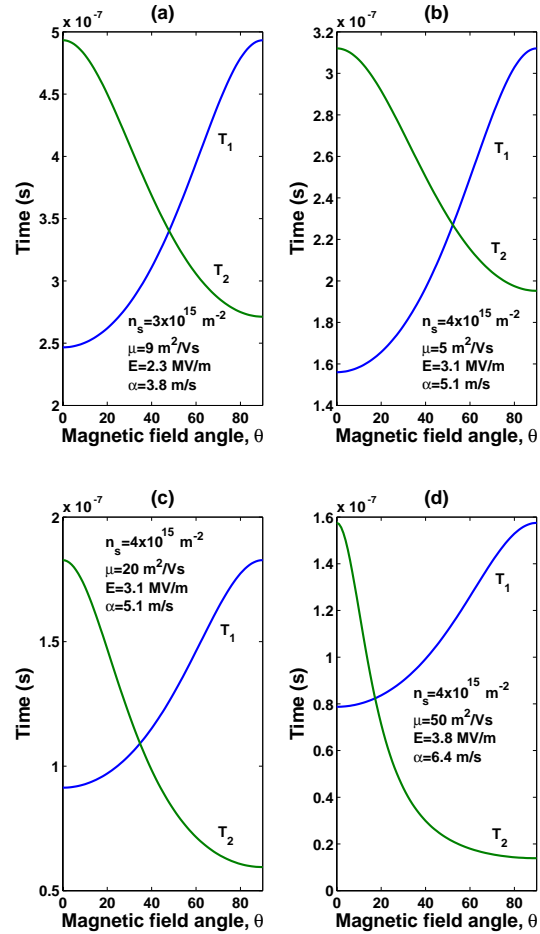

Figure 3.2: Spin relaxation times as a function of static magnetic field direction (where $\theta = 0$ is perpendicular to the 2DEG plane) for specific values of 2DEG density and Rashba asymmetry. The quantum well is assumed to be completely donor-layer populated and as such, $\alpha$ is calculated directly with Eq. 3.5 as a function of the 2DEG density.

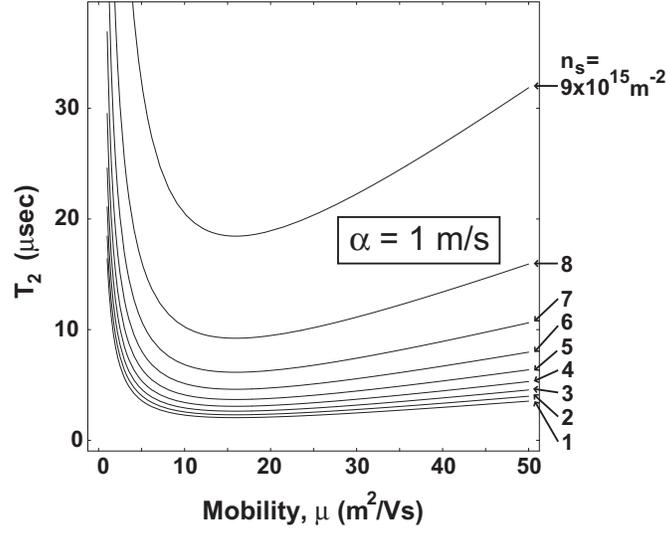

Figure 3.3: ESR linewidth lifetime $T_2$ from Eq. 3.17 for constant asymmetry coefficient, $\alpha = 1$ m/s, as a function of 2DEG mobility, $\mu$, and density, $n_s$. For donor-layer populated quantum wells, divide the times listed by $\alpha^2$: $T_2(\alpha) = T_2(\alpha = 1)/\alpha^2$. The magnetic field is assumed to be $B = 0.33$ Tesla, perpendicular to the plane of the 2DEG.

where

$$H_1^{int}(t) = \exp(iH_0t/\hbar)H_1\exp(-iH_0t/\hbar). \tag{3.10}$$

This equation can be integrated to give

$$\rho^{int}(t) = \rho^{int}(0) + \frac{i}{\hbar}\int_0^t \left[\rho^{int}(t'), H_1^{int}(t')\right]dt', \tag{3.11}$$

and this can be solved interatively, which in second order gives

$$
\begin{aligned}
\frac{d\rho^{int}(t)}{dt} &= \frac{i}{\hbar}\left[\rho^{int}(0), H_1^{int}(t)\right] \\
&+ \left(\frac{i}{\hbar}\right)^2\int_0^t dt'\left[\left[\rho^{int}(0), H_1^{int}(t^{'})\right], H_1^{int}(t)\right].
\end{aligned}
\tag{3.12}
$$

For example, let the steady field be in the $z$-direction, so that $H_0 = \hbar\omega_c\sigma_z/2$. The fluctuating field $\left[H_1^{int}(t)\right]_{ss''} = \sum_{i=x,y} h^i(t)\sigma_{ss''}^i$ is in the transverse direction. The first order matrix element vanishes and we are left, in second order, with

$$
\begin{aligned}
\left(\frac{i}{\hbar}\right)^{-2}\left(\frac{d\rho^{int}_{ss'}}{dt}\right)_2 &= \sum_{s''s'''} \rho_{ss''}(0)\left[H^{int}_1(t')\right]_{s''s'''}\left[H^{int}_1(t)\right]_{s'''s'} e^{i(s''-s''')t'}e^{i(s'''-s')t} \quad (3.13)\\
&+ \sum_{s''s'''}\left[H^{int}_1(t)\right]_{ss''} e^{i(s-s'')t}\left[H^{int}_1(t)\right]_{s''s'''} e^{i(s''-s''')t'}\rho_{s'''s'}(0)\\
&- \sum_{s''s'''}\left[H^{int}_1(t')\right]_{ss''} e^{i(s-s'')t'}\rho^{int}_{s''s'''}(0)\left[H^{int}_1(t)\right]_{s'''s'} e^{i(s'''-s')t}\\
&- \sum_{s''s'''}\left[H^{int}_1(t)\right]_{ss''} e^{i(s-s'')t}\rho^{int}_{s''s'''}(0)e^{i(s'''-s')t'}\left[H^{int}_1(t')\right]_{s'''s'},
\end{aligned}
$$

where we define $e^{st} = e^{\epsilon_s t}$. Averaging the random field, we find

$$
\begin{aligned}
\overline{\left[H^{int}_1(t)\right]_{ss''}\left[H^{int}_1(t')\right]_{s''s'''}} &= \sum_{ij}\overline{h^i(t)h^j(t')}\sigma^i_{ss''}\sigma^j_{s''s'''}\\
&= \sum_{i=x,y,z}\chi^i(t-t')\sigma^i_{ss''}\sigma^i_{s''s'''}, \quad (3.14)
\end{aligned}
$$

where $\tau = t - t'$ and we define $\chi^i(\tau) = \overline{h^i(t)h^i(t')}$. In the geometry with a nonzero component of the external field parallel to the layer, the relaxation tensor can be non-diagonal as far as symmetry is concerned, and there can be three relaxation times.[69] In the case under consideration, where the Dresselhaus cubic term is absent, the only fluctuating field is the Rashba field. As long as the scattering is s-wave and cyclotron motion effects are ignored, as we assume, the off-diagonal (e.g. $x-y$) components of the correlation function vanish, in a coordinate system that is fixed in the frame of the sample. In this frame the relaxation tensor is diagonal and there are only two experimentally accessible relaxation times at any fixed $\theta$, where $\theta$ is the angle of the field with respect to the growth axis.

We now substitute expression 3.14 into Eq. 3.13 and do the matrix algebra. Eq. 3.13 can then be integrated in the limit where $t$ is large. We neglect the oscillating terms, which yields

$$
\left(\frac{i}{\hbar}\right)^{-2}\left\langle\frac{d\rho_{++}}{dt}\right\rangle = \left[\chi^x(\omega_L) + \chi^y(\omega_L)\right]\left[\rho_{++}(0) - \rho_{--}(0)\right]
$$

and

$$
\left(\frac{i}{\hbar}\right)^{-2}\left\langle\frac{d\rho_{+-}}{dt}\right\rangle = \rho_{+-}(0)\left[\chi^x(\omega_L) + \chi^y(\omega_L) + 2\chi^z(0)\right]
$$

To obtain the relaxation times we must consider the equation for the spin:

$$
\begin{aligned}
\frac{d \langle \sigma^z \rangle}{dt} &= \frac{d \left[ Tr \left( \sigma^z \rho \right) \right]}{dt} = \frac{d}{dt} (\rho_{++} - \rho_{--}) \\
&= 2 \left( \frac{i}{\hbar} \right)^2 \left[ \chi^x(\omega_L) + \chi^y(\omega_L) \right] \left[ \rho_{++}(0) - \rho_{--}(0) \right] \\
&= -2 \left( \frac{1}{\hbar} \right)^2 \left[ \chi^x(\omega_L) + \chi^y(\omega_L) \right] \langle \sigma_z \rangle
\end{aligned}
\tag{3.15}
$$

and so

$$
1/T_1 = 2 \left[ \chi^x(\omega_L) + \chi^y(\omega_L) \right] / \hbar^2.
$$

Also

$$
\begin{aligned}
\frac{d \langle \sigma^x \rangle}{dt} &= \frac{d Tr \left( \sigma^x \rho \right)}{dt} = \frac{d}{dt} (\rho_{+-} + \rho_{-+}) \\
&= - \left[ \rho_{+-}(0) + \rho_{-+}(0) \right] \left[ \chi^x(\omega_L) + \chi^y(\omega_L) + 2\chi^z(0) \right] / \hbar^2 \\
&= - \left[ \chi^x(\omega_L) + \chi^y(\omega_L) + 2\chi^z(0) \right] \langle \sigma_x \rangle / \hbar^2,
\end{aligned}
$$

which gives

$$
1/T_2 = \left[ \chi^x(\omega_L) + \chi^y(\omega_L) + 2\chi^z(0) \right] / \hbar^2.
$$

This gives a relation

$$
1/T_2 = 1/2T_1 + 2\chi^z(0)/\hbar^2.
$$

We now wish to specialize to the case of a DP mechanism in a 2DEG. The main point is that the static field may be in any direction, while the fluctuationg field is in the plane. Consider first the special case that the static field is along $\hat{z}$. Then

$$
1/T_2 = 1/2T_1 = \left[ \chi^x(\omega_L) + \chi^y(\omega_L) \right] / \hbar^2.
$$

Now consider a general direction, say $\vec{B}$ along the direction $B_x \hat{x} + B_z \hat{z} = \sin \theta \hat{x} + \cos \theta \hat{z}$, so that $\theta$ is the angle to the normal. Then the longitudinal fluctuations $\chi^{\parallel}$, which are quadratic in the field, are proportional to $\sin^2\theta$ and the transverse ones $\chi^\perp$ to $\cos^2\theta$. Thus

$$1/T_1(\theta) = 2\left[\cos^2\theta\,\chi^x(\omega_L) + \chi^y(\omega_L)\right]/\hbar^2 \tag{3.16}$$

while

$$1/T_2(\theta) = \left[\cos^2\theta\,\chi^x(\omega_L) + \chi^y(\omega_L) + 2\sin^2\theta\,\chi^x(0)\right]/\hbar^2. \tag{3.17}$$

For the DP mechanism in a 2DEG the random field is constant in magnitude, but random in direction. The statistics of this field are Poisson: namely that if the $x$-component the field at time $t = 0$ is $h^x$, then the chance of it remaining at $h^x$ decays as $\exp(-t/\tau_p)$. Hence

$$\overline{h^i(0)h^i(\tau_p)} = \langle h_x^2\rangle\, e^{-t/\tau} = \frac{1}{2}\alpha^2 p_F^2 e^{-t/\tau_p},$$

and

$$\chi^{x,y}(\omega) = \frac{\alpha^2 p_F^2 \tau_p^2}{1 + \omega^2\tau_p^2}.$$

In these formulas, $\tau_p$ is the momentum relaxation time. Note that these formulas assume s-wave scattering. Finally, we are left with

$$\frac{1}{T_1} = 2\alpha^2 p_F^2\tau\left[\frac{\cos^2(\theta)+1}{1+\omega_L^2\tau^2}\right]/\hbar^2$$

and

$$\frac{1}{T_2} = \alpha^2 p_F^2\tau\left[\frac{\cos^2(\theta)+1}{1+\omega_L^2\tau^2} + 2\sin^2\theta\right]/\hbar^2.$$

The zero-frequency limit of these formulas agrees with the recent results of Burkov and MacDonald [8] [see their Eq.(17)] . They do not agree with the formulas in Wilamowski *et. al.* [88], [see, e.g., their Eq.(3)] who state that the relaxation from the DP mechanism should vanish when $\theta = 0$. This is not consistent with our results.

The DP mechanism has the nice feature that it is relatively easy to isolate experimentally. It is strongly anisotropic in the direction of the applied field compared to other mechanisms. To illustrate this we plot the ESR linewidths as a function of field angle in Fig. 3.2. What is most striking is the opposite dependence on angle for the rates $1/T_1$ and $1/T_2$, with $1/T_2$ maximized when the field is in the plane of the 2DEG, while $1/T_1$ is maximized when the field is perpendicular to the plane of the 2DEG. Physically, this comes from the fact that the electric field is perpendicular, so that the fluctuations of the effective magnetic field are in the plane. Longitudinal relaxation ($T_1$) is due to fluctuations perpendicular to the steady field, while transverse relaxation ($T_2$) is due to fluctuations both perpendicular and parallel to the steady field. This mechanism has the characteristic that the change in $1/T_1$ as the field is rotated through 90 degrees is always a factor of two. The change in $1/T_2$ is frequency- and lifetime-dependent, with the anisotropy increasing as the mobility increases.

The DP relaxation also has the counter-intuitive inverse dependence of the spin relaxation time on the momentum relaxation time: $1/T_{1,2} \propto \tau_p$, for small $\tau_p$ (or zero field), typical for motional narrowing. We plot the dependence of $T_2$ on the mobility in Fig. 3.3. At high mobilities and high frequencies $\omega >> 1/\tau_p$, we find $1/T_{1,2} \propto 1/\tau_p$.

## 3.5 Discussion and Implications

We have calculated the transverse and longitudinal relaxation times of a silicon 2DEG in an arbitrary static magnetic field. To test our calculations, we compare them to known ESR data.[45, 84, 88, 82, 37] We limit ourselves to low temperatures, $\epsilon_F \sim 10 - 15$ K, and realistic material parameters for state-of-the-art heterostructures.

The most robust prediction of the theory is the anisotropy, particularly that of $T_1$, which is completely independent of all parameters. The only measurement, in Ref. [3], gives satisfactory agreement for $T_1$ : $T_1^{B\|z}/T_1^{B\perp z} = 0.67$ as opposed to the prediction 0.5. Furthermore, the anisotropy of $1/T_2$ goes in the opposite direction, as it should. The magnitude of this anisotropy is measured to be $T_2^{B\|z}/T_2^{B\perp z} = 12.5$ which is about a factor of six larger than the theory predicts for the quoted mobility. The relaxation times, as far as can be determined by the range set by the uncertainty in silicon band parameters, are in rough agreement with what one gets from estimates assuming that this well is a device of type 1. The anisotropy of the 2DEG ESR linewidth is independent of $\alpha$ and is indeed only dependent on one free variable: the momentum relaxation time $\tau_p$ which we assume is directly proportional to the mobility. The magnitude of the relaxation time, on the other hand, is set by the Rashba coefficient together

| Source | Linewidth | Anisotropy |
|---|---|---|
| Ref. [88] <br> 5-30 K <br> $\mu \sim 20$ m²/V-s <br> $n_s \sim 4\times10^{15}$ m⁻² <br> CW-ESR <br> **donor** populated | Exp.: <br> $T_2^{B\parallel z} = 420$ ns (0.15 G) <br> $T_2^{B\perp z} = 140$ ns (0.45 G) <br> Pred.: <br> $T_2^{B\parallel z} = 191$ ns <br> $T_2^{B\perp z} = 60$ ns | Exp.: <br> $T_2^{B\parallel z}/T_2^{B\perp z} = 3$ <br><br> Pred.: <br> $T_2^{B\parallel z}/T_2^{B\perp z} = 3.2$ <br> $T_1^{B\parallel z}/T_1^{B\perp z} = 0.5$ |
| Ref. [84] <br> 5 K <br> $\mu \sim 9$ m²/V-s <br> $n_s \sim 3\times10^{15}$ m⁻² <br> Pulsed-ESR <br> **light** populated | Exp.: <br> $T_2^{B\parallel z} = 3\ \mu$s <br> $T_2^{B\perp z} = 0.24\ \mu$s <br> $T_1^{B\parallel z} = 2\ \mu$s <br> $T_1^{B\perp z} = 3\ \mu$s <br> Pred.: <br> $T_2^{B\parallel z} = 502$ ns <br> $T_2^{B\perp z} = 272$ ns <br> $T_1^{B\parallel z} = 251$ ns <br> $T_1^{B\perp z} = 502$ ns | Exp. <br> $T_2^{B\parallel z}/T_2^{B\perp z} = 12.5$ <br> $T_1^{B\parallel z}/T_1^{B\perp z} = 0.67$ <br><br> Pred.: <br> $T_2^{B\parallel z}/T_2^{B\perp z} = 1.8$ <br> $T_1^{B\parallel z}/T_1^{B\perp z} = 0.5$ |
| Ref. [45] <br><br> $\mu \sim 10$ m²/V-s <br> $n_s \sim 3\times10^{15}$ m⁻² <br> **light/gate** populated | Exp.: <br> $T_2^{B\parallel z} = 12\ \mu$s <br> $T_2^{B\perp z} = 500$ ns <br> Pred.: <br> $T_2^{B\parallel z} = 480$ ns <br> $T_2^{B\perp z} = 193$ ns | Exp.: <br> $T_2^{B\parallel z}/T_2^{B\perp z} = 24$ <br><br> Pred.: <br> $T_2^{B\parallel z}/T_2^{B\perp z} = 1.9$ <br> $T_1^{B\parallel z}/T_1^{B\perp z} = 0.5$ |
| Ref. [82] <br><br> $\mu \sim 5$ m²/V-s <br> $n_s \sim 4\times10^{15}$ m⁻² <br> CW-ESR <br> **donor** populated | Exp.: <br> $T_2^{B\parallel z} = 50$ ns (0.13 G) <br> $T_2^{B\perp z} = 30$ ns (0.215 G) <br> Pred.: <br> $T_2^{B\parallel z} = 31$ ns <br> $T_2^{B\perp z} = 20$ ns | Exp.: <br> $T_2^{B\parallel z}/T_2^{B\perp z} = 1.65$ <br><br> Pred.: <br> $T_2^{B\parallel z}/T_2^{B\perp z} = 1.6$ <br> $T_1^{B\parallel z}/T_1^{B\perp z} = 0.5$ |
| Ref. [37] <br> 4.2 K <br> $\mu \sim 9$ m²/V-s <br> $n_s \sim 4\times10^{15}$ m⁻² <br> ED-ESR <br> **donor** populated | $T_2^{B\parallel z} = 105$ ns (0.6 G) <br> $T_2^{B\perp z} = 49$ ns (1.3 G) <br><br> Pred.: <br> $T_2^{B\parallel z} = 212$ ns <br> $T_2^{B\perp z} = 115$ ns | Exp.: <br> $T_2^{B\parallel z}/T_2^{B\perp z} = 2.2$ <br><br> Pred.: <br> $T_2^{B\parallel z}/T_2^{B\perp z} = 1.8$ <br> $T_1^{B\parallel z}/T_1^{B\perp z} = 0.5$ |

Table 3.1: We calculate relaxation times assuming that the QW is completely populated by the donor layer. Accurate analysis is made difficult due to lack of precise values for mobility and density, which are often not measured directly (or reported) for the specific sample addressed with ESR. The anisotropy does not depend on the Rashba coefficient. Note also that there is some disagreement in the literature as to how to convert from linewidth to a relaxation time, we use the equations derived by Poole in Ref. [62], $T_2 = 2\hbar/\left(\sqrt{3}g\mu_B\Delta H_{pp}^0\right)$, but others may differ by a factor of up to $2\pi$.

with the Fermi momentum.

Table 3.1 details results for four other samples as well, also nominally of type 1 for which measurements of $T_2$ have been performed. Agreement is good for the sample of Ref. [88], particularly for the anisotropy of $T_2$. This is a donor-layer-populated sample measured with CW-ESR. Ref. [37], measured via electrically-detected ESR (ED-ESR), also is in good agreement with the anisotropy predicted by the theory. Comparison with the sample of Ref. [82] also seems to be very good. This is an IBM 2DEG with a density of roughly $4 \times 10^{15}$ m$^{-2}$ and a quantum well thickness of 10 nm, fully donor-layer populated. Agreement is considerably less good for the sample in Ref. [45]. Here the mobility is not well-known, and the sample is partially populated by illumination. So we are lead to believe that there is some difference between the two sets of samples that causes one set to have a larger $T_2$ anisotropy than our theory predicts, even for very similar material parameters (density and mobility). Further experimental work along the lines of measuring the density, mobility, $T_1$, and $T_2$ on the same samples is needed.

The magnitude of the predicted relaxation times is in general well predicted by the theory, but there is a range of error. The position of the ionized centers which populate the well is important for the calculation of the electric field which is thus hard to characterize. Photoelectrons created with light at the bandgap energy may also have symmetry changing effects. This may explain why the Rashba coefficient derived from varying the 2DEG density by light in Ref. [88] appears to be independent of density. It is also important to point out that parallel conductivity (current paths through both the 2DEG and the donor-layer for example) is a common problem in today's SiGe quantum wells and may effect the transport measurements of density and especially mobility making comparison with theory difficult.

Other mechanisms may become important as we leave the parameter range considered in this paper. Electron-electron collisions which do not greatly affect the mobility at low temperatures may start to contribute at higher temperatures and mobilities, as they appear to do in GaAs quantum wells.[35] These collisions will also relax the spin, but the relation between momentum relaxation and spin relaxation is not expected to be the same as for the elastic collisions considered here. At higher magnetic fields, the cyclotron motion of the electrons is important. In the semiclassical picture, when $\omega_c \tau_p \geq 1$, the average value of the momentum perpendicular to the magnetic field shrinks, reducing or even eliminating DP spin relaxation, as has been considered for III-V semiconductors in the fixed magnetic field case.[93, 42] This effect may become important at high mobilities and would be dependent on magnetic field angle, increasing the anisotropy predicted here while also increasing the relaxation times. Quantum effects may become important in this regime however. The wave vector dependence of the conduction band electron g-factor may also lead to relaxation, as has been pointed out recently,[7] but hasn't been considered here. Finally, the addition of details related to the presence of two conduction-band valleys may differentiate further the case of Si from that of GaAs. Golub and Ivchenko [36] have considered spin relaxation in symmetrical ($\alpha$=0) SiGe QWs, where valley domains (even or odd monolayer regions of the QW) may have influence over spin dynamics. Random spin-orbit coupling due to variations in the donor layer charge distribution may also be important in symmetric quantum wells.[74, 73]

Long spin relaxation times on the order of hundreds of nanoseconds to microseconds, found in presently available SiGe quantum wells, hold great promise for both quantum information processing and spintronics. Our results demonstrate that decreasing the reflection asymmetry within the device will appreciably decrease the Rashba coefficient and the consequent spin relaxation at low temperatures. This can be achieved by a symmetric doping profile or using an external electric field to cancel out the field of the ions. They further show that the anisotropy of the ESR linewidth as a function of angle may be a good indicator of 2DEG quality (mobility) independent of transport measurements. As higher mobility and more exotic SiGe heterostructures are grown and characterized, new physics may emerge.

# Chapter 4

# Orbital and spin relaxation in Si Quantum Dots

Quantum computation with solid-state qubits is fundamentally a non-equilibrium undertaking. Information stored in the excited states of quantum systems will eventually relax, corresponding with the emission of energy to the environment. Here we address directly the phonon-mediated relaxation of excited quantum dot states, with or without a spin-flip, across the same valley state in [001] strained silicon lateral quantum dots. Knowing these time scales is important for assessing quantum dot quantum computing feasibility and for the interpretation of transport experiments on present-day silicon quantum dot devices. We find a number of new results directly relevant to spin qubit decoherence and optical readout and initialization schemes.

Our calculations show that orbital relaxation in strained silicon is much faster than bulk silicon in some important cases. In fact, the rate of spontaneous decay from the first excited orbital state in silicon quantum dots is comparable to that of GaAs quantum dots, which are commonly expected to relax more efficiently due to the crystal's piezoelectric nature. This has important implications for optically-induced motional spin-charge transduction (readout) and optical pumping (initialization) of spin qubits, making the former harder and the latter easier. We build on this theory to calculate for the first time the dominant spin relaxation mechanism in lateral silicon quantum dots.

Spin-flip relaxation, or the qubit $T_1$ time, in silicon quantum dots at quantum computing temperatures is as in the donor case due to single phonon emission, mediated by spin-orbit coupling (SOC). In silicon quantum wells, spin-orbit coupling is mainly due to structural inversion asymmetry (or Rashba SOC) from the large electric field in the growth direction. The form of this potential induces spin relaxation even when the magnetic field is perpendicular to the quantum well. By contrast, the dominant bulk-silicon SOC contributions predict infinite $T_1$ times in this most important case for quantum dot quantum computing. Relevant to experimental concerns, we find that the spin-flip rate is proportional to the seventh power of the magnetic field unlike for GaAs quantum dots where the rate as the fifth power. Typical $T_1$ times in lateral silicon quantum dots due to one-phonon processes are found to be *seconds* or much longer at milli-Kelvin temperatures, well suited for quantum information processing applications. We estimate spin relaxation due to two-phonon processes at high temperatures (5 K) to be in the microsecond range.

## 4.1 Introduction

The energy relaxation of localized electronic states acts as a constraint on many schemes for quantum nanodevices. In quantum dot quantum computing (QD QC), for example, single- or few-electron states can act as storage for quantum information while the ability to entangle spatially-separated states may allow for full quantum information processing (QIP).[55] But in order for QIP devices to work, they must operate faster than the fundamental energy and information dissipation mechanisms in the given architecture.[79] Electron spins in silicon seem ideal for such a device due to the availability of isotopically purified $Si^{28}$ (reducing hyperfine-induced decoherence[17]) and silicon's inherently low spin-orbit coupling.[81] As a result, excited spin-states can be extremely robust.[83] Several proposals for spin-based quantum computation in silicon heterostructures have already been proposed.[48, 87, 32]

These lifetimes set an upper bound on qubit coherence and are important for other quantum control processes. In the nomenclature of NMR, these are $T_1$ times and correspond with loss of energy (usually through phonons) to the environment. While the phase coherence time $T_2$, which can be dissipation-less, is often considered the more important single-spin parameter for QC, $T_1$ is on the order of $T_2$ in cases where spin-orbit coupling dominates decoherence processes (in first order). [59][Loss] In addition to decoherence considerations, the lifetime of excited orbital states are relevant to optical pumping schemes, many-phonon relaxation calculations, and other areas of quantum control. Understanding the relaxation mechanisms of excited quantum dots states is key to designing and controlling promising quantum devices in these systems.

## 4.2   Silicon Quantum Well Quantum Dot states

In a biaxially-strained silicon QW, the number of states is doubled relative to GaAs, as both the $\pm z$ conduction band minima are populated but not the $\pm x$ or $\pm y$ minima.[80] The electron wave functions are superpositions of Bloch states at the bottom of the CB(s) of the host semiconductor, so the many-valley nature of silicon (as opposed to a single $\Gamma$-valley in GaAs) takes on great importance. Hence, in the absence of magnetic fields and valley-splitting effects, the electronic ground state in this system is fourfold degenerate. When potentials or boundary conditions that mix the two valleys are present, as they always will be to some extent (see Chapter **??**), there will be excited valley states corresponding to different linear combinations of the valley minima. So each valley state has its own identical set of orbital and spin states and an additional quantum number is needed to specify which valley state the electron occupies. We restrict the present work to physical setups and approximations relevant to quantum computing situations. The system in question is a SiGe-Si-SiGe quantum well (QW) which has been formed into a quantum dot by Schottky top-gates or etched side-gates. (See Chapter 3 for more information on the specific material system.)

A magnetic field splits the degeneracy of the spin states. The valley degeneracy is split by the hard confinement of the QW interfaces (or the impurity potential in that case) and is influenced by a number of parameters, especially the magnitude of the electric field in the growth direction. We assume for simplicity that any static magnetic field is directed perpendicular to the plane of the QW and that the well walls are smooth. In these circumstances the wave function of the QD can be separated as $\psi_{QD}(x, y, z) = \psi_{xy}(x, y)\psi_z(z)$. The wavefunctions can be written in the common Kohn-Luttinger approximation as

$$\psi_{QD}^{\pm} \approx \psi_{xy} \frac{1}{\sqrt{2}} F_z(z) \left[ \alpha_z^{\pm} u_z(z) e^{ik_z z} + \alpha_{-z}^{\pm} u_{-z}(z) e^{-ik_z z} \right], \tag{4.1}$$

where $\psi_{xy}$ is the 2D wave function of the electron, the $\pm$ refer to the symmetric (even) and anti-symmetric (odd) valley states (which one is the ground state is not immediately obvious), $F_z(\mathbf{r})$ is the envelope function satisfying the effective mass equation in the growth direction, $u_j(\mathbf{r}) \exp(i\mathbf{k}_j \cdot \mathbf{r})$ is the Bloch wave at the conduction band minimum $\mathbf{k}_j$, and the $\alpha_i$ are in general complex numbers that give the proportion of the electron wave function in each valley (which is 50% in this case, $|\alpha_i|^2 = 1/2$). In the infinite square well, zero electric field limit, Eq. 4.1 takes on a simpler form: $\psi_{QD}^{+} \approx \sqrt{2} \cos(\pi/L) \cos(k_z z) \psi_{xy}$ and $\psi_{QD}^{-} \approx \sqrt{2} \cos(\pi/L) \sin(k_z z) \psi_{xy}$, excluding the Bloch oscillations. The envelope function is defined by the width of the quantum well and the fast oscillations are determined by the location of the minima in $k$-space. The wavefunctions for a more realistic device, calculated in a tight-binding theory of Ref.[6], are shown in Figure 4.1.

## 4.3   Phonons in strained silicon: deformation potentials

Phonons are quanta of lattice vibrations, that is, mechanical stress. This time-dependent stress, like its static counterpart, alters the band structure by shifting band energies and lifting degeneracies [63, 92]. It is typically assumed that this effect does not change the band curvature (effective masses) but does shift the energy states of interest. The shift in energy of the band edge per unit elastic strain is called the *deformation potential* and is common to all semiconductors and solids. In polar crystals, distortion of the lattice can also create large internal electric fields which affect the electron. This is the piezoelectric interaction. Ionic crystals like GaAs suffer from piezo-phonons, which are often very efficient at electron scattering; silicon, being non-polar and centrosymmetric, has none. While optical strain also exists in materials with two atoms per unit cell, such as silicon, optical phonons in silicon have a narrow bandwidth centered at a much higher energy than a quantum computer would operate. Here, we follow the theory of the electron-phonon deformation potential interaction outlined by Herring

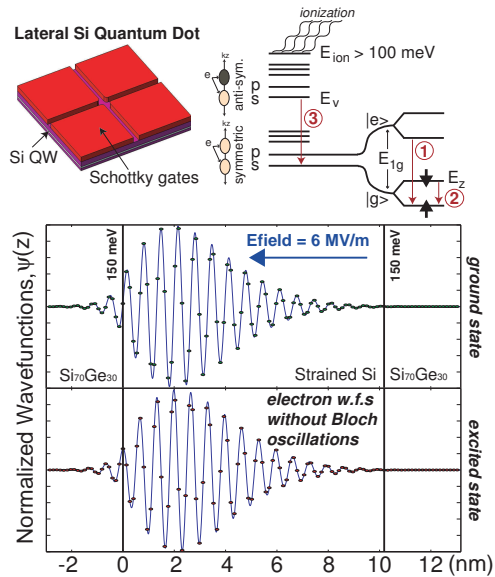

Figure 4.1: The circled numbers represent relevant relaxation processes: (1) orbital relaxation across the first energy gap; (2) spin relaxation of the electron qubit; and (3) valley relaxation. The $z$-component of the electron dot wave function is the output of a 2-band tight-binding calculation (points) which has been interpolated (line) for a typical SiGe heterostructure with a quantum well of 10 nm, barriers of 150 meV, and a large growth direction electric field due to space-charge separation from the donor layer of $6 \times 10^6$ V/m.

| Constant | Value |
|---|---|
| $e$ | $1.6 \times 10^{-19}$ C |
| $\hbar$ | $1.05 \times 10^{-34}$ J s |
| $c$ | $3 \times 10^{8}$ m/s |
| $\epsilon_0$ | $8.85 \times 10^{-12}$ C$^2$/Nm$^2$ |
| $\epsilon_{Si}$ | 11.8 |
| $g_\parallel$(Si) | 1.999 |
| $g_\perp$(Si) | 1.998 |
| $\Xi_u$(Si) | 9.29 eV=1.49E-18 J |
| $\Xi_d$(Si) | -10.7 eV=-1.71E-18 J |
| $\rho$(Si) | 2330 kg/m$^3$ |
| $v_l$(Si) | 9330 m/s |
| $v_t$(Si) | 5420 m/s |
| $k_B$ | 1.38E-23 J/K |

Table 4.1: Physical constants and materials parameters for silicon.

and VogleV [40] and apply it to silicon in the tradition of others [39, 12].

The energy shift of a non-degenerate band edge due to strain is given by

$$H_{eL} = \sum_{\alpha,\beta} U_{\alpha\beta}\Xi_{\alpha\beta}^{(i)},\qquad(4.2)$$

where

$$U_{\alpha\beta} = \frac{1}{2}\left(\frac{\partial u_i}{\partial x_i} + \frac{\partial u_j}{\partial x_i}\right) = U_{\alpha\beta}$$

is the strain tensor and $\mathbf{\Xi}^{(i)}$ is the deformation potential tensor for the $i$th silicon valley. Since our electron is confined to massively strained [001] silicon, it exists only in the [001] valleys which have deformation potential tensors given by

$$\mathbf{\Xi}^{\pm z} = \Xi_d\delta_{\alpha\beta} + \Xi_u K_\alpha^{(i)} K_\beta^{(i)} = \begin{pmatrix} \Xi_d & 0 & 0 \\ 0 & \Xi_d & 0 \\ 0 & 0 & \Xi_d + \Xi_u \end{pmatrix},\qquad(4.3)$$

where $\Xi_d$ relates to pure dilation and $\Xi_u$ is associated with shear strains. For silicon under compressive stress along [001], opposing valleys move in energy identically. In order to associate phonon modes with strain, we can expand the unit cell displacement $\mathbf{u}(\mathbf{r})$ in plane waves,

$$\mathbf{u}(\mathbf{r}) = \sum_{q,\mathbf{s}} \left[ \mathbf{e}_\mathbf{s}(\mathbf{q}) a_{qs} e^{i\mathbf{q}\cdot\mathbf{r}} + \mathbf{e}_\mathbf{s}^*(\mathbf{q}) a_{qs}^* e^{-i\mathbf{q}\cdot\mathbf{r}} \right],$$

and associate each phonon of wave vector $\mathbf{q}$ and polarization $\mathbf{e}_s$ (2 transverse and 1 longitudinal) with its corresponding component in this Fourier expansion. This results in a strain tensor due to a phonon,

$$U(t)_{\alpha\beta} = \frac{i}{2} \left[ (\mathbf{e}(s)_\alpha q_\beta + \mathbf{e}(s)_\beta q_\alpha) a_{qs}^* e^{-i\mathbf{q}\cdot\mathbf{r}} + (\mathbf{e}(s)_\alpha q_\beta + \mathbf{e}(s)_\beta q_\alpha) a_{qs} e^{i\mathbf{q}\cdot\mathbf{r}} \right]. \tag{4.4}$$

The amplitudes $a_q$ and $a_q^*$ are now phonon operators with matrix elements (normalization)

$$\begin{aligned} \langle n_q - 1 | \, a_q \, | n_q \rangle &= \sqrt{\hbar n_q / 2 M_c \omega_q}, \\ \langle n_q + 1 | \, a_q^* \, | n_q \rangle &= \sqrt{\hbar \left( n_q + 1 \right) / 2 M_c \omega_q}, \end{aligned}$$

where $M_c = \rho V$ is the mass of the crystal and $n_q = 1/\left( e^{\hbar\omega_\mathbf{q}/kT} - 1 \right)$ is the phonon occupation number of the mode with wave number $\mathbf{q}$. The Fourier component of the energy shift due to the $\mathbf{q}$th mode with polarization $\mathbf{e}_s$ is then given by

$$\begin{aligned} \delta E_{\mathbf{q},\mathbf{s}} &= \left\{ \Xi_d \left( \mathbf{e}_s \cdot \mathbf{q} \right) + \frac{1}{2} \Xi_u \left[ \left( \mathbf{q} \cdot \mathbf{K}_i \right) \left( \mathbf{e}_s \cdot \mathbf{K}_i \right) + \left( \mathbf{q} \cdot \mathbf{K}_j \right) \left( \mathbf{e}_s \cdot \mathbf{K}_j \right) \right] \right\} a_{qs} e^{i\mathbf{q}\cdot\mathbf{r}} \\ &\quad + \left\{ \Xi_d \left( \mathbf{e}_s \cdot \mathbf{q} \right) + \frac{1}{2} \Xi_u \left[ \left( \mathbf{q} \cdot \mathbf{K}_i \right) \left( \mathbf{e}_s \cdot \mathbf{K}_i \right) + \left( \mathbf{q} \cdot \mathbf{K}_j \right) \left( \mathbf{e}_s \cdot \mathbf{K}_j \right) \right] \right\} a_{qs}^* e^{-i\mathbf{q}\cdot\mathbf{r}}, \end{aligned}$$

where $\mathbf{K}_i$ and $\mathbf{K}_j$ are unit vectors to the $i$th and $j$th valleys. The complete electron-phonon Hamiltonian must be summed over phonon modes and polarizations. For a [001] strained-silicon quantum well, the full electron-phonon Hamiltonian can be written succinctly as

$$H_{ep}(\mathbf{q}, s) = \sum_{s=1}^{3} \sum_{\mathbf{q}} i \left[ a_{qs}^* e^{-i\mathbf{q}\cdot\mathbf{r}} + a_{qs} e^{i\mathbf{q}\cdot\mathbf{r}} \right] q \left( \Xi_d \hat{e}(t)_x \hat{q}_x + \Xi_d \hat{e}(t)_y \hat{q}_y + \left( \Xi_d + \Xi_u \right) \hat{e}(t)_z \hat{q}_z \right), \tag{4.5}$$

where the hat indicates a unit vector.

We are especially concerned with the anisotropic effects due to the massive strain of the system in question. As can be seen from Eq.4.3 and the deformation constant values in Table 4.1, the shift in energy of a specific valley due to an acoustic phonon is very anisotropic. In the case of bulk silicon, the six conduction band minima are equidistant from the $\Gamma$-point and thus form an isotropic response to phonon deformations. This means essentially that transverse phonons will not contribute to the relaxation times for intervalley transitions of the same symmetry (that is, $\alpha_n^{(i)} = \alpha_n^{(j)}$ for initial state $i$ and final state $j$). Another way to see this is to consider the electron-lattice matrix element between different states on the same minima (Equation 3.29 of Ref. [39] or Chapter 2),

$$\left\langle \psi_m^{(i)} \left| H_{eL} \right| \psi_n^{(i)} \right\rangle_{qt} = a_{qt} \left[ i\mathbf{e}_t(\mathbf{q}) \cdot \Xi^{(i)} \cdot \mathbf{q} \right] f_{mn}^{(i)}(\mathbf{q}) + c.c.$$

where $f_{mn}^{(i)}(\mathbf{q}) = \int F_m^{(i)}(\mathbf{r}) e^{i\mathbf{q}\cdot\mathbf{r}} F_n^{(i)} d\mathbf{r}$. The matrix element between two complete states is then

$$\left\langle \sum \alpha_m^{(i)} \psi_m^{(i)} \left| H_{eL} \right| \sum \alpha_n^{(i)} \psi_n^{(i)} \right\rangle_{qt} = a_{qt} \left[ i\mathbf{e}_t(\mathbf{q}) \cdot \sum \alpha_m^{(i)} \alpha_n^{(i)} \Xi^{(i)} \cdot \mathbf{q} \right] f_{mn}^{(i)}(\mathbf{q}) + c.c.$$

|         | Longitudinal $(s = l)$ | Transverse $(s = t_1)$ | Transverse $(s = t_2)$ |
|---------|:---------------------:|:----------------------:|:----------------------:|
| $e_x$   | $\sin\theta\cos\phi$  | $\sin\phi$             | $-\cos\theta\cos\phi$  |
| $e_y$   | $\sin\theta\sin\phi$  | $-\cos\phi$            | $-\cos\theta\sin\phi$  |
| $e_z$   | $\cos\theta$          | $0$                    | $\sin\theta$           |

Table 4.2: Polarization components.

It's easy to see from the above equation that if $\sum \alpha_m^{(i)} \alpha_n^{(i)} \Xi^{(i)}$ is proportional to the identity matrix (assuming $\alpha_m = \alpha_n = 1$), then the transverse phonon matrix elements must be zero since $\mathbf{e}_{t1}(\mathbf{q}) \perp \mathbf{e}_{t2}(\mathbf{q}) \perp \mathbf{q}$. The point is that in strained silicon, only the $\pm z$ minima are occupied so unlike the bulk silicon case, transverse phonons will contribute. This turns out to be very important in relaxation calculations since the velocity of transverse phonons in silicon is half that of longitudinal phonons. More on this below.

## 4.4 Orbital relaxation in strained Si Quantum Dots

Spontaneous emission of a single acoustic phonon is the dominant relaxation mechanism for excited electronic states in silicon at QC temperatures (<100s mK) (by empirical evidence for silicon donors [28, 29]). In standard deformation potential theory, the electron-phonon interaction is constructed by relating the change in electronic energies to applied strains due to long-wavelength phonons.[40] The phonon-induced energy shift is very anisotropic for a silicon conduction band valley. Because of this the results we will obtain for strained silicon are quite different from those of bulk silicon obtained previously.[1]

To re-cap, in the bulk, the six conduction band minima are equidistant from the $\Gamma$-point and thus form an isotropic response to phonon deformations. So the angular integral over transverse phonons averages to zero. For example, the transition from the ground, 1s-like symmetric state to the 2p-like, symmetric state in bulk silicon has no transverse phonon contribution.[1] The same transition in strained silicon does, because the cubic symmetry of the six minima has been broken by strain. This greatly increases the relaxation rate since the velocity comes into the rate equations with a very high power as we will now show.

Fermi's Golden Rule, between states of arbitrary spin,

$$\Gamma = \frac{2\pi}{\hbar} \left| \langle n\delta_n | H_{ep} | m\delta_m \rangle \right|^2 \delta \left( E_{ph} - E_{n\delta_n m\delta_m} \right),$$ (4.6)

is the basis for our relaxation rate calculations. Here, $|n\delta_n\rangle$ is the state of the electron in the dot on level $n$ with spin state $\delta_n$, including the effect of a magnetic field. We assume an isotropic phonon spectrum such that the energy of the phonon is $E_{ph} = \hbar\omega_{\mathbf{q}s}$, where $\omega_{\mathbf{q}s} = v_s \mathbf{q}$ and $v_s$ is the velocity of the mode $s$. Setting $\delta_n = \delta_m$ for orbital relaxation without a spin-flip, $E_{n\delta_n m\delta_m} = E_{mn}$, $V$ to the electron-phonon interaction of Equation 4.5, summing over phonon modes, and using $\hat{q} = (\sin\theta\cos\phi, \sin\theta\sin\phi, \cos\theta)$ and $\hat{e}_l = \hat{q} \perp \hat{e}_{t1} \perp \hat{e}_{t2}$ (see Table 4.2) for the wave vector and polarization vectors and using the electric dipole (ED) approximation, $e^{i\vec{q}\cdot\vec{r}} \approx 1 + i\vec{q}\cdot\vec{r}$, we find for the phonon-relaxation rate,

$$\Gamma_{mn}^{ED} = \frac{|E_{mn}|^5}{\hbar^6 \pi \rho_{Si}} \left\{ \left( |M_x|^2 + |M_y|^2 \right) \Upsilon_{xy} + |M_z|^2 \Upsilon_z \right\} (n_q + 1),$$ (4.7)

| | Si QD (ED) | Si QD (exact) | GaAs QD | Bulk Si Donor (ED) |
|---|---|---|---|---|
| $E_{1g}$ | Eq. 4.7 | Eq. 4.13 | Ref. [49] | 2p($A_1$)→1s($A_1$) : Ref. [1] |
| 0.1 meV | $3 \times 10^{-9}$ s | $4 \times 10^{-8}$ s | $3 \times 10^{-9}$ s | $3 \times 10^{-7}$ s |
| 0.5 meV | $5 \times 10^{-12}$ s | $3 \times 10^{-10}$s | $2 \times 10^{-8}$ s | $1 \times 10^{-10}$ s |
| 1 meV | $3 \times 10^{-13}$ s | $6 \times 10^{-11}$ s | $3 \times 10^{-8}$ s | $4 \times 10^{-12}$ s |
| 3 meV | $4 \times 10^{-15}$ s | $2 \times 10^{-11}$ s | $1 \times 10^{-7}$ s | $1 \times 10^{-14}$ s |

Table 4.3: Comparison of like-spin, orbital relaxation times for several fundamental dot sizes at 100 mK.

where

$$\Upsilon_{xy} = \frac{35\Xi_d^2 + 14\Xi_d\Xi_u + 3\Xi_u^2}{210v_l^7} + \frac{2\Xi_u^2}{105v_t^7},$$

$$\Upsilon_z = \frac{35\Xi_d^2 + 42\Xi_d\Xi_u + 15\Xi_u^2}{210v_l^7} + \frac{\Xi_u^2}{35v_t^7},$$

and the matrix elements are $\vec{M} = \langle m|\vec{r}|n\rangle$. The single electron wavefunctions can be calculated by solving the Poisson and Schroedinger equations directly as in Ref. [32] or, as is normally done, can be approximated by parabolic states defined by the fundamental energies $E_{21} = \hbar\omega_{x,y}$. These would give $M_{x,y}^{21} = \sqrt{\hbar/m^*\omega_{x,y}} = \sqrt{\hbar^2/m^*E_{1g}}$. Because of the fifth power dependence on the energy gap, $E_{mn}$, of these equations, an accurate value for this parameter is much more important than equivalent accuracy in the wave function matrix elements. The dipole approximation is valid until roughly $ql_0 >> 1$ ($l_0$ is the size of the dot in largest dimension) when the relaxation rate starts to decrease due to phonon bottleneck effects. We treat this case explicitly in the next section.

Table ?? compares the output of Eq.4.7 for orbital relaxation in lateral silicon quantum dots to the times for GaAs quantum dots and bulk silicon for typical fundamental dot energies. Excited orbital states in strained silicon typically relax in nanoseconds or faster, corresponding to a level broadening of a micro-eV or wider (assuming $E = h/\Gamma$).

## 4.4.1 Beyond the electric dipole approximation: the phonon bottleneck effect

The basic reason for vanishing relaxation when $l_0$ is reduced is the impossibility of satisfying simultaneously energy and momentum conservation during an electron-acoustic-phonon scat-tering event [2]. Mathematically, the phonon bottleneck effect is due to the fast oscillating exponential terms in the matrix elements of the Golden rule transition equation. The momentum matrix element between the localized wave functions vanishes beyond $q \approx 2\pi/L_z$ [3]. Physically, this corresponds to the wavelength of the emitted phonon, $\lambda = 2\pi\hbar v_s/E_{mn}$, being smaller than the dimensions of the electron wave function. The momentum conservation condition is embodied in this electron-phonon matrix element. The matrix element also begins to diminish when $ql_0 \gg 1$, as the exponential begins to oscillate appreciably in this case. For more information see the seminal papers on the phonon bottleneck effect [4, 3, 5, 2]. Table 4.4 charts this transition for silicon quantum dots. Note that in a lateral quantum dot, unlike excitonic quantum dots, Auger processes and electron-hole scattering do not play a role in negating phonon bottleneck. Assuming parabolic dots, only dots with fundamental energies $E_{1g} = \Delta \approx 10$ meV are small enough (and virtually impossible to construct with laterally-gated devices) for phonon-bottleneck effects to make a significant impact on increasing the orbital relaxation times.

| $E_{1g}$ (meV) | $l_0$ (nm) - Si | $\lambda$ (nm) - Si($v_t$) |
|:---:|:---:|:---:|
| 0.05 | 63 | 447 |
| 0.1 | 45 | 223 |
| 0.5 | 20 | 47 |
| 1 | 14 | 22 |
| 2 | 10 | 11 |
| 3 | 8.0 | 7.4 |
| 10 | 4.5 | 2.2 |

Table 4.4: Numbers relevant to the phonon bottleneck effect assuming a quantum dot of lateral size $l_0 = \hbar/\sqrt{2m^* E_{1g}}$ .

To calculate Eq. 4.6 exactly, we begin with the full expression for the electron-phonon matrix element between the electronic silicon crystal states (Eq. 4.1),

$$
\begin{aligned}
M_{mn} &= \langle n | H_{ep} | m \rangle = \int \psi_m^* (H_{ep}) \psi_n dV \\
&= \sum_{i,j} \alpha_m^i \alpha_n^j \left[ \Xi_d (\mathbf{e}_s \cdot \mathbf{q}) + \Xi_u (\mathbf{q} \cdot \mathbf{K}_i)(\mathbf{e}_s \cdot \mathbf{K}_i) \right] a_q^* \int F_i^* F_j u_{k_i}^* u_{k_j} e^{-i(-\mathbf{k}_i + \mathbf{k}_j + \mathbf{q}) \cdot \mathbf{r}} dV \quad (4.8)
\end{aligned}
$$

We proceed rigorously following the derivation by Castner [12]. This derivation will be useful when valley relaxation is considered. A function which is periodic with the period of the lattice may be expanded in a Fourier series in the reciprocal lattice vectors $\mathbf{K}_\nu$, so

$$u_{k_i}^*(\mathbf{r})u_{k_j}(\mathbf{r}) \quad = \quad \sum_\nu C_{k_i}^\nu e^{i\mathbf{K}_\nu \cdot \mathbf{r}},$$

and the integral in $M_{mn}$ becomes

$$\sum_\nu C_{k_i}^\nu \int |F_i|^2\, e^{-i(-\mathbf{k}_i+\mathbf{k}_j+\mathbf{q}-\mathbf{K}_\nu)\cdot \mathbf{r}} d\mathbf{r}.$$

The envelope probability can be Fourier transformed,

$$|F_i(\mathbf{r})|^2 \quad = \quad \frac{1}{(2\pi)^3}\sum_{\mathbf{k}''} f(\mathbf{k}'')e^{i\mathbf{k}''\cdot \mathbf{r}}.$$

Plugging this into the integral in $M_{mn}$ gives

$$\sum_\nu C_{k_i}^\nu \int \frac{1}{(2\pi)^3}\sum_{\mathbf{k}''} f(\mathbf{k}'')e^{-i\left(-\mathbf{k}_i+\mathbf{k}_j+\mathbf{q}-\mathbf{K}_\nu-\mathbf{k}''\right)\cdot \mathbf{r}} d\mathbf{r}$$

which equals

$$\sum_\nu C_{k_i}^\nu \frac{1}{(2\pi)^3}\sum_{\mathbf{k}''} f(\mathbf{k}'')(2\pi)^3\delta\left(\mathbf{k}_i-\mathbf{k}_j-\mathbf{q}+\mathbf{K}_\nu+\mathbf{k}''\right) = \sum_\nu C_{k_i}^\nu f^{ij}(-\mathbf{k}_i+\mathbf{k}_j+\mathbf{q}-\mathbf{K}_\nu).$$

Finally, the matrix element is given by

$$M_{mn} \quad = \quad \sum_{i,j} \alpha_m^i \alpha_n^j \left[\Xi_d\left(\mathbf{e}_s\cdot \mathbf{q}\right)+\Xi_u\left(\mathbf{q}\cdot \mathbf{K}_i\right)\left(\mathbf{e}_s\cdot \mathbf{K}_i\right)\right] a_q^* \sum_\nu C_{k_i}^\nu f^{ij}(-\mathbf{k}_i+\mathbf{k}_j+\mathbf{q}-\mathbf{K}_\nu) \tag{4.9}$$

We are calculating an intra-valley scattering process (orbital relaxation across the lowest valley state) so $\mathbf{K}_\nu = 0$ is the dominant term, $\mathbf{k}_i = \mathbf{k}_j$, and $\alpha_m = \alpha_n$ which gives

$$M_{mn} \quad = \quad \sum_{i,j} \alpha_m^i \alpha_n^j \left[\Xi_d\left(\mathbf{e}_s\cdot \mathbf{q}\right)+\Xi_u\left(\mathbf{q}\cdot \mathbf{K}_i\right)\left(\mathbf{e}_s\cdot \mathbf{K}_i\right)\right] a_q^* C_{k_i}^0 f^{mn}(\mathbf{q}),$$

and for the most relevant transition,

$$
\begin{aligned}
M_{21} &= \sum_{i,j} \alpha_m^i \alpha_n^j \left[ \Xi_d q_l + \Xi_u e_z q_z \right] a_q^* C_{k_i}^0 f^{12}(\mathbf{q}) \\
&= \left[ \Xi_d q_l + \Xi_u e_z q_z \right] a_q^* C_{k_i}^0 f^{12}(\mathbf{q}),
\end{aligned}
$$

where $q_l = q_\Delta$ for longitudinal phonons but zero for transverse phonons ($q_\Delta = \Delta/hv_s$). $C_{k_i}^0$ is the first coefficient in the Bloch wave expansion and is typically assumed to be 1 in this treatment.

The envelope function of the ground state QD wave function in the absence of a magnetic field in the lowest approximation is a multiple of Gaussians,

$$
F^{(1)}(\mathbf{r}) = F(x, y, z) = F(x)F(y)F(z),
$$

where

$$
F(x) = (2/\pi)^{1/4} \, x_0^{-1/2} \exp\left(-x^2/x_0^2\right),
$$

$$
F(y) = (2/\pi)^{1/4} \, y_0^{-1/2} \exp\left(-y^2/y_0^2\right),
$$

and

$$
F(z) = (2/\pi)^{1/4} \, z_0^{-1/2} \exp\left(-z^2/z_0^2\right).
$$

The excited state, if $y_0 >> x_0 >> z_0$, is

$$
F^{(2)}(\mathbf{r}) = F(x, y, z) = F(x)F^{(2)}(y)F(z),
$$

where

$$
F^{(2)}(y) = \left(2/\sqrt{y_0^3}\right) (2/\pi)^{1/4} \, y \exp\left(-y^2/y_0^2\right).
$$

Then, the overlap integral is given by

$$
f^{(12)}(\mathbf{q}) = \int F^{(2)}(\mathbf{r}) e^{i\mathbf{q}\cdot\mathbf{r}} F^{(1)}(\mathbf{r}) d\mathbf{r},
$$

$$
\begin{aligned}
f^{(1)}(q_x)f^{(1)}(q_y)f^{(1)}(q_z) &= \int F^{(1)}(x)^2 e^{iq_x x} dx \int F^{(2)}(y) e^{iq_y y} F^{(1)}(y) dy \int F^{(1)}(z)^2 e^{iq_z z} dz \\
&= \exp\left(-\frac{1}{8}x_0^2 q_x^2\right)\frac{iy_0}{2}q_y \exp\left(-\frac{1}{8}y_0^2 q_y^2\right)\exp\left(-\frac{1}{8}z_0^2 q_z^2\right).
\end{aligned}
$$

Plugging these into the Golden Rule, we find that

$$
\begin{aligned}
\Gamma_{21} &= \frac{2\pi}{\hbar}\sum_{\mathbf{q},s}|M_{21}|^2\,\delta\left(\Delta - \hbar\omega_{\mathbf{q},s}\right), &&(4.10)\\
&= \frac{2\pi}{\hbar}\sum_s \frac{V}{(2\pi)^3}\int_0^\infty q^2 dq \int d\Omega\,|M_{21}|^2\,\delta\left(E_{21} - \hbar v_s q\right) &&(4.11)\\
&= \frac{2\pi}{\hbar}\sum_s \frac{V}{(2\pi)^3}\int_0^\infty q^2 dq \int \sin\theta d\theta d\phi\,|M_{21}|^2\,\frac{1}{|-\hbar v_s|}\delta\left(q - \frac{E_{21}}{\hbar v_s}\right), &&(4.12)
\end{aligned}
$$

continuing,

$$
\begin{aligned}
\Gamma_{21} &= \frac{2\pi}{\hbar^2}\sum_s \frac{V}{(2\pi)^3}\int_0^\infty q^2 dq \int \sin\theta d\theta d\phi \frac{1}{v_s}\left|\left[\Xi_d q_l + \Xi_u e_z q_z\right]a_q^* C_{k_i}^0 f^{12}(\mathbf{q})\right|^2\delta\left(q - q_\Delta\right)\\
&= \frac{V}{(2\pi)^2\hbar^2}\left(C_{k_i}^0\right)^2\sum_s \int_0^\infty q^2 dq \int \sin\theta d\theta d\phi \frac{1}{v_s}\left(\frac{\hbar\left(n_q+1\right)}{2\rho_{Si}V\omega_{q,s}}\right)\left|\left[\Xi_d q_l + \Xi_u e_z q_z\right]f^{12}(\mathbf{q})\right|^2\delta\left(q - q_\Delta\right)\\
&= \frac{\left(n_q+1\right)\left(C_{k_0}^0\right)^2}{2(2\pi)^2\rho_{Si}\hbar}\sum_s \int_0^\infty q \int \sin\theta \frac{1}{v_s^2}\left|\left[\Xi_d q_l + \Xi_u e_z q_z\right]f^{12}(\mathbf{q})\right|^2\delta\left(q - q_\Delta\right)d\theta d\phi dq\\
&= \frac{\left(n_q+1\right)\left(C_{k_0}^0\right)^2}{2(2\pi)^2\rho_{Si}\hbar}\sum_s q_{\Delta s}\int \sin\theta \frac{1}{v_s^2}\left|\left[\Xi_d q_{\Delta l} + \Xi_u e_z q_{\Delta s}\hat{q}_z\right]f^{12}(q_{\Delta s}\hat{\mathbf{q}})\right|^2 d\theta d\phi\\
&= \frac{\left(n_q+1\right)\left(C_{k_0}^0\right)^2}{2(2\pi)^2\rho_{Si}\hbar}\sum_s q_{\Delta s}I_s.
\end{aligned}
$$

We are left with calculating the three angular integrals $I$ which have units kg$^2$/s$^2$,

$$
\begin{aligned}
I_s &= \int \sin\theta \frac{1}{v_s^2}\left|\left[\Xi_d q_{\Delta l} + \Xi_u e_z q_{\Delta s}\hat{q}_z\right]f^{12}(q_{\Delta s}\hat{\mathbf{q}})\right|^2 d\theta d\phi\\
&= \int \sin\theta \frac{1}{v_s^2}\left[\Xi_d q_{\Delta l} + \Xi_u e_z q_{\Delta s}\hat{q}_z\right]^2 \exp\left(-\frac{1}{4}x_0^2 q_x^2\right)\frac{y_0^2}{4}q_y^2 \exp\left(-\frac{1}{4}y_0^2 q_y^2\right)\exp\left(-\frac{1}{4}z_0^2 q_z^2\right)d\theta d\phi.
\end{aligned}
$$

We can immediately point out that $I_{t_2} = 0$ because $e_z(t_1) = 0$. Then,

$$
\begin{aligned}
I_s \;=\; & \int \sin\theta \frac{q_{\Delta s}^2 y_0^2}{4 v_s^2} \left[\Xi_d q_{\Delta l} + \Xi_u e_{zs} q_{\Delta s} \cos\theta\right]^2 \exp\left(-\frac{1}{4} x_0^2 q_{\Delta s}^2 \sin^2\theta \cos^2\phi\right) \\
& \times \sin^2\theta \sin^2\phi \exp\left(-\frac{1}{4} y_0^2 q_{\Delta s}^2 \sin^2\theta \sin^2\phi\right) \exp\left(-\frac{1}{4} z_0^2 q_{\Delta s}^2 \cos^2\theta\right) d\theta d\phi.
\end{aligned}
$$

If we assume an approximately circular dot in $x$ and $y$, then using $\cos^2\phi + \sin^2\phi = 1$ we can do the $\phi$ integral easily,

$$
\begin{aligned}
I_s \;=\; & \pi \exp\left(-\frac{1}{4} x_0^2 q_{\Delta s}^2\right) \int \frac{q_{\Delta s}^2 y_0^2}{4 v_s^2} \left[\Xi_d q_{\Delta l} + \Xi_u e_{zs} q_{\Delta s} x\right]^2 (1 - x^2) \exp\left(\frac{1}{4}(x_0^2 - z_0^2) q_{\Delta s}^2 x^2\right) dx, \\
I_l \;=\; & \pi \exp\left(-\frac{1}{4} x_0^2 q_{\Delta s}^2\right) \frac{q_{\Delta l}^2 y_0^2}{4 v_l^2} \left[\Xi_d^2 q_{\Delta l}^2 (A_l^0 - A_l^2) + \Xi_d \Xi_u q_{\Delta l}^2 (A_l^2 - A_l^4) + \Xi_u^2 q_{\Delta l}^2 (A_l^4 - A_l^6)\right], \\
I_{t_2} \;=\; & \pi \exp\left(-\frac{1}{4} x_0^2 q_{\Delta t}^2\right) \frac{q_{\Delta t}^2 y_0^2}{4 v_t^2} \Xi_u^2 q_{\Delta t}^2 \left[A_t^2 - 2 A_t^4 + A_t^6\right],
\end{aligned}
$$

where

$$
A_s^n \;=\; \int_1^{-1} x^n \exp\left(\frac{1}{4}(x_0^2 - z_0^2) q_{\Delta s}^2 x^2\right).
$$

Finally, the orbital relaxation rate for a parabolic dot (in all three dimensions) from its first excited state is given by (with $E_{21} = \Delta$, the common notation)

$$
\begin{aligned}
\Gamma_{21} \;=\; & \frac{(n_q + 1)\left(C_{k_0}^0\right)^2}{2(2\pi)^2 \rho_{Si} \hbar} \frac{\pi y_0^2}{4} \left\{ \frac{\exp\left(-\frac{1}{4} x_0^2 q_{\Delta l}^2\right)}{v_l^2} \frac{\Delta^5}{\hbar^5 v_l^5} \left[\Xi_d^2 (A_l^0 - A_l^2) + \Xi_d \Xi_u (A_l^2 - A_l^4) + \Xi_u^2 (A_l^4 - A_l^6)\right] \right. \\
& \left. + \frac{\exp\left(-\frac{1}{4} x_0^2 q_{\Delta t}^2\right)}{v_t^2} \frac{\Delta^5}{\hbar^5 v_t^5} \Xi_u^2 \left[A_t^2 - 2 A_t^4 + A_t^6\right] \right\}.
\end{aligned}
\tag{4.13}
$$

The results are plotted in Figure 4.3. The exact solution for the orbital relaxation rate begins to diverge from the electric-dipole approximation early on and never falls below $10^{-11}$ seconds or so. Despite this, the electric-dipole approximation holds well for small energy gaps, $0.1 - 2$ meV, where our quantum computer will most likely operate. This figure shows that there is no significant benefit in going beyond a few meV. Only around 10 meV (impractical for a real device) does the relaxation time start to increase, and slowly.

## 4.5 Spin relaxation of a Si Quantum Dot qubit, $T_1$

Our expressions for orbital relaxation in strained silicon can be extended to the case of a spin-flip transition due to spin-orbit coupling (SOC) in a QW. Again, relaxation via a phonon is the dominant cooling mechanism, in this case mediated by SOC which mixes pure spin states. We expect that the hyperfine interaction with spin-1/2 nuclei is not significant for $T_1$ processes [89, 50] (all the more certain in isotopically purified $Si^{28}$). Spin-flips at QC temperatures are thought to be dominated by phonon emission and absorption mediated by SOC (and known to be for donors). Structural inversion asymmetry or Rashba SOC is the dominant SOC in silicon quantum wells and we will now calculate its contribution, which is nonzero, to $T_1$. Rashba SOC in silicon quantum wells is due to the large electric field common to state-of-the-art SiGe heterostructures. The nature of this SOC has been well described elsewhere (Ref. [81] and Chapter 3). We are fortunate in that its magnitude is quite small (about a thousand times smaller than GaAs) so its effect on energy level splitting is negligible. This validates our use of perturbation theory with electron wavefunctions and energy levels from SOC-free Hamiltonians.

To date there is no finite spin-flip time prediction for lateral silicon quantum dots when $B$ is parallel to $z$. Previous theories for $T_1$ in silicon have been based on the two dominant mechanisms relevant to P:Si donors: the valley-repopulation mechanism (bulk SOC mixing with the six nearby 1s-like states) and the one-valley mechanism (bulk SOC mixing with continuum states).[39, 66, 65] Both mechanisms are independent of the size and shape of the localized electron wave function. The former, we showed rigorously in Chapter 2 and Ref. [80], goes away with [001] strain while the latter is slightly modified [34] and goes to zero for certain directions of the static magnetic field, particularly the [001] direction (the most relevant to QC), for both one and two-phonon processes.

### 4.5.1 Roth-Hasegawa valley mechanisms

For donor states in bulk silicon, Roth and Hasegawa [39, 66, 65] identified two mechanisms that have been confirmed experimentally up to 2 Kelvin which describe spin relaxation via a modulation of the system's g-factor by acoustic phonons. [89] Both mechanisms are direct single-phonon processes. The g-factor for a given conduction band state can be written as a sum over the g-factors at each conduction band minima,

$$\mathbf{g} = \sum_i C_i \mathbf{g}_i,$$

where $C_i^2$ is the population of the single electron at the $i$th valley and $\mathbf{g}^{(i)} = g_\perp \delta_{\alpha\beta} + (g_\parallel - g_\perp) K_\alpha^{(i)} K_\beta^{(i)}$. The *valley-repopulation mechanism* is due to mixing between the symmetric ground state of the donor electron and the split-off doublet state where a phonon changes the $C_i's$. The *one-valley mechanism* is due to phonon-induced modulation of the $\mathbf{g}_i$ themselves and subsequent mixing with nearby conduction bands which are coupled through an inter-band deformation potential.

In Ref. [80], we showed rigorously how the valley-repopulation contribution to the spin-flip time goes away with increasing [001] compressive strain. This can be seen easily qualitatively. The population sets $\alpha$ describing the lowest six conduction states are given by [89]

$$
\begin{aligned}
Singlet: \quad & \alpha_{11} = \tfrac{1}{\sqrt{6}}(1,\, 1,\, 1,\, 1,\, 1,\, 1) \\
Doublet: \quad & \alpha_{21} = \tfrac{1}{\sqrt{12}}(-1,-1,-1,-1,2,2) \\
& \alpha_{22} = \tfrac{1}{\sqrt{4}}(1,\, 1,\, -1,\, -1,\, 0,\, 0) \\
Triplet: \quad & \alpha_{31} = \tfrac{1}{\sqrt{2}}(1,-1,\, 0,\, 0,\, 0,\, 0) \\
& \alpha_{32} = \tfrac{1}{\sqrt{4}}(0,\, 0,\, 1,\, -1,\, 0,\, 0) \\
& \alpha_{33} = \tfrac{1}{\sqrt{4}}(0,\, 0,\, 0,\, 0,\, 1,\, -1),
\end{aligned}
$$

in the basis $(x, -x, y, -y, z, -z)$. SOC mixes the singlet ground state only with one of the doublet states. With compressive strain in the $z$ direction, the electron increasingly exists only in the $\alpha_{11}$ and $\alpha_{33}$ states—the symmetric and antisymmetric valley states—which are not mixed, giving a vanishing matrix element. So, in the quantum well limit, the valley-repopulation contribution to the spin-flip rate is negligible. It is important to note that we have only considered mixing to the six lowest s-like states of the donor. Higher excited states have been disregarded due to their distance in energy.

The one-valley mechanism, however, is relevant to the QD case. Roth, based on her work calculating the $g$-tensor in silicon, showed that in bulk silicon, the contribution from mixing with nearby bands is described by the Hamiltonian

$$H_{one-valley}^{bulk} = A\beta(U_{xy}(\sigma_x H_y + \sigma_y H_x) + c.p.),$$

where c.p. stands for cyclic permutations. This Hamiltonian can be derived from the method of invariants as well. She also showed, through perturbation and group theory, that the dominant contribution to $A$ comes from mixing with the nearby $\Delta_2'$ and $\Delta_5$ bands and is given by

$$
\begin{aligned}
A &= \frac{2i\beta}{3m} \frac{\langle\Delta_{2'}|p_z|\Delta_{2'}\rangle\langle\Delta_{2'}|D_{xy}|\Delta_1\rangle}{E_{12'}^2 E_{15}} \{\langle\Delta_1|p_x|\Delta_{5'}^x\rangle\langle\Delta_5^x|h_x|\Delta_{2'}\rangle + \langle\Delta_1|h_x|\Delta_5^y\rangle\langle\Delta_5^y|p_x|\Delta_{2'}\rangle\} \\
&\times \Delta g_\perp \frac{E_{15}}{E_{12}^2}\langle\Delta_{2'}|D_{xy}|\Delta_1\rangle,
\end{aligned}
$$

where $\mathbf{h} = \nabla V \times \mathbf{P}$ is the usual crystal SO vector, $E_{ij}$ are the energy gaps to the relevant bands, and $D$ is the inter-band deformation potential. $H_{one-valley}^{bulk}$ represents a sum over the six minima in the bulk case, but in a QWQD the dominant contribution comes only from the $\pm z$ minima and was determined recently by Glavin and Kim [34] to be

$$H_{one-valley}^{QD} = A\beta U_{xy}(\sigma_x H_y + \sigma_y H_x).$$

The constant $A$ was experimentally determined by Wilson and Feher [89] to be $A = 0.44$. The $T_1$ time due to $H_{one-valley}^{QD}$ can be readily calculated and is given by [34]

$$\frac{1}{T_1^{one-valley}} = \frac{2\pi^4 A^2 \hbar}{5g^2\rho v_t^5}\left(\frac{g\mu B}{2\pi\hbar}\right)^5 (1 + n(g\mu B))\sin^2\theta(\cos^2 2\phi + \cos^2\theta\sin^2 2\phi),$$

where $(\theta, \phi)$ define the angle of the magnetic field relative to [001]. It is evident that this equation produces infinite relaxation times if the magnetic field points along the [001] or [011] axes. Figure 4.2 plots the one-valley relaxation rate as a function magnetic field direction.

### 4.5.2 Rashba mixing mechanism

Now we will consider spin mixing with higher, non-s-like states of the quantum dot. Because a quantum dot is so much more shallow than a donor, there are near-lying p-, d-, etc.-like

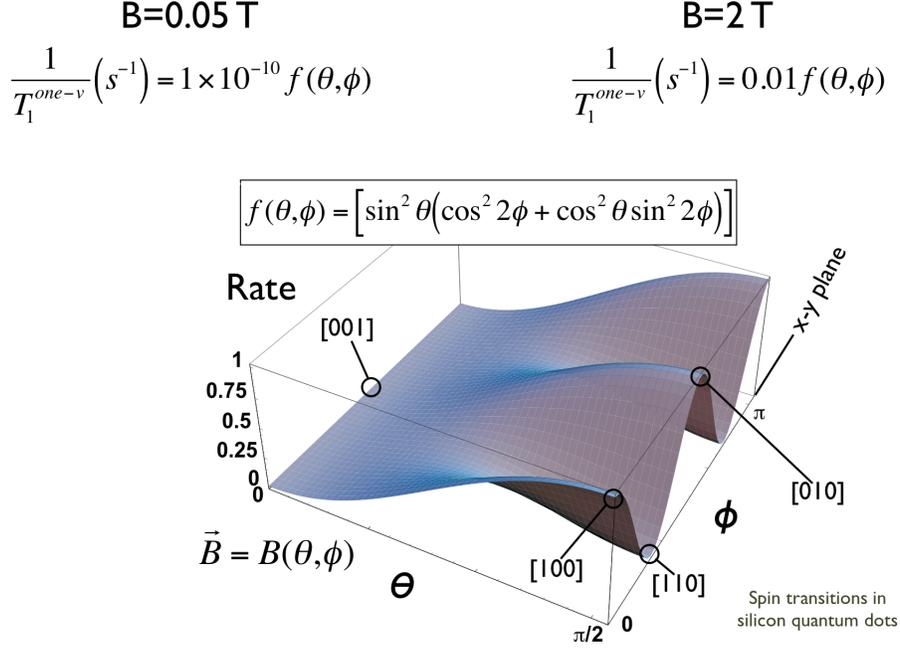

$$B=0.05\text{ T} \qquad\qquad B=2\text{ T}$$

$$\frac{1}{T_1^{one-v}}\left(s^{-1}\right) = 1\times10^{-10}\,f(\theta,\phi) \qquad\qquad \frac{1}{T_1^{one-v}}\left(s^{-1}\right) = 0.01\,f(\theta,\phi)$$

$$f(\theta,\phi) = \left[\sin^2\theta\left(\cos^2 2\phi + \cos^2\theta\sin^2 2\phi\right)\right]$$

Figure 4.2: Spin relaxation due to the one-valley mechanism.[34]

states which become important sources for virtual transitions. SOC mixes spin up and down states allowing a phonon to make a spin-forbidden transition and relax the excited spin-qubit state. Rashba SOC has the same effect, as has been pointed out by Khaetskii and Nazarov (KN). [49] In their analysis of GaAs spin-relaxation, they ignore the Rashba effect because of the dominance of Dresselhaus SOC in GaAs. In silicon this isn't true and we follow KN's approach to calculate the relaxation rate from Rashba admixture. Our results differ from those of the GaAs analog due to the many-valley nature of silicon and the dominance of acoustic over piezo-phonons in silicon.

We assume that SO coupling between the states of our dot is dominated by Rashba structural inversion asymmetry and has the matrix elements between two dot states of

$$(H_{so})_{kn}^{\uparrow\downarrow} = \alpha\frac{imE_{kn}}{\hbar}\left(x_{kn}\sigma_y^{\uparrow\downarrow} - y_{kn}\sigma_x^{\uparrow\downarrow}\right),$$

where we have used the trick $(p_x)_{kn} = imE_{kn}x_{kn}/\hbar$, have chosen a particular spin transition for simplicity, and where $\hat{\sigma}_x^{\uparrow\downarrow} = -\cos\varphi\cos\vartheta - i\sin\varphi$ and $\hat{\sigma}_y^{\uparrow\downarrow} = -\sin\varphi\cos\vartheta + i\cos\varphi$. SOC makes the spin-flip matrix element nonzero,

$$\langle n \uparrow |V|m \downarrow \rangle = \sum_{r \neq m} \frac{V_{nr}(H_{so})^{\uparrow\downarrow}_{rm}}{E_{mr} + g\mu B} + \sum_{r' \neq n} \frac{(H_{so})^{\uparrow\downarrow}_{nr'} V_{r'm}}{E_{nr'} - g\mu B} = V^{(1)}_{n\uparrow m\downarrow} + V^{(2)}_{n\uparrow m\downarrow} + \dots,$$

where we have expanded around $g\mu B$ with

$$\frac{1}{E_{nk} \pm g\mu B} = \frac{1}{E_{nk}} \left( 1 \mp \frac{g\mu B}{E_{nk}} + \dots \right).$$

Thus,

$$V^{(1)}_{n\uparrow m\downarrow} = \alpha \frac{im}{\hbar} \left[ -(Vx)_{nm} \sigma^{\uparrow\downarrow}_y + (Vy)_{nm} \sigma^{\uparrow\downarrow}_x + (xV)_{nm} \sigma^{\uparrow\downarrow}_y - (yV)_{nm} \sigma^{\uparrow\downarrow}_x \right],$$

and

$$V^{(2)}_{n\uparrow m\downarrow} = \alpha \frac{im}{\hbar} g\mu B \left[ \sum_{r \neq m} \frac{V_{nr}}{E_{mr}} \left( x_{rm}\sigma^{\uparrow\downarrow}_y - y_{rm}\sigma^{\uparrow\downarrow}_x \right) - \sum_{r' \neq n} \left( x_{mr'}\sigma^{\uparrow\downarrow}_y - y_{mr'}\sigma^{\uparrow\downarrow}_x \right) \frac{V_{r'n}}{E_{nr'}} \right].$$

Notice that $\langle n \uparrow |V|m \downarrow \rangle$ is zero if $B = 0$ and $m = n$. Also, if $m = n$ and $V_{nr} = V_{rn}$, then $V^{(1)}_{n\uparrow m\downarrow}$ is zero because $(H_{SO})^{\downarrow\uparrow}_{kn} = -(H_{SO})^{\downarrow\uparrow}_{kn}$ and the overall matrix element is reduced by roughly $g\mu B/\hbar\omega_0$, where $\omega_0$ is the fundamental energy of the dot. This is the manifestation of the so-called Van Vleck cancellation.

We are concerned with spin flip transitions, $\delta_m \neq \delta_n$, on the same dot level, $m = n$. Applying the rate and perturbation equations, we find that

$$\langle n \downarrow |H_{ep}(\mathbf{q},t)|n \uparrow \rangle = 2g\mu B \sum_{k \neq n} \left[ \frac{(H^{\mathbf{q}t}_{ep})_{nk} (H_{SO})^{\downarrow\uparrow}_{kn}}{(E_{nk})^2} \right]$$

and

$$V^{(2)}_{n\uparrow n\downarrow} = \alpha \frac{2g\mu B}{\hbar} im \sum_r \frac{(H^{\mathbf{q}t}_{ep})_{rn}}{E_{nr}} \left( x_{nr}\sigma^{\uparrow\downarrow}_y - y_{nr}\sigma^{\uparrow\downarrow}_x \right),$$

where we have defined a polarization tensor

$$\xi^{(n)}_{ik}(B) = -2e^2 \sum_{m \neq n} \frac{(x_i)_{nm} (x_k)_{mn}}{E_{nm}}.$$

Neglecting terms in $z$,

$$\frac{1}{T_1} = \frac{(g\mu B)^7 (m\alpha)^2}{2\rho(2\pi)^2 \hbar^8 e^4}$$
$$\times \sum_t \frac{1}{v_t^7} \int \sin\theta d\theta d\phi \left(\hat{\mathbf{n}} \cdot \Xi^z \cdot \hat{\mathbf{q}}\right)^2 \left[\hat{q}_x \left(\sigma_y^{\uparrow\downarrow}\xi_{xx}^{(n)} - \hat{\sigma}_x^{\uparrow\downarrow}\xi_{xy}^{(n)}\right) + \hat{q}_y \left(\sigma_y^{\uparrow\downarrow}\xi_{yx}^{(n)} - \hat{\sigma}_x^{\uparrow\downarrow}\xi_{yy}^{(n)}\right)\right]^2,$$

for an arbitrarily shaped dot and magnetic field direction. We work in the electric dipole approximation, which is very good for common Zeeman energies ($<< 1$ meV) as we showed in the previous section on the phonon bottleneck effect. Performing the integral gives

$$\frac{1}{T_1} = \Upsilon_{xy} \quad \frac{(g\mu B)^7 (m\alpha)^2}{4\rho\pi\hbar^8 e^4} \left[\left(\xi_{xx}^2 + \xi_{xy}^2 + \xi_{yx}^2 + \xi_{yy}^2\right)(3 + \cos[2\theta])\right.$$
$$\left. + 2\left(\xi_{xx}^2 - \xi_{xy}^2 + \xi_{yx}^2 - \xi_{yy}^2\right)\cos[2\phi]\sin^2[\theta]\right. \tag{4.14}$$
$$\left. + 4\left(\xi_{xx}\xi_{xy} + \xi_{yx}\xi_{yy}\right)\sin^2[\theta]\sin[2\phi]\right]. \tag{4.15}$$

The noteworthy differences from the classic results of Roth and Hasegawa are the seventh power dependence on the magnetic field and the proportionality to the square of the Rashba coefficient. The latter indicates that this mechanism will nominally disappear in a perfectly symmetric device ($\alpha \to 0$). More importantly, the spin-flip rate is *nonzero for all magnetic field directions*, including $B \parallel z \parallel [001]$. Additionally, as is clear from the structure of the polarization tensors, this Rashba mechanism is very dependent on the shape and size of the specific dot. Just *decreasing the size of the dot will diminish the spin qubit relaxation* (by increasing the energy level spacings and thus decreasing the amount of spin-mixing). In contrast, the Roth/Hasagawa mechanisms have no dependence on the dot geometry.

The seventh power dependence on the Zeeman splitting as compares to fifth power in GaAs QDs [49] is a consequence of the absence of piezoelectricity in silicon. The contribution due to acoustic phonons in GaAs is also seventh power. It is commonly neglected however because the piezoelectric interaction is usually more efficient. Now, when comparing to bulk Si, which also has a fifth-power dependence on the magnetic field, we must remember that in that case, spin mixing occurs across all s-like lowest excited states. Here, those states are unoccupied and the mixing occurs with the nearest p-like and higher excited states, increasing the order in perturbation theory.

As we said before, the polarization tensors (matrix elements) can be calculated numerically, or in the small magnetic field limit (where the first excitation energy is much less than the Landau energy and electron wave functions are largely unaltered) the zero B-field parabolic matrix elements can be used as a decent approximation. We can include the magnetic field in a circular dot explicitly if $B||z$ by utilizing the Fock-Darwin states. In this case, $x = y$, $\langle 00|\,x\,|01\rangle = \langle 00|\,x\,|10\rangle = \left|\sqrt{2}l_0/2\right|$, where $\hbar\omega_0$ is the fundamental energy of the dot, $l_0 = (\hbar/eB)/\sqrt{1 + 4\omega_o^2/\omega_c^2}$, $\omega_c = eB/m$, $E_{00,01} = \hbar\omega_-$, $E_{00,10} = \hbar\omega_+$, and $\omega_{\pm} = \sqrt{\omega_0^2 + \omega_c^2/4}\pm\omega_c/2$. Then, taking into account dipole selection rules (only transitions $n'_{\pm} = n_{\pm} \pm 1$ are allowed),

$$\xi_{xx} = \xi_{yy} = -e^2 l_o^2 \left[\frac{1}{\hbar\omega_-} + \frac{1}{\hbar\omega_+}\right] \text{ and } \xi_{xy} = \xi_{yx} = 0,$$

which, in the $B = 0$ limit, reduces to $\xi_{xx} = \frac{e^2}{m\omega_x^2}$ and $\xi_{yy} = \frac{e^2}{m\omega_y^2}$ for an elliptical dot. Finally, for a circular dot in a small magnetic field (such that the electron wavefunctions are largely unchanged),

$$\frac{1}{T_1} \approx \Upsilon_{xy}\frac{(g\mu B)^7(m\alpha)^2}{2\rho\pi\hbar^8 e^4}\xi_{xx}^2\left[3+\cos\left(2\theta\right)\right] \approx \frac{(g\mu B)^7\alpha^2}{\rho\pi\hbar^4\left(\hbar\omega_0\right)^4}\frac{\Xi_u^2}{105v_t^7}\left[3+\cos\left(2\theta\right)\right], \quad (4.16)$$

having dropped the much smaller $v_l$ term.

The outputs from Equation 4.16 for typical dot sizes are shown in Figures 4.3 and 4.5.

### 4.5.3 Two-phonon Orbach process

Relevant to recent ESR experiments at higher temperature (4-10 K), it's interesting to consider the two-phonon contribution from Rashba mixing in silicon quantum dots. In analogy with donors where Orbach processes seem to dominate over 2 K, we can use the methods proposed by Castner to estimate the Rashba+Orbach spin relaxation path. We arrive at

$$\frac{1}{T_1^{Orbach}} \approx M_{SO}^2\Gamma_{1\rightarrow g}n\left(E_{1g}\right)$$

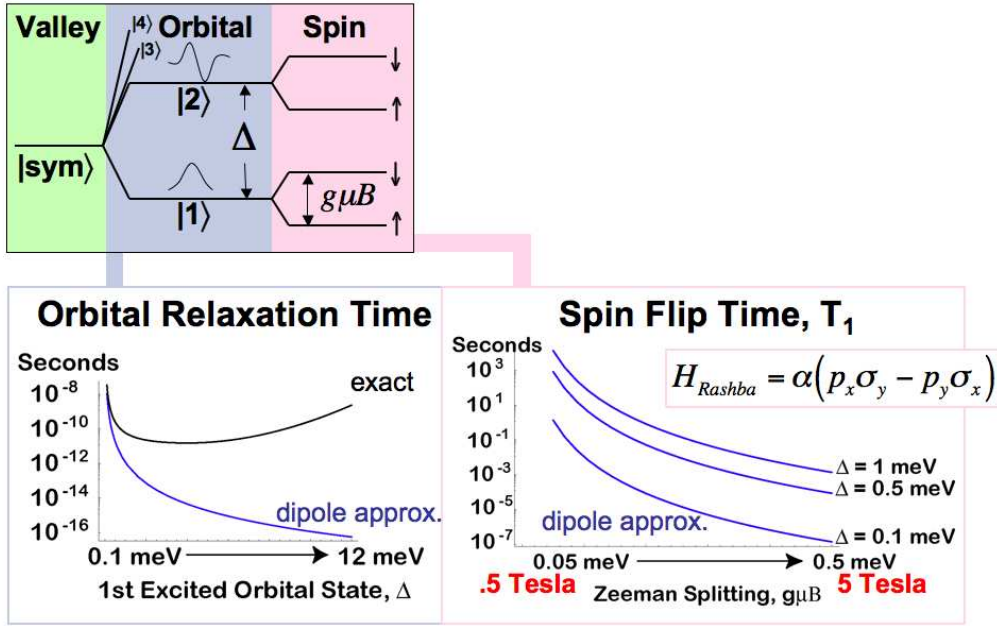

Figure 4.3: Orbital and spin relaxation of first excited state in parabolic, asymmetric quantum dot.

where we estimate for a circular dot that $M_{SO} \approx \langle 1 \uparrow | x | g \downarrow \rangle / \langle 1 | x | g \rangle \approx 2\alpha \sqrt{m^*/E_{1g}}$ as the spin-mixing of the two states (which compares well with $\Delta g$) and $\Gamma$ is the orbital relaxation time across the first energy gap. So we come to, for a circular, parabolic dot,

$$\frac{1}{T_1^{Orbach}} \approx \Upsilon_{xy} \frac{8\alpha^2 |E_{1g}|^3}{\hbar^4 \pi \rho_{Si}} \exp\left[-|E_{1g}|/kT\right]. \tag{4.17}$$

One interesting point is that Orbach dominated ESR will give us a spectroscopy of the first energy splitting in silicon quantum dots as it does quite accurately for donors.

## 4.6   Discussion and Implications

We now consider the ramifications for electron spin-based quantum dot quantum computing. We began by considering phonon relaxation of excited orbital states across the same valley state in a lateral silicon QD. We found that cooling could be dramatically faster in biaxially strained silicon than in the bulk. As will be seen in the next chapter, this speeds up spin qubit initialization via optical pumping schemes.[33] The relatively large matrix element connecting the ground state to higher orbital states also increases leakage of the qubit Hilbert space as temperature increases, making dilution refrigeration of the QC all the more necessary. The phonon bottleneck effect will eventually decrease the orbital relaxation rate but only for unrealistically small dots.

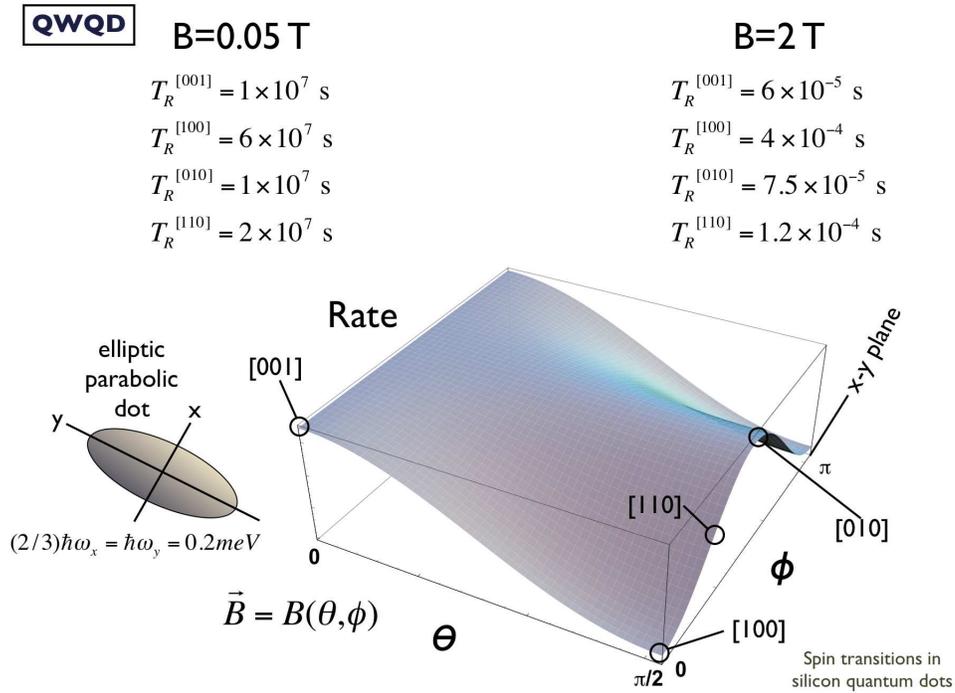

Figure 4.4: Angular dependence of spin relaxation due to the Rashba mechanism.

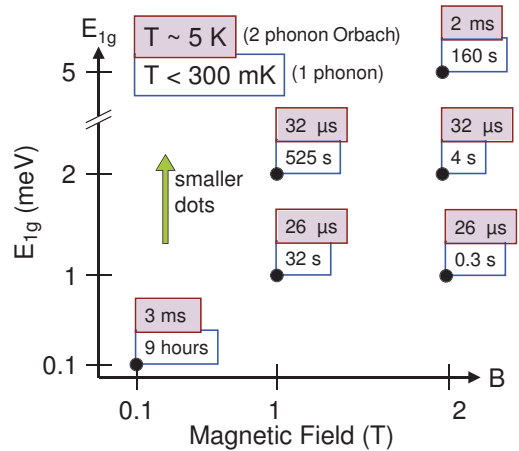

Figure 4.5: Orbital and spin relaxation of first excited state in parabolic, asymmetric quantum dot.

As to spin-based QD QC, we next considered spin relaxation due to Rashba spin-orbit coupling. Our calculations give a spin-flip time which, for the first time, is not predicted to be infinite for $B \parallel z$ in a strained-silicon QD, and so is the dominant mechanism. We find that $T_1$ is dependent on the size and shape of the dot and is still very long, on the order of seconds for typically envisioned dot sizes. In silicon, where isotopic purification may mean that phonons could dominate $T_2$ spin decoherence, this would serve as a good estimate for $T_2$ as well. Our calculations also guide experiment in constructing longer-lived spin qubits: make the QDs smaller and the QW more symmetric to get longer decoherence times. Our predictions also have a key dependence on magnetic field, $1/T_1 \propto B^7$ which can be confirmed by transport experiments in silicon quantum dots like those done in GaAs [43]. For experiments on quantum dots at higher temperatures, we have estimated $T_1$ relaxation due to Orbach processes and shown how it can be used to characterize the size (fundamental energy) of the quantum dots being measured.

# Chapter 5

# Spin readout and initialization in a semiconductor Quantum Dot

The magnetic moment of one electron spin, our qubit, is $\mathbf{m} = 9.3 \times 10^{-24}$ J/T, roughly 19 orders of magnitude weaker than a typical refrigerator magnet. Reliably and quickly reading out the state of a single spin qubit may be the most challenging technological goal in quantum dot quantum computing. Precisely what gives magnetic spins their long decoherence times—exceptional decoupling from the environment—also makes manipulation and single spin measurement by electrical means difficult. Several promising schemes incorporate ancillary tunnel couplings that may provide unwanted channels for decoherence. Here, we propose a novel spin-charge transduction scheme, converting spin information to orbital information within a single quantum dot by microwave excitation. The same quantum dot can be used for rapid initialization, gating, and readout. We analyze the performance of such devices for feasibility.

In this chapter we introduce our original scheme for readout and initialization via a spin-flip microwave excitation. We then revisit the estimated time scales used to assess the device based on the more in-depth calculations of the previous chapter. A new scheme is proposed which addresses some of the speed and device limitations of the original proposal. We end by considering limitations of our analysis and considerations for future work in this area.

## 5.1    Introduction

Measurements of spin qubits pose a special challenge. On the one hand, qubits should be well isolated from their environment to avoid decoherence. On the other hand, it is necessary to individually couple the qubits to an external measurement device. Qubit initialization involves an additional dissipative coupling to the environment. For quantum computing, we must initiate such couplings selectively, and with sufficient strength to perform the operations quickly. Indeed, scalable quantum computing relies on fault-tolerant quantum error correction algorithms, involving frequent, parallel measurements, and a steady supply of initialized qubits [79]. Fortunately, rapid and sensitive quantum measurement techniques involving radio frequency single electron transistors (rf-SETs) have been developed [24]. rf-SETs have been used to detect the tunneling of individual electrons in semiconductor devices.

We have previously designed a quantum dot architecture specifically for the purpose of manipulating electron spins for fast and accurate two-qubit operations that serve as universal gates for quantum computations [32]. Recent experimental results have shown that decoherence does not pose a fundamental problem for such gate operations [83]. Using special qubit geometries [31], it should be possible to perform reliable gate operations in silicon quantum dots at rates between about 1 MHz and 1 GHz. The question is now whether similar speeds and reliable operation can be achieved for measurement and initialization operations. We show how to extend our architecture to enable the readout and initialization of the state of a single spin. The general scheme does not assume a specific material system. However, here we focus upon silicon-based devices because of their desirable coherence properties.

Our method relies on the idea of converting spin information to charge information discussed by Loss and DiVincenzo [55], and using single electron transistors to read out the resulting spin state as proposed by Kane et al. [48], who posit spin-dependent charge motion onto impurities in Si. Martin et al. [56] have proposed a scheme for single spin readout that also converts spin information into charge information in an electron trap near a conducting channel. The resistance of the channel depends on the occupation of the trap, which in turn can be made to depend on the spin. Our proposed device has the additional feature in that it enables both readout and rapid initialization of the spin state. Rapid initialization (as opposed to initializa-tion by thermalization) is just as essential as readout, and just as difficult to carry out. Our design is simple in that it obviates the need for spin-polarized leads or ancillary qubits. We go beyond conceptual design—the full three-dimensional system is simulated self-consistently and the operating parameters are thereby optimized. This enables us to demonstrate quantitatively that the device is in a working regime.

## 5.2    Motional spin-charge transduction

Our architecture is the basic Si-SiGe quantum well quantum dot heterostructure as previously described.[32] The main point for present purposes is that the active layer is pure strained Si, which minimizes decoherence from spin-phonon coupling [14] and nuclear spins. The qubits are gated quantum dots which hold one electron, and the gate geometry is the key to our scheme. Specifically, the electrons are confined in asymmetric lateral wells, such that orbital excitation results in lateral center-of-charge (COC) movement. Figure 1 shows the first two orbital states of an electron confined to an asymmetric box, with COC positions varying by a distance $y$. If an external radiation source of frequency $E_{12}/h$ drives the system between the ground and first excited states, $n = 1$ and $n = 2$, the COC of the electron will oscillate in time at the Rabi frequency $\nu_R$. This motion can be detected by a sensitive electrometer, such as a single electron transistor or a quantum point contact. With the system placed in a magnetic field, charge motion can be generated in a *spin-dependent* fashion. We define the logical 0 and 1 of the qubit as the spin-up and spin-down states of the electron in its orbital ground state and use the orbital excited state only during initialization and readout. The system is now driven at the readout frequency $\nu_{12}(B) = E_{12}/h - g\mu_B B/h$. This causes charge motion only if the qubit was in the down state at the time of measurement. Although this transition is forbidden at the electric dipole level because of the spin-flip, spin-orbit coupling allows for a nonzero transition rate, as described below.

Crucially, this architecture also allows for rapid initialization of the qubit. Consider exposing a random ensemble of qubits to radiation of frequency $\nu_{12}$. An electron in the $(1, \downarrow)$ state will be excited to the level $(2, \uparrow)$ and experience a relatively fast relaxation to the ground state $(1, \uparrow)$ , compared to a spin-flip relaxation to the level $(1, \downarrow)$. The net result is to rapidly polarize and thereby initialize the qubits. By varying the gate voltages (and thus $\nu_{12}$) on individual dots we ensure that only desired qubits are brought into resonance. Clearly, understanding the competition between the three time scales, $1/\nu_R$, $1/\Gamma_I$, and $T_1$, as a function of material parameters and gate potentials is the key to utilizing this device for readout and initialization. Note that $T_1$ represents a thermal initialization time from $(1, \downarrow)$ to $(1, \uparrow)$. Since $T_1$ is a decoherence time, we require $1/\nu_R << T_1$. The Rabi oscillation frequency $\nu_R$ depends on the incident intensity and is therefore controllable, within limits. Robust measurement requires that many Rabi oscillations occur before orbital decay: $1/\nu_R \leq 1/\Gamma_I$.

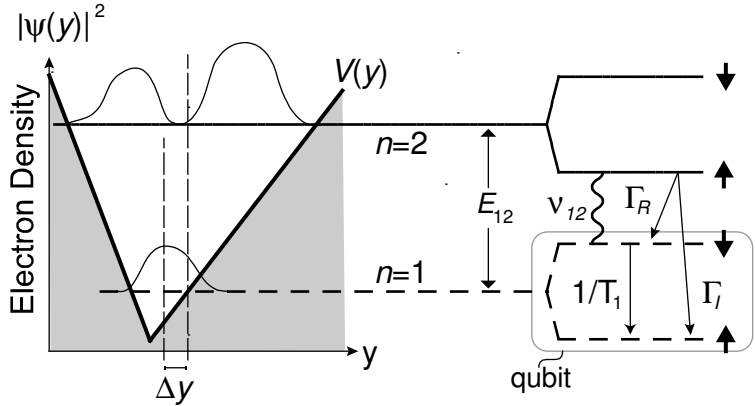

Figure 5.1: Schematic example of a one-dimensional, asymmetric confining potential in a constant magnetic field. Quantum dot energy states and transition rates for readout and initialization are shown. Microwave excitations from $n = 1$ to $n = 2$ cause oscillations in the electron's center-of-charge.

### 5.2.1 Time scale estimates

In order to understand the competition of time scales, we introduce an rms interaction energy, time averaged (as indicated by $\{\}$) over an optical cycle, which expresses the strength of the interaction in the electric dipole approximation. This defines the Rabi oscillation frequency [77]: $|h\nu_R|^2 = \{|V^{E1}|^2\}$. Here, $V^{E1} = (-e\hbar E_0/m^* E_{12})\hat{\epsilon} \cdot \mathbf{p}$ is the dipole term in the Hamiltonian, $|E_0|^2$ is twice the mean value of $|\mathbf{E}(t)|^2$ averaged in time, and $\hat{\epsilon}$ is the polarization unit vector. The electric field $E_0$ inside the semiconductor with dielectric constant $\varepsilon = \varepsilon_r \varepsilon_0$ is related to the intensity of the external radiation $I$ by $E_0 = \sqrt{2I/c\varepsilon_0 \sqrt{\varepsilon_r}}$.

The dipole Hamiltonian does not flip the spin directly, but spin-orbit coupling causes each qubit state to be an additive mixture of up and down spin. In the 2D limit, the spin-orbit (so) Hamiltonian is dominated by the bulk [Dresselhaus (D)] and structural [Rashba (R)] inversion asymmetry terms, $H_{SO} = H_D + H_R$, where $H_D = \beta(p_y\sigma_y - p_x\sigma_x)$ and $H_R = \alpha(p_x\sigma_y - p_y\sigma_x)$. Including $H_{SO}$ perturbatively gives a nonzero dipole matrix element, and for light polarized in the $y$ direction with $B||z$, the readout frequency is given by

$$|\nu_R| \simeq \frac{eE_0}{2\pi h \nu_{12}(B)}\sqrt{\alpha^2 + \beta^2}\,|\langle 2|\,y\partial_y\,|1\rangle|\,. \tag{5.1}$$

Note that the applied radiation need not be circularly polarized for this readout scheme. The Dresselhaus and Rashba parameters $\alpha$ and $\beta$ depend on intrinsic material properties, device design, and external electric field. For GaAs, both theoretical and experimental values vary widely [16]: $\alpha = 1 - 1000$ m/s and $\beta = 1000 - 3000$ m/s. In a centrosymmetric crystal such as silicon which has no bulk inversion asymmetry, $\beta = 0$. The one known data point for a SiGe two-dimensional electron gas gives $\alpha = 8$ m/s [88], which we use in our estimates below.

The relaxation of our quantum dot to its ground state enables spin polarization, but limits or even inhibits readout if it occurs too quickly. This problem has been addressed by Khaetskii and Nazarov in GaAs quantum dots [49]. In silicon, where there is no piezoelectric interaction, we calculate the relaxation rate via the golden rule with the usual deformation potential electron-phonon Hamiltonian [63]. At sufficiently low temperatures $T < 1$ K, optical polar phonons and multi-phonon processes do not contribute. By considering only longitudinal phonons, with dispersion $\omega = v_l q$, and using the long-wavelength approximation $e^{i\mathbf{k}\cdot\mathbf{r}} \simeq 1 + i\mathbf{k}\cdot\mathbf{r}$, we obtain the orbital decay rate due to electron-lattice coupling:

$$\Gamma_I = \frac{(E_{12})^5}{6\pi\hbar^6 v_l^7 \rho}\,(\Xi_d + \Xi_u/3)^2 \sum_i |\langle 2|\,x_i\,|1\rangle|^2\,, \tag{5.2}$$

where $\rho$ is the mass density, and $\Xi_d$ and $\Xi_u$ are the deformation constants. In strained systems, transverse phonons can also be important, as we have discussed in the more detailed derivation in Chapter 4.

## 5.2.2   Performance

We perform a numerical analysis to obtain performance characteristics for the proposed measurement scheme. The numerical techniques used are an extension of those used in Refs. [32, 31]. The gate potentials, the electronic orbitals, and their corresponding image potentials (arising predominantly from the metallic gates) are computed self-consistently by a combination of three-dimensional finite element and diagonalization techniques. Specifically, we determine the readout oscillation frequency Eq. 5.1, the orbital decay rate Eq. 5.2, and the coupling sensitivity of the qubit electron to an integrated SET. We have modeled numerous devices to find an optimal configuration. One of the most promising is shown in Fig. 2. This is a 6 nm strained silicon quantum well sandwiched between layers of strain-relaxed silicon-germanium ($Si_{85}Ge_{15}$). The bottom barrier (30 nm) separates the quantum well from a grounded metallic back gate. The top barrier (30 nm) separates the quantum well from lithographically patterned Schottky top gates, whose voltages can be controlled independently. We consider negative gate potentials, which provide lateral confinement of the quantum dot through electrostatic repulsion. The main features of the quantum dot are that it is narrow, long, and slightly asymmetric. The *narrow* feature ensures that the excited orbitals are non-degenerate, so a microwave field with narrow linewidth will not induce unwanted transitions. The *long* feature ensures that the energy splitting $E_{12}$ (and thus $\Gamma_I$ in the electric-dipole approximation) will be small. The *asymmetry* ensures that excited orbitals will provide charge motion. Figure 3 shows the confinement potential for the device of Fig. 2; the asymmetry of the dot potential is apparent.

Our most interesting results are obtained in the regime where the gate and image potentials are comparable in size. The effective confinement potential, including images, is rather complicated due to the inhomogeneous gate arrangement. Indeed, it is a very poor approximation to neglect screening in this system. The computed wavefunctions for our dot are shown in Fig. 3.

We have also incorporated a SET into our proposed readout device, as shown in Fig. 2. The island of the SET is sheathed by a thin, 2-nm layer of silicon dioxide, and tunnel coupled to adjoining source and drain gates. A third, capacitively coupled gate is placed nearby to provide full control over both the charge and the potential of the island. When the device is operated in the Coulomb blockade regime, it becomes a sensitive electrometer [19]. In the gate arrangement of Fig. 2, the SET plays two roles. First, voltage control of the island allows us to vary the size of the dot and thus the energy splitting $E_{12}$. Second, capacitive coupling to the quantum dot enables detection of its orbital state. The scheme works as follows. Since the dot orbitals are spatially distinct, they induce different amounts of charge on the SET island. Consequently, transport currents through the SET will reflect the orbital states of the dot. The device exhibits optimal sensitivity if biased at the half maximum of the conductance peak. The third SET gate has been introduced specifically to adjust this working point. As expected for our geometry, we find that the SET couples most strongly to the excited electronic orbital.

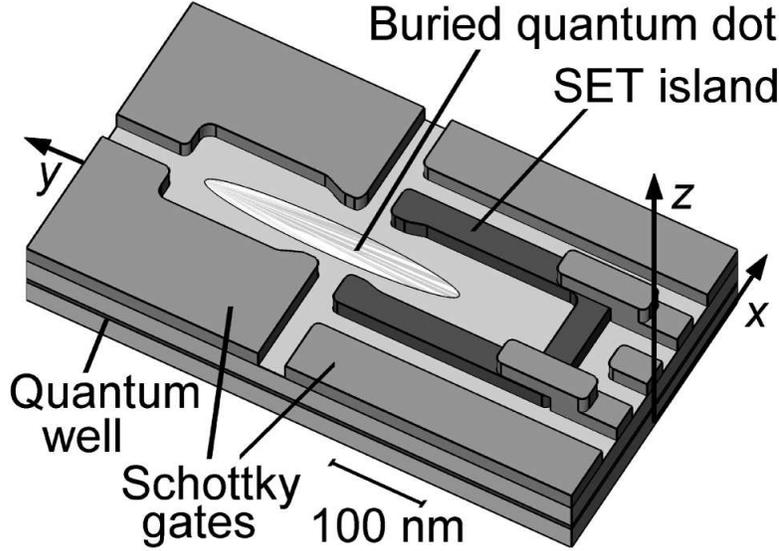

Figure 5.2: Proposed and simulated readout device.

For the device shown in Fig. 2 we obtain an energy splitting of $E_{12} = 0.129\,\mathrm{meV} = 31.2\,\mathrm{GHz}$ between the two lowest orbital states, and the dominant matrix elements $\langle 2|\, y\, |1\rangle = 48$ nm and $\langle 2|\, y\partial_y\, |1\rangle = 3.6$. From these results we obtain the readout oscillation frequency $\nu_R = 5.5 \times 10^5 \sqrt{I}$ Hz (for microwave intensity $I$ in units of $\mathrm{W/m^2}$), and the orbital decay rate for spontaneous phonon emission $\Gamma_I = 12.7$ MHz. The orbital decay rate for emission of a photon is very much less:

$$\Gamma_\gamma = \frac{e^2 E_{12}^3\, |\langle 2|\, y\, |1\rangle|^2\, \sqrt{\varepsilon}}{3\pi \hbar^4 c^3 \varepsilon_0^{3/2}} = 6.5\ \mathrm{Hz},$$

and therefore is not a limiting factor in this scheme. All results are obtained at an ambient temperature of $T = 100$ mK and a magnetic field of $H = 0.05$ T corresponding to a frequency shift of 1.40 GHz. The dielectric constant of Si is $\varepsilon = 11.9\varepsilon_0$. Note that the natural widths of the states are small compared to their separation. The charge on the SET island was computed by integrating $\mathbf{n} \cdot \mathbf{D}$ over the surface. The excess or induced charge of the excited quantum dot orbital, relative to its ground state, was found to be $\Delta Q = 0.052e$. The corresponding center-of-charge motion in the dot is $\Delta y = 4.3$ nm. However, the charge motion does not track closely with $\Delta Q$; the latter is determined primarily by the capacitive coupling between the dot and the SET. Finally, we find that by changing the bias voltage on the SET by 20% (thus reducing the dot size), the excitation resonance frequency $\nu_{12}$ changes by 8 GHz.

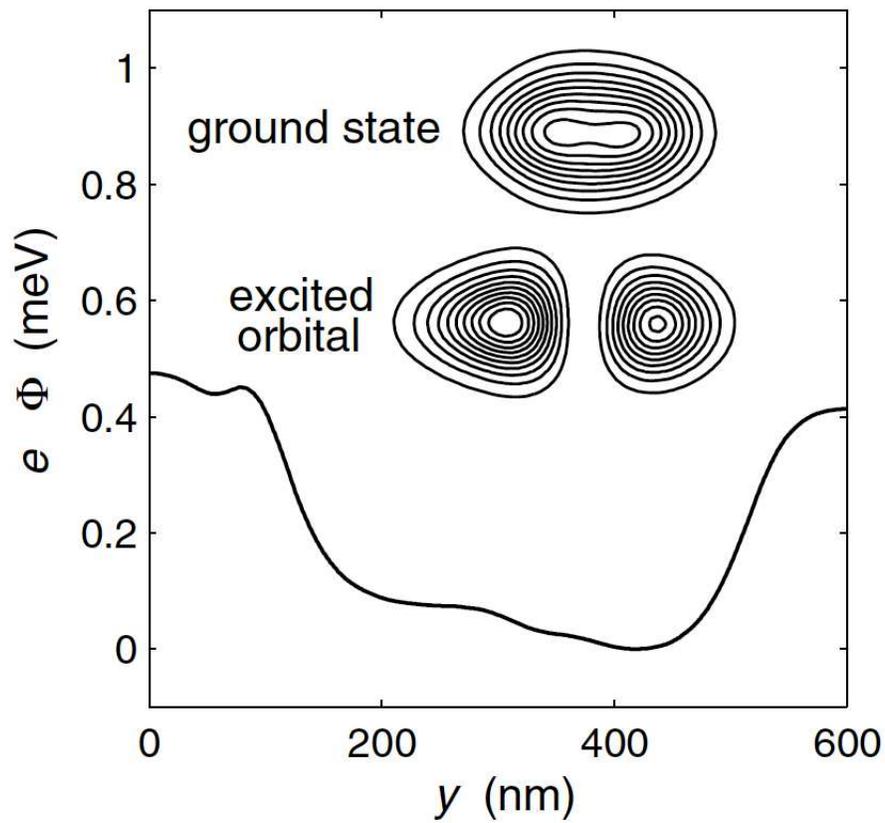

Figure 5.3: Electrostatic confinement potential and qubit electron wave functions. The bare potential shown here is obtained in the quantum well along the device symmetry line, $x = 0$. The contour plots show the electron probability densities in the $x - y$ plane.

### 5.2.3 Analysis of original scheme

The prospects for initialization and readout in this scheme are partially summarized in Fig. 4. Theoretical considerations of shot noise [19, 53] place an upper bound of about $4 \times 10^{-6} e/\sqrt{\text{Hz}}$ on the detection sensitivity for charge induced on the island of an optimized rf-SET, as a function of the measurement bandwidth. Similarly, the decay rate of the excited electronic orbital places a lower bound on the readout oscillation frequency $\nu_R$. The latter is a function of microwave intensity. By directing 70 pW microwave power onto a dot of size $0.1 \times 0.1$ $\mu\text{m}^2$ (or 7000 W/m$^2$), as consistent with low-temperature transport experiments, we would obtain the $\nu_R$ working point marked in Fig. 4, with $1/\nu_R \lesssim 1/\Gamma_i$. Sample heating outside the dot can be minimized by using integrated on-chip antennas to focus the power [27]. We must also satisfy the requirement $T_s = 1/\Gamma_I$ ns where $T_s$ is the spin decoherence time. Theoretical estimates for $T_1$ in Si quantum dots exceed 1 ms [80]. Experiments on dots have yet to be performed, but spins of electrons in impurity donor states should have similar lifetimes. Recent work on electron spins in $^{28}$Si:P [83] would imply that $T_s > 60$ ms. Hence the crucial hierarchy of time scales $1/\nu_R \lesssim 1/\Gamma_I \ll T_s$ should be achievable.

### 5.2.4 Time scales revisited

Many of the time scales estimated in our original analysis of readout feasibility have now been supplanted by the new derivations of quantum dot electronic relaxation times in Chapter 4.

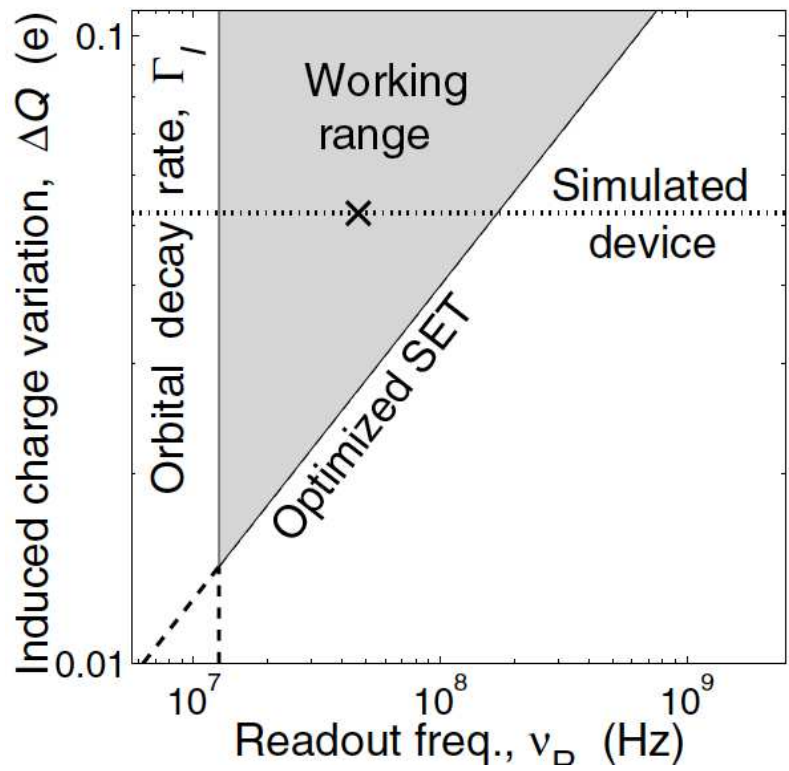

Figure 5.4: Electrostatic confinement potential and qubit electron wave functions. The bare potential shown here is obtained in the quantum well along the device symmetry line, $x = 0$. The contour plots show the electron probability densities in the $x - y$ plane.

Most importantly, the new analysis takes into account the effects of transverse phonons, which substantially increases the initialization rate, and goes beyond the electric-dipole approximation, allowing exploration of the phonon blockade regime. We also now have an accurate calculation for the $T_1$ time of the qubit. We incorporate these results into our analysis and take the opportunity to re-derive the readout oscillation frequency in an off-axis magnetic field, which has experimental implications for understanding and extracting Rashba spin-orbit coupling in a silicon quantum dot.

**Relaxation rates**

The analytical results obtained in the electric dipole approximation for orbital relaxation in strained silicon are quite general in that they can accept wavefunctions and energy gaps of any type, making them a good fit for the simulation results. We found in Chapter 4 for the spontaneous emission phonon relaxation rate from the first excited orbital to the ground state for our readout dot to be

$$\Gamma_{21} \quad = \quad \frac{|E_{12}|^5}{\hbar^6 \pi \rho_{Si}} |M_y|^2 \, \Upsilon_{xy},$$  (5.3)

where

$$\Upsilon_{xy} \quad = \quad \frac{35\Xi_d^2 + 14\Xi_d\Xi_u + 3\Xi_u^2}{210 v_l^7} + \frac{2\Xi_u^2}{105 v_l^7}.$$

Using the same device numbers as our readout dot above, we find that the orbital relaxation rate is 2.7 ns or $\Gamma_I = 0.37$ GHz, roughly thirty times faster than our original estimate. This pushes the limit of the workable device range. Unfortunately, our exact calculation of spontaneous emission beyond the electric-dipole approximation in Chapter 4 showed that the phonon bottleneck effect won't help us in this case. Only with energy gaps bigger than 10 meV does the orbital relaxation rate begin to be slower than that of our readout dot as defined.

Our assumption that the qubit spin-flip time was in essence infinite in relation to the time scale of orbital relaxation was a good one. As calculated in Chapter 4, the qubit spin-flip time for our asymmetric readout dot is given in an arbitrarily directed magnetic field by

$$\frac{1}{T_1} = \Upsilon_{xy} \quad \frac{(g\mu B)^7 (m\alpha)^2}{4\rho\pi\hbar^8 e^4} \left[ \left( \xi_{xx}^2 + \xi_{xy}^2 + \xi_{yx}^2 + \xi_{yy}^2 \right) (3 + \cos[2\theta]) \right.$$

$$+ 2 \left( \xi_{xx}^2 - \xi_{xy}^2 + \xi_{yx}^2 - \xi_{yy}^2 \right) \cos[2\phi] \sin^2[\theta] \tag{5.4}$$

$$\left. + 4 \left( \xi_{xx}\xi_{xy} + \xi_{yx}\xi_{yy} \right) \sin^2[\theta] \sin[2\phi] \right], \tag{5.5}$$

where $\xi_{xx} = \xi_{yy} = -e^2 l_o^2 \left[ \frac{1}{\hbar\omega_-} + \frac{1}{\hbar\omega_+} \right]$ and $\xi_{xy} = \xi_{yx} = 0$, assuming parabolic potentials. In the small magnetic field in which the readout dot operates, $\xi_{xx} = \frac{e^2}{m\omega_x^2}$ and $\xi_{yy} = \frac{e^2}{m\omega_y^2}$. We approximate the lower limit of $T_1$ by assuming a circular dot and using $\alpha = 6$ m/s. We find that the $T_1$ time due to Rashba mediated SOC is $4 \times 10^7$ seconds. In this regime of very small magnetic field, the Rashba mechanism is ineffective. Our very crude two-phonon Orbach estimate from Chapter 4 gives $T_1^{Orbach} \approx 300$ s at 100 mK and $T_1^{Orbach} \approx 3$ ms at 1 K. This implies that Orbach spin relaxation dominates in this regime. This is easily characterizable in experiments by changing dot size or temperature and looking for the characteristic energy dependence. It also suggests that revisiting two-phonon relaxation may be worthwhile. In any case, these time scales are long enough for readout.

**Derivation of spin-flip transition frequencies in an arbitrarily directed small magnetic field**

We consider the interactions of a quantum dot electron with a classical radiation field. A monochromatic, linearly polarized radiation source of constant intensity is assumed. The energy separation between the two orbital states, the ground, $|1\rangle$, and first, $|2\rangle$, (envelope wavefunction) excited state, is of order 0.5 meV. Therefore, the radiation frequency needed to excite this transition will be $\omega_0 = 0.5$ meV$/\hbar = 0.5$ meV$/6.6 \times 10^{-13}$ meV $\cdot$ s $= 7.6 \times 10^{11}$ Hz, i.e., microwave radiation ($\lambda = \frac{2\pi}{\omega} \frac{c}{n} = \frac{2\pi}{\omega} \frac{c}{\sqrt{\epsilon_{Si}}} = \frac{2\pi}{\omega} \frac{c}{\sqrt{11.8}} = 7.22 \times 10^{-4}$ m, so actually it's nearer infra-red in silicon). The basic light Hamiltonian, with second order terms omitted, is

$$H = \frac{\mathbf{p}^2}{2m_e} + U(\mathbf{r}) - \frac{e}{m_e c} \mathbf{A} \cdot \mathbf{p} = H_0 - \frac{e}{m_e c} \mathbf{A} \cdot \mathbf{p}.$$

The last term, the interaction term, $V = -\frac{e}{m_e c}\mathbf{A}\cdot\mathbf{p}$, perturbs the system. Written as a plane wave,

$$
\begin{aligned}
\mathbf{A} &= 2A_0\hat{\varepsilon}\cos\left(\frac{\omega}{c}\hat{n}\cdot\mathbf{r}-\omega t\right) \\
&= A_0\hat{\varepsilon}\left[e^{i(\omega/c)\hat{n}\cdot\mathbf{r}-i\omega t}+e^{i(\omega/c)\hat{n}\cdot\mathbf{r}+i\omega t}\right],
\end{aligned}
$$

and for the absorption case specifically,

$$
V^{\dagger}=\frac{-eA_0}{m_e c}\left(e^{i(\omega/c)\hat{n}\cdot\mathbf{r}}\hat{\varepsilon}\cdot\mathbf{p}\right).
$$

The wave has polarization $\hat{\varepsilon}$ (where the hat indicates a unit vector), propagation direction $\hat{n}$, and frequency $\omega$. An expansion of the exponential, $e^{i(\omega/c)\hat{n}\cdot\mathbf{r}}=1+i\frac{\omega}{c}\hat{n}\cdot\mathbf{r}+\ldots$, gives the multi-pole interaction terms:[1]

$$
\begin{aligned}
V &= \frac{-eA_0}{m_e c}\left[\hat{\varepsilon}\cdot\mathbf{p}+i\frac{\omega}{c}\hat{n}\cdot\mathbf{r}\hat{\varepsilon}\cdot\mathbf{p}+\ldots\right] \\
&= V^{E1}+V^{M1}+V^{E2}+\ldots.
\end{aligned}
$$

If we look at the individual components of the vector potential $\mathbf{A}$ (and $A_0=\frac{c}{\omega}E_0$),

$$
\mathbf{E}(\mathbf{r},t)=-\frac{\partial\mathbf{A}}{\partial t}=\hat{\varepsilon}E_0\cos(\hat{n}\cdot\mathbf{r}-i\omega t)
$$

and

$$
\mathbf{B}(\mathbf{r},t)=\nabla\times\mathbf{A}=-\frac{1}{c}(\hat{n}\times\hat{\varepsilon})E_0\cos(\hat{n}\cdot\mathbf{r}-i\omega t),
$$

the electric dipole interaction (E1), magnetic dipole interaction (M1), and electric quadrupole interaction (E2) can be written as:

$$
\begin{aligned}
V^{E1} &= e\mathbf{r}\cdot\mathbf{E}(\mathbf{r},t)=-\mathbf{d}\cdot\mathbf{E}(\mathbf{r},t), \\
V^{M1} &= -\mathbf{M}\cdot\mathbf{B}(\mathbf{r},t)=\frac{e}{2m_e c}(\mathbf{L}+g\mathbf{S})\cdot\mathbf{B}(\mathbf{r},t),
\end{aligned}
$$

---

[1] The first term is the electric dipole term. The second term can be written as $i\frac{\omega}{c}\hat{n}\cdot\mathbf{r}\hat{\varepsilon}\cdot\mathbf{p}=i\frac{\omega}{c}\frac{1}{2}\hat{\varepsilon}\cdot\mathbf{p}\hat{n}\cdot\mathbf{r}+i\frac{\omega}{c}\frac{1}{2}(\hat{n}\times\hat{\varepsilon})\cdot(\mathbf{r}\times\mathbf{p})$ where the first term of this expression is the electric quadrupole term and the second is the magnetic dipole term.

$$V^{E2} \;=\; -\frac{1}{6}\mathbf{Q}^{(2)} : \nabla E(\mathbf{r}, t),$$

where $\mathbf{d}$ is the dipole moment. Atoms (and presumably artificial atoms like quantum dots) don't have permanent electric dipole moments, but they do have transient transition moments which are responsible for radiative excitation.

For microwave radiation, wavelengths are much larger than atomic (or dot) dimensions, so that the interaction ordinarily is, to a good approximation, that between a spatially uniform (but time varying) electric field and an atomic dipole moment.[77] In that case, we can truncate our perturbation potential to just $V^{E1}$. We assume that the matrix representation of $V$ has no diagonal elements.[2] We can now introduce an rms interaction-energy, time averaged (indicated by $\{\}$) over an optical cycle, which expresses the strength of the interaction and defines a characteristic frequency, $\nu$:[77]

$$|\hbar\nu|^2 = \{|V^{E1}|^2\} = 2\{|\mathbf{d} \cdot \mathbf{E}(t)|^2\} = |\mathbf{d} \cdot \varepsilon|^2 |E_0|^2.$$

Here, $\mathbf{d}$ is the *transition* dipole moment ($\langle 1|\mathbf{d}|g\rangle$ in our case) and $E_0$ is related to the intensity of the radiation (traveling trough a solid remember) by

$$E_0 = \sqrt{\frac{2I}{c\epsilon_0 \sqrt{\epsilon_r}}}.$$

Thus, for coherent excitation *exactly at resonance*, we can relate the radiation intensity and position matrix element (using $\mathbf{d} = -e\mathbf{r}$) with the Rabi frequency,

$$\nu_{\text{no spin}} = \frac{e}{\hbar}\sqrt{\frac{2I}{c\epsilon_0 \sqrt{\epsilon_r}}}\,\hat{\varepsilon} \cdot \langle 2|\mathbf{r}|1\rangle.$$

We must be careful in this assignment because the Rabi frequency refers to a transition dipole element between distinct states, that is, non-degenerate states. Degeneracy introduces different orientations of the dipole moment with respect to the vector $\hat{\varepsilon}$, and so introduces different Rabi frequencies for different initial states. In our case, an asymmetric quantum dot in a magnetic field, there is no relevant degeneracy.

---

[2]For electric-dipole induced transitions amongst bound atomic states of an isolated atom the matrix representation of V(t) usually has no diagonal elements. This rule is based on the fact that the dipole moment connects states of opposite parity. It is possible to have basis states with mixed parity however, such as atoms in static electric fields (as occur in solids). In all such cases there may be diagonal elements of the dipole moment. When the field is static any diagonal elements produce an energy shift that is linear in the electric field strength. Periodic diagonal elements, when present have little effect except at very high frequencies. We assume no diagonal components in our calculation.[77]

The task then becomes to calculate the transition matrix element between the ground and first excited state with a spin-flip, $\langle 2, \uparrow | V | 1, \downarrow \rangle$. We assume a small magnetic field. We want to calculate a matrix element between unlike spin levels on different orbitals. The energy levels $n = 1$ and $n = 2$ are not pure spin states but in fact mixed states. These states are written in perturbation theory as

$$|1, \downarrow\rangle^{so} = |1, \downarrow\rangle + \sum_{n\delta} \frac{|n\delta\rangle\langle n\delta|H_{so}|1, \downarrow\rangle}{E_{n\delta, 1\downarrow}}$$

and

$$|2, \uparrow\rangle^{so} = |2, \uparrow\rangle + \sum_{n\delta} \frac{|n\delta\rangle\langle n\delta|H_{so}|2, \uparrow\rangle}{E_{n\delta, 2\downarrow}}.$$

Then, the spin-flip matrix element is given by

$$\langle 2 \uparrow | V | 1 \downarrow \rangle = \sum_n \frac{V_{2n}(H_{so})^{\uparrow\downarrow}_{n1}}{E_{n1} + g\mu B} + \sum_{n'} \frac{(H_{so})^{\downarrow\uparrow}_{2n'} V_{n'1}}{E_{n'2} - g\mu B},$$

where for arbitrarily directed spins and for silicon (the Dresselhaus term $\beta = 0$) we rewrite the SO Hamiltonian as

$$(H_{so})^{\uparrow\downarrow}_{kn} = \alpha \frac{im E_{kn}}{\hbar} \left( \sigma_y^{\uparrow\downarrow} x_{kn} - \sigma_x^{\uparrow\downarrow} y_{kn} \right).$$

Using the electric-dipole approximation in a more general form,

$$V_{kn} = \frac{-e E_0}{m\omega} \hat{\varepsilon} \cdot \mathbf{p}_{kn} = \frac{-e E_0}{m\omega} \frac{\hbar}{i} \hat{\varepsilon} \cdot \vec{\nabla}_{kn},$$

and expanding around $g\mu B$, we find that the matrix element

$$\langle 2 \uparrow | V | 1 \downarrow \rangle = \alpha \frac{-e E_0}{\omega} \hat{\varepsilon} \cdot \left\{ \left[ \sum_{n \neq 1} \vec{\nabla}_{2n} \left( \sigma_y^{\uparrow\downarrow} x_{n1} - \sigma_x^{\uparrow\downarrow} y_{n1} \right) - \sum_{n' \neq 2} \left( \sigma_y^{\downarrow\uparrow} x_{2n'} - \sigma_x^{\downarrow\uparrow} y_{2n'} \right) \vec{\nabla}_{n'1} \right] \right.$$
$$\left. + g\mu B \left[ -\sum_{n \neq 1} \vec{\nabla}_{2n} \left( \sigma_y^{\uparrow\downarrow} x_{n1} - \sigma_x^{\uparrow\downarrow} y_{n1} \right) \frac{1}{E_{n1}} - \sum_{n' \neq 2} \left( \sigma_y^{\downarrow\uparrow} x_{2n'} - \sigma_x^{\downarrow\uparrow} y_{2n'} \right) \vec{\nabla}_{n'1} \frac{1}{E_{n'2}} \right] \right\}$$

$$\approx \quad \alpha \frac{-eE_0}{\omega} \hat{\varepsilon} \cdot \left[ \sum_{n \neq 1} \vec{\nabla}_{2n} \left( \sigma_y^{\uparrow\downarrow} x_{n1} - \sigma_x^{\uparrow\downarrow} y_{n1} \right) - \sum_{n' \neq 2} \left( \sigma_y^{\downarrow\uparrow} x_{2n'} - \sigma_x^{\downarrow\uparrow} y_{2n'} \right) \vec{\nabla}_{n'1} \right]$$

$$= \quad \alpha \frac{-eE_0}{\omega} \hat{\varepsilon} \cdot \left[ \left( \vec{\nabla} x \right)_{21} \sigma_y^{\uparrow\downarrow} - \left( \vec{\nabla} y \right)_{21} \sigma_x^{\uparrow\downarrow} - \left( x \vec{\nabla} \right)_{21} \sigma_y^{\downarrow\uparrow} + \left( y \vec{\nabla} \right)_{21} \sigma_x^{\downarrow\uparrow} \right]$$

$$= \quad \alpha \frac{-eE_0}{\omega} \left[ \hat{\varepsilon}_x \left( -\langle 2|x \frac{d}{dx}|1 \rangle \sigma_y^{\downarrow\uparrow} + \langle 2|y \frac{d}{dx}|1 \rangle \sigma_y^{\uparrow\uparrow} \right) + \hat{\varepsilon}_y \left( -\langle 2|x \frac{d}{dy}|1 \rangle \sigma_y^{\downarrow\uparrow} + \langle 2|y \frac{d}{dy}|1 \rangle \sigma_x^{\downarrow\uparrow} \right) \right].$$

Finally, we arrive at the readout frequency in an arbitrarily-oriented magnetic field,

$$|\nu_R| = \frac{\alpha e E_0}{2\pi \left( \hbar \nu_{12} \right)} f(\theta, \phi) \left( \hat{\varepsilon}_x \left| \langle 2| \, y \partial_x \, |1 \rangle \right| + \hat{\varepsilon}_y \left| \langle 2| \, y \partial_y \, |1 \rangle \right| \right),$$

where

$$f(\theta, \phi) = \sqrt{\cos^2 \phi \cos^2 \theta + \sin^2 \phi}$$

We see strong anisotropy dependence on the direction of the magnetic field. With the B-field in the $\pm x$ direction in the $x - y$ plane, the excitation rate goes to zero in first order (this is for an elliptical dot with $y > x$). This anisotropy is characteristic of the Rashba SOC Hamiltonian and may provide a good experimental test of our theory.

## 5.3  Motional spin-charge transduction with no spin-flip required

We now present a modified scheme which increases readout speed over 1000 times by causing spin-dependent charge motion *without a spin-flip*. The greatest limitation of our original readout proposal was the bottleneck during the spin-flip excitation. Since the spin-orbit mixing in silicon is so small, this transition rate was quite slow, of order MHz in the best design. Even in GaAs quantum wells, which have almost 100 times the magnitude of SOC by some estimates, the spin-flip excitation may be a bottleneck if it is slower than the detection speed of the rf-SET.

Imagine though if the Zeeman splittings of the ground and first excited states were different. In this case, the non-spin-flip microwave excitation between either pair of levels (up-up or down-down) would be different in magnitude. Thus, for example, a resonant microwave pulse of $\nu_{up}$ (see Figure 5.6) would cause charge motion *only* if the electron was measured to be in the up

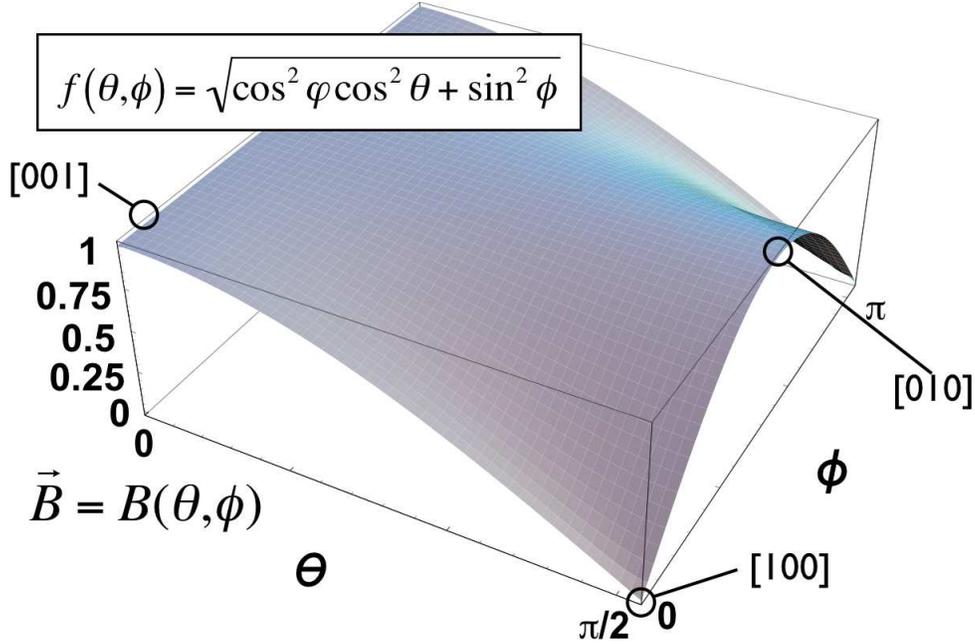

$$f(\theta,\phi) = \sqrt{\cos^2\varphi\cos^2\theta + \sin^2\phi}$$

Figure 5.5: Magnetic angular dependence of readout speed for an asymmetric dot with $y > x$. As seen from the figure, there will be suppressed spin-flip excitation (due to Rashba spin-orbit coupling) if the magnetic field is along the $x$ direction.

state originally. This design allows for selective readout without a spin-flip.

There are two ways the Zeeman splittings of the two lowest orbital states can be different. One possibility is that the electron g-factors could be different; the other possibility is that the magnetic fields are different. The latter is the simplest approach, and the one we take. As depicted in Figure 5.6, by adding a lateral magnetic field gradient (by a local wire, for example) we can use the asymmetry of our quantum dot and note that the COC of each orbital level feels a different average magnetic field strength.

In the new readout design, qubit initialization is not automatic. If, as we have assumed, $\nu_{up}$ is our readout probe frequency, there are two possible situations to consider. Say the qubit is measured in its ground, *up*, state. In this case, the new scheme operates as the old, and fast, like-spin orbital relaxation initializes the qubit after the readout has occurred. When the qubit is in its excited spin state, *down*, however, the qubit is left (strongly measured) in that state, where the long $T_1$ time means it will stay there for some time. Therefore, although the spin is well known by the experimenter in this case, to fully achieve polarization a spin-rotation or optical pumping initialization (as in the old scheme) must then be performed. This adds another step to spin polarization in the new readout design.

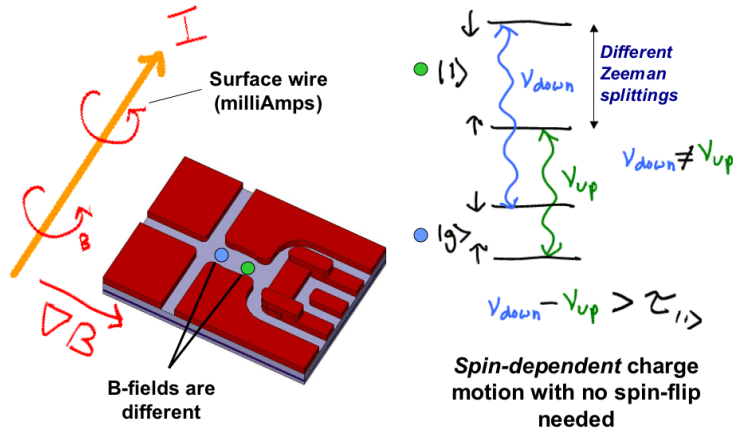

Figure 5.6: Modified readout scheme. A lateral magnetic field gradient (created by a local wire for example) causes different Zeeman splittings in the ground and first excited state of the dot. This makes the spin up-to-up and down-to-down transition frequencies different, allowing for spin-dependent charge motion and readout.

## 5.4  Future work needed

We have considered in first order the effects of a radiation field on quantum dot states to justify our device proposal. But our analysis is naive from the perspective of the established fields of quantum optics and atomic physics. Although our derivations for spin-flip excitation represent a new contribution to the field of condensed matter physics, they are only a first step in analyzing the feasibility of our readout/initialization concepts. For example, our derivation of the readout (Rabi) frequency is only accurate in a very weak field where the AC Stark effect is negligible. We also neglect to properly deal with the interaction of the light field with the full four (or more) level system, including decay rates. This would normally be done in a rate equation or density matrix formulation. If we are serious about understanding the interaction of quantum dot states in a strong microwave field for qubit control we must take thes higher order physics into account.

### 5.4.1   The Optical (AC) Stark effect

Let's consider the interaction of a two level quantum system with an arbitrarily strong field. This is a common example given in quantum optics texts [57]. But it has significant relevance to our original readout scheme so it is worth re-deriving here. We show below how the two electron levels of an atom in an intense field split into four levels.

We begin with the Schroedinger equation for a single electron in an atom or quantum dot,

$$H_0 \psi_n(\mathbf{r}) = \hbar \epsilon_n \psi_n(\mathbf{r}),$$

with

$$H_0 = -\frac{\hbar^2 \nabla^2}{2m_0} - \frac{e^2}{r}.$$

The light field introduces time-dependent changes of the wave function,

$$i\hbar \frac{\partial \psi(\mathbf{r}, t)}{\partial t} = [H_0 + H_I]\psi(\mathbf{r}, t),$$

with (in the electric dipole approximation),

$$H_I(t) = -exE(t) = -dE(t).$$

Expanding the electron wave function in the eigenfunctions of the bare Schroedinger equation,

$$\psi(\mathbf{r}, t) = \sum_m a_m(t) e^{-i\epsilon_m t} \psi_m(\mathbf{r}),$$

we can derive the series of coupled, differential equations

$$i\hbar \frac{da_n}{dt} = -E(t) \sum_m e^{-i\epsilon_{mn} t} \langle n|d|m \rangle a_m.$$

Returning to our two level system, for $\epsilon_2 > \epsilon_1$, we see that

$$i\hbar \frac{da_1}{dt} = -E(t) e^{-i\epsilon_{21} t} d_{12} a_2$$

and

$$i\hbar \frac{da_2}{dt} = -E(t)e^{-i\epsilon_{21}t}d_{21}a_1,$$

where we used $d_{ii} = 0$. Assuming a simple monochromatic field of the form

$$E(t) = \frac{1}{2}E(\omega)[e^{-i\omega t} + c.c.],$$

we rewrite to find

$$i\hbar \frac{da_1}{dt} = -d_{12}\frac{E(\omega)}{2}\left[e^{-i(\omega+\epsilon_{21})t} + e^{i(\omega-\epsilon_{21})t}\right]a_2$$

and

$$i\hbar \frac{da_2}{dt} = -d_{21}\frac{E(\omega)}{2}\left[e^{-i(\omega+\epsilon_{21})t} + e^{i(\omega-\epsilon_{21})t}\right]a_1.$$

In the rotating wave approximation we are left with

$$i\hbar \frac{da_1}{dt} = -d_{12}\frac{E(\omega)}{2}e^{i(\omega-\epsilon_{21})t}a_2$$

and

$$i\hbar \frac{da_2}{dt} = -d_{21}\frac{E(\omega)}{2}e^{i(\omega-\epsilon_{21})t}a_1.$$

Differentiating and substituting, we find that

$$\frac{d^2a_2}{dt^2} = i\frac{d_{21}E(\omega)}{2\hbar}\frac{da_1}{dt} = -\left|\frac{d_{21}E(\omega)}{2\hbar}\right|^2 a_2 = -\frac{\omega_R^2 a_2}{4},$$

where $d_{21} = d_{21}^*$ and

$$\omega_R = \frac{|d_{21}E_0|}{\hbar}$$

is the characteristic Rabi frequency that we have calculated before (again assuming zero detuning). The solution is of the form

$$a_2(t) = a_2(0)e^{\pm i\omega_R t/2}$$

and the same for $a_1$. Finally, inserting these into our expansion, we get

$$\psi(\mathbf{r}, t) = a_1(0)e^{-i(\epsilon_1 \pm \omega_R/2)t}\psi_1(\mathbf{r}) + a_2(0)e^{-i(\epsilon_2 \pm \omega_R/2)t}\psi_2(\mathbf{r}).$$

So the original frequencies $\epsilon_1$ and $\epsilon_2$ have been split into two levels each, separated by energy $\omega_R$. We write then the solution for finite detuning for reference

$$\epsilon_1 \rightarrow \epsilon_1 + \frac{\nu}{2} \pm \frac{1}{2}\sqrt{\nu^2 + \omega_R^2}$$

and

$$\epsilon_2 \rightarrow \epsilon_2 - \frac{\nu}{2} \pm \frac{1}{2}\sqrt{\nu^2 + \omega_R^2}$$

where $\nu = \epsilon_{21} - \omega$.

It is evident from this derivation that the AC Stark effect comes into play when $\omega_R$ is greater than the natural linewidth of the levels in question. We have already calculated these time scales for readout. For the like-spin transition, $\omega_R = 2\pi\frac{e}{\hbar}\sqrt{\frac{2I}{c\epsilon_0\sqrt{\epsilon_r}}}\langle 2|y|1\rangle \approx 6\sqrt{I} \times 10^9$ Hz, giving an AC Stark shift splitting of $E_R \approx \sqrt{I} \times 4$ $\mu$eV. For the original (with spin-flip) scheme, $E_R \approx 0.3$ meV. This is not only bigger than the natural lineiwdth of the level, it's also bigger than the splitting of the ground and first excited state in the proposed dot! Since both of these levels are split by this amount, this may not be disastrous. The bigger worry is that even with a detuning of $g\mu B$, the like-spin transition is much bigger than that spin-flip transition. Spin-dependent charge motion would then be impossible. The modified scheme, at first glance, is immune to these considerations. But it is clear that a quantum analysis of the full four-level system interacting with a strong light field is needed to make a convincing case for the scheme's feasibility.

## 5.5  Discussion and Implications

We have proposed to use microwave excitation to initialize and readout a spin qubit housed in a lateral silicon quantum dot. The scheme has yet to be verified experimentally. The idea of using optical pumping to initialize quantum systems is an old one in the atomic community and applied to condensed matter systems will undoubtedly find much use in the field of quantum information processing. Our calculations show that qubit polarization by optical pumping in silicon systems is very efficient and may provide a good source of initialized qubits for quantum computation and error correction.

Our novel proposal to use optical transitions to induce charge motion rests on less solid footing. Although preliminary analysis of transition frequencies and the amount of charge induced on a nearby rf-SET indicate that the scheme is workable, a more serious quantum treatment is needed to accurately predict the effects of an intense light field on the four level quantum dot system. This remains to be done.

# Chapter 6

# Conclusions and Outlook

In 1957 Abrahams [1], Pines, Bardeen, and Slichter [61]—no slouches in semiconductor spin physics—calculated the dominant spin relaxation mechanism for donors in bulk silicon. When the experiments were eventually done by Feher and colleagues a few years later [28, 29, 89], those predictions were shown to be off by more than a few orders of magnitude. The electron's complex environment in a semiconductor matrix leads to any number of possible decoherence mechanisms, which may or may not dominate in a certain working regime. Only tightly-coupled theoretical and experimental investigations will give us a clear picture of the spin qubit dynamics in a potential silicon quantum computer. A measurement of single spin decoherence (either $T_1$ or $T_2$) in a single or few-electron silicon quantum dot has yet to be achieved.

This work has taken the philosophy of justifying our theoretical calculations by building off experimentally confirmed theories in similar systems, namely P donors in bulk silicon, GaAs quantum dots, and silicon 2DEGs. Our results are very promising for spin-based quantum dot quantum computing in silicon. We have shown that the dominant spin relaxation mechanism in silicon donors—due to phonons and bulk spin-orbit coupling, both quite weak to begin with—is greatly diminished with large [001] compressive strain, exactly as is found in silicon quantum wells. Using electron spin resonance experiments in silicon 2DEGs, we were able to identify and quantify a new spin-orbit coupling term, parameterized by the Rashba or structural inversion asymmetry coefficient, and use it to develop a semi-classical theory for spin relaxation in silicon 2DEGs. Although this theory is incomplete and some samples are still not well explained, we believe it is good enough to extract a reasonable estimate of the Rashba coefficient in typical silicon 2DEG structures. The magnitude of the Rashba spin-orbit coupling allowed us to calculate the additional source of spin relaxation it created. At the present time, this is the dominant spin-flip mechanism in these devices. We found $T_1$ times for typical proposed quantum dot qubits that ranged from microseconds to minutes. Our theory makes key, experimentally-verifiable predictions for the $T_1$ dependence on magnetic field strength and quantum dot size. We found that simply making the quantum dot smaller will significantly increase spin relaxation time. In the process of these calculations we also developed a more complete theory for the relaxation of excited orbital states in strained silicon devices. These results are important for calculating higher order decoherence and control processes in silicon quantum dots. Finally, we proposed a scheme for readout and fast initialization in a semiconductor quantum dot using microwave excitation. It has the benefits of needing no external qubits and providing a very fast stream of polarized qubits. This has yet to be demonstrated experimentally.

Many open questions remain in the study of qubit decoherence in semiconductor architectures. Our analysis has focused on relaxation processes which involve emission of a phonon to the crystal environment. While in some cases this $T_1$ time may be of the same order as the $T_2$ time [59, 86], generally the phase decoherence time $T_2$ will be shorter. For fault tolerant quantum computing time scales, $T_2$ is the more important single spin parameter. $T_1$, however, has more immediate experimental relevance in that it is easier to measure in single quantum dot transport measurements. It also plays an important role in many qubit control dynamics. There is reason to hope that $T_1 \sim T_2$ in isotopically-purified silicon quantum dots. Decoherence mechanics due to Si$^{29}$ nuclear flip-flopping has been well characterized recently for GaAs quantum dots and silicon donor qubits [17]. It is clear that this mechanism goes away as the percentage of spin-1/2 nuclei goes to zero. After this, it is unknown what dominates $T_2$ decoherence. Bulk experiments of silicon donors in isotopically-purified silicon suggest that indeed $T_1 \sim T_2$ across a wide temperature range until $T_2$ pulls away [83] at a critical temperature. In this case, magnetic dipole-dipole coupling becomes dominant, as the donors are still quite near each other. This coupling is well characterized and can be refocused or even utilized for computation. Other mechanisms may appear. Charge fluctuations in the top-gates or $1/f$ noise from shifting surface states may be culprits. If, in the end, spin-orbit coupling together with phonons turns out to be the dominant decoherence mechanism in ultra-pure silicon quantum dots, our calculations will have direct relevance. Present knowledge indicates that individual silicon spins are excellent candidates for quantum storage.

The interaction of a light field with an electron-populated quantum dot is uncharted territory in solid state physics. We began a consideration of the interaction terms involved in Chapter 5. In the same chapter we proposed a novel scheme for spin-charge transduction that may find use in single-spin measurement and quantum computer readout in the future. More work is needed if microwave control is to be a viable qubit control mechanism. Issues of quantum optics, heating, and focusing must all be addressed.

Looking forward, we must go beyond the single spin picture in our decoherence analysis. A good first step is considering the relaxation properties of two coupled spins during an exchange operation. The singlet-triplet relaxation time as well as the orbital relaxation of a two quantum dot charge qubit in silicon can be calculated with relatively straight forward extensions to the theory developed here. Experiments to corroborate these predictions are on the horizon, first in GaAs devices and eventually in silicon. These considerations lead naturally to the greater question of how multiple, entangled qubits decohere. As we noted in Chapter 2, the density matrix that describes a quantum computer of $n$ qubits will generally have unique decoherence times for each of the off-diagonal elements. It is an open experimental question at present whether these are related or are much shorter than the single qubit phase coherence times of the individual spins and whether any error correction scheme can deal with this or not. We should build a quantum computer and find out.